\documentclass[%
preprintnumbers,
nofootinbib,
 amsmath,amssymb,
 aps,
prd,
]{revtex4-1}

\usepackage{float}
\usepackage{mathrsfs}
\usepackage{amssymb} 
\usepackage{comment}
\usepackage{graphicx,amssymb,slashed}
\usepackage{dsfont}
\usepackage{subcaption}
\usepackage[justification=raggedright]{caption}
\usepackage{dcolumn}
\usepackage{bm}
\usepackage{hyperref}
\hypersetup{
  colorlinks   = true, 
  urlcolor     = blue, 
  linkcolor    = blue, 
  citecolor   = red 
}

\usepackage{slashed}
\usepackage{parskip}

\def\be{\begin{equation}}
\def\ee{\end{equation}}
\def\bea{\begin{eqnarray}}
\def\eea{\end{eqnarray}}
\def\bear{\begin{array}}
\def\ear{\end{array}}
\def\bfig{\begin{figure}}
\def\efig{\end{figure}}
\def\bcen{\begin{center}}
\def\ecen{\end{center}}
\def\bi{\begin{itemize}}
\def\ei{\end{itemize}}

\usepackage{bm}
\usepackage{xcolor}
\usepackage{multirow}
\usepackage{slashed}
\usepackage{physics}
\usepackage{makecell}

\newcommand{\mpi}{m_{\pi}}

\begin{document}

\title{Electromagnetic isovector form factors of the transition from the $N^*(1520)$ to the nucleon}

\author{Di An}
\author{Stefan Leupold}%
\affiliation{%
Institutionen f\"or fysik och astronomi, Uppsala universitet, Box 516, S-75120 Uppsala, Sweden 
}%

\date{\today}

\begin{abstract}
Dispersion theory is used to provide a model-independent low-energy representation of the three electromagnetic isovector transition form factors $N^{*}(1520)\to N$. At low energies the virtual photon couples dominantly to a pion pair. Taking the very well understood pion vector form factor and pion re-scattering into consideration, the determination of the transition form factors is traced back to the determination of pion-baryon scattering amplitudes. Their low-energy aspects are parametrized by baryon exchange, accounting for the main decay channels of the $N^*(1520)$. Short-distance physics is encoded in subtraction constants that are fitted to data on space-like form factors and hadronic decays. It is shown that a limitation in the determination of the subtraction constants lies in the fact that isovector form factors require sufficient information about the differences between protons and neutrons. In particular, this calls for improvements in the form factor extraction from the electroproduction of the $N^*(1520)$ on the neutron. 
Via the dispersion relations, space- and time-like regions are naturally connected from first principles. This allows to predict the time-like form factors that enter the Dalitz decays $N^*(1520)\to N e^- e^+$ and $N^*(1520)\to N \mu^- \mu^+$. Under the assumption of the dominance of the isovector over the isoscalar channel, the Dalitz decay distributions are predicted. 

\end{abstract}

\maketitle


\section{Introduction}

Electromagnetic form factors (FFs) for elastic and transition reactions provide important structural information on the nucleon and its excited states \cite{Pascalutsa:2006up,Aznauryan:2011qj,Punjabi:2015bba,Eichmann:2018ytt,Lin:2021umz,Ramalho:2023hqd}. The present work focuses on the isovector transition FFs (TFFs) from the nucleon to its first negative-parity spin 3/2 excitation, the $N^*(1520)$. A generalization of the methods presented here to other nucleon excitations is straight forward. Besides the quest to explore the structure of the nucleon, a second motivation to focus on the $N^*(1520)$ is its importance for the description of electromagnetic radiation from hot and dense strongly interacting matter \cite{Peters:1997va,Rapp:1999ej,Krivoruchenko:2001jk,Zetenyi:2001fu,Post:2003hu,Rapp:2009yu,Salabura:2020tou}. 

To achieve a proper qualitative understanding of the FFs, one needs to identify first the 
relevant degrees of freedom and then reveal 
their dynamics. Quantitatively, a full understanding requires the use of first-principle methods related to QCD or to quantum field theory in general. Perturbative QCD, lattice QCD, dispersion theory, and effective field theories are among such methods. In the following, we shall put these and other methods into the context of the determination of FFs.

At high energies, the virtual photon resolves the quark structure of the hadrons. Due to asymptotic freedom, 
it is the minimal quark content that is probed in the exclusive reactions that provide the FFs \cite{brodskey}. Based on our understanding of perturbative QCD and factorization, the high-energy region is much better understood than the low-energy region. Therefore, the focus of the present work is on the low-energy region.

As a result of confinement and dynamical chiral symmetry breaking, the relevant degrees of freedom at (very) low energies are the baryons and the Goldstone bosons. The latter are the very light pions (if one focuses on hadrons built from the lightest two quark flavors). Based on the Goldstone theorem, their interaction with baryons and with themselves scales with their momenta and is therefore suppressed at low energies. This allows for a systematic derivative expansion, which gives rise to chiral perturbation theory (ChPT) as the effective field theory of low-energy QCD \cite{Weinberg:1978kz,Gasser:1984gg,Gasser:1983yg,Scherer:2012xha}. 
The dynamics of the Goldstone bosons is well established by the framework of a nonlinear realization \cite{Coleman:1969sm,Callan:1969sn} of the chiral symmetry of QCD. ChPT has been proven to be very successful phenomenologically \cite{Gasser:1984gg,Gasser:1983yg,Scherer:2012xha,Scherer:2002tk}. Indeed, at very low energies it can provide a decent description of the FFs\cite{Gasser:1987rb,Kubis:2000aa,Scherer:2012xha} and TFFs \cite{Hilt:2017iup,Unal:2021byi} for the lowest-lying spin 1/2 and spin 3/2 states. 

Moving towards higher energies, the strength of the pionic interactions grows more and more. At some point, one leaves the applicability regime of the derivative expansion of ChPT. As a very important effect, related to confinement, the quarks build hadronic resonances, which often dominate reaction amplitudes. The importance of resonances for interactions with and between hadrons gave rise to the concept of resonance saturation and to the development of isobar models; see, e.g., \cite{Ecker:1988te,Donoghue:1988ed,Anisovich:2004zz,Hanhart:2012wi} and references therein. Concerning electromagnetic FFs, the virtual photon can couple to single hadrons with vector quantum numbers. For isovector FFs these are the $\rho$ meson and its excited states. Vector-meson dominance is the concept that stresses the importance of resonances for the case of electromagnetic interactions \cite{feynmanVMD,sakuraiVMD}. Various hadronic models are based on this concept \cite{Friman:1997tc,Peters:1997va,Krivoruchenko:2001jk,Zetenyi:2001fu,Zetenyi:2012hg}. They are complemented by other methods, e.g.\ QCD sum rules \cite{sumrule,sumrule2,sumrule3}, quark models \cite{PDG,quarkmodelold1,Ramalho:2013mxa,Ramalho:2016zgc,Ramalho:2023hqd}, and functional methods \cite{Eichmann:2016yit}, which all became very useful in understanding hadrons and their FFs at medium energies.

Are there model-independent alternatives that allow to include the physics of the $\rho$ meson into the description of vector-isovector FFs? This is indeed possible based on dispersion theory, a method that relates via the optical theorem unknown to known reaction amplitudes \cite{Donoghue:1990xh,Donoghue:1996kw,Niecknig:2012sj,Kang:2013jaa,Leupold:2017ngs}. In fact, the $\rho$ meson is an elastic resonance. It couples practically to 100\% to two pions. Therefore, the pertinent pion phase shift contains the information about the $\rho$ meson in a model-independent way. This pion p-wave phase shift and also the coupling of pions to virtual photons is very well known \cite{Colangelo:2001df,GarciaMartin:2011cn,Hanhart:2012wi}. 

Note that the situation is far less satisfying for the higher-lying energy regime of the excited vector mesons (above 1 GeV). This regime is governed by three- and more-particle states and by many coupled channels. The corresponding many-particle amplitudes, which would be of interest for a dispersive treatment, are just not available. Therefore our focus is on the intermediate energy regime where the $\rho$ meson matters, i.e.\ beyond pure ChPT, but where the details of multi-particle and multi-channel effects do not play a role yet. In the spirit of scale separation (the underlying concept of renormalization and of effective field theory), we will encode these unresolvable higher-energy effects in subtraction polynomials of our dispersion relations.

A complementary first-principle approach is lattice QCD \cite{Gattringer:2010zz}. This most popular non-perturbative method has made significant progress in recent years. But it faces challenges in accurately describing FFs for the transition of an unstable resonance to a ground-state baryon, in particular at physical pion masses \cite{Leinweber:1992pv,Alexandrou:2005em}. A proper description of an unstable state requires not only an interpolating current with its minimal quark content, but also the inclusion of the multi-particle states to which the unstable state decays \cite{Luscher:1990ux,Dudek:2012xn,Polejaeva:2012ut,Mai:2021lwb}. For the envisaged $N^*(1520)$ this includes the rather important \cite{PDG} three-particle channels of $\pi$-$\pi$-$N$. The corresponding inelastic five-point functions needed for the TFFs are numerically and conceptually beyond present-day capabilities. 
These problems appear on top of the usual Euclidean-space limitation that makes it impossible to calculate directly FFs in the time-like region.

On the other hand, starting with the optical theorem, one can derive dispersion relations that relate scattering amplitudes and FFs among different kinematical regions.
\begin{figure}[h!]
\centering
    \includegraphics[height=0.15\textheight]{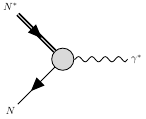}

    \caption{Illustration of an electromagnetic TFF for the transition of the $N^{*}$(1520) resonance to the nucleon. The excitation of the nucleon is denoted by $N^*$, and the virtual photon that couples to the TFF is denoted by $\gamma^*$.}
    \label{fig:N1520 TFFs}
\end{figure}

In the present work, we follow to quite some extent the formalism established by the Uppsala group for hyperon TFFs \cite{Granados:2017cib,Junker:2019vvy}. The nucleon FFs have been addressed in \cite{Leupold:2017ngs,Alvarado:2023loi}. The nucleon-to-Delta TFFs have been explored in \cite{Aung:2024qmf}. All these studies concerned the respective ground-state baryons of spin 1/2 and 3/2. In this work we address transitions of the nucleon to a nucleon excitation, the $N^*(1520)$. A diagrammatic representation of the TFFs is shown in Fig.\ \ref{fig:N1520 TFFs}. For quark models studying the transition $N^{*}(1520)\to N \gamma^{*}$, we refer to \cite{Ramalho:2013mxa,Ramalho:2016zgc,tailandgroup,Santopinto:2012nq,Aiello:1998xq,Aznauryan:2017nkz}, and for vector meson dominance models to \cite{Krivoruchenko:2001jk,Zetenyi:2001fu,Zetenyi:2012hg}.

On the experimental side, there are four distinct regions that are kinematically allowed. The corresponding dependence of the TFFs on the photon virtuality $q^2$ can be explored in different experiments \cite{Aznauryan:2011qj,Aznauryan:2012ba}: 
\begin{enumerate}
  \item Space-like $q^2<0$:\\
Accessible in the $e^{-}N\to e^{-} N^{*}$ process, which is explored, e.g., at Jlab \cite{CLAS:2009ces,Mokeev:2015lda,jlabdata3}. 
\item Photon-point at $q^2=0$: \\
Accessible in the photo-production process $\gamma N \to N^*$, measured, e.g., at ELSA \cite{CB-ELSA:2004sqg,GDH:2002pkk,CBELSATAPS:2015kka,Dugger:2007bt}.
\item Time-like $q^2>0$:
  \begin{enumerate}
   \item Dalitz decay region  $(m_{N^{*}}-m_N)^2>q^2>4m_\ell^2$; accessible in $N^{*}\to \ell^{+} \ell^{-} N$ processes. Here $\ell$ denotes an electron or muon. The Dalitz decay region can be explored by the HADES experiment \cite{Ciepal:2024yub}.
    \item scattering region $q^2>(m_{N^{*}}+m_N)^2$; accessible in $e^{+}e^{-}\to \gamma^{*}\to X$ processes, where $x=N^* \bar{N}, N \bar{N^*}$. This region is accessible with BESIII \cite{BESIII:2020nme} and Belle II \cite{Belle-II:2018jsg}.
 \end{enumerate}
\end{enumerate}

Based on dispersion relations, we 
will present, firstly, a decent description of the current space-like 
TFF data from
Jlab. Secondly, this work will provide a model-independent prediction for the TFFs of $N^{*}(1520)\to N \gamma^{*}$ in the time-like region using hadronic input and the constraints from the space-like region imposed on the dispersion relations.

The paper is structured as follows.
In Section \ref{sec:Transition form factor definitions and constraints}, the definitions of the unconstrained TFFs $F_{i}$ will be introduced and their relation to helicity amplitudes $H_{0,\pm 1}$ will be derived. The differential decay width of the reaction $N^*(1520) \to N \, l^+ l^-$ (with $\ell = e,\mu$) and the real photon decay width $\Gamma_{N\gamma}$ will be given in terms of the helicity amplitudes $H_{0,\pm 1}$. 
The dispersive framework will be presented in Section \ref{section:disp machinery}. It relates the TFFs to the pion vector FF and pion-baryon scattering amplitudes. Re-scattering among the pions will also be taken care of by dispersion theory. 
In practice, those pion-baryon amplitudes will be constructed from one-baryon exchange processes (plus pion re-scattering). 
In Section \ref{sec: hadronic input} we will introduce the corresponding phenomenological Lagrangians for the exchanged baryons, i.e.\ we will provide the necessary three-point interactions of $N^*(1520)$-$\pi$-$N$ and $N^*(1520)$-$\pi$-$\Delta(1232)$. All other three-point interactions are specified by ChPT input, which we will also review for completeness.
In Section \ref{sec:Projector formalism}, a projector formalism will be constructed where the general decay amplitudes $N^*(1520)\to N \pi^+ \pi^-$ will be decomposed into 4 hadronic helicity amplitudes. 
The partial wave projection and the analytic continuation of the partial-wave amplitudes will be discussed in Section \ref{sec: Partial-wave projection}. The final ingredient are the subtraction constants of the dispersive integrals. They encode the information about the physics that is not resolved by our framework. In Section \ref{sec: results}, we will determine these subtraction constants and the three-point couplings by fits to data on hadronic two-body decays and space-like TFFs. In the same section we will present our final results for the TFFs in the time- and space-like region. Summary and outlook will be provided in Section \ref{sec:summary}. Several technical aspects will be covered by appendices.

\section{Transition form factors, helicity amplitudes and kinematical constraints}
\label{sec:Transition form factor definitions and constraints}

$N^{*}(1520)$ is an excited state of the nucleon, obtained by an angular-momentum excitation of a quark. The quantum numbers are $I(J^{P})=1/2(3/2^{-})$. In this section, we introduce our definition of the TFFs and the corresponding helicity amplitudes together with their constraints. For completeness, we discuss also the high-energy behavior of these quantities. We start with the time-like region of Dalitz decays and turn then to the space-like region of electro-production. 

\subsection{Dalitz decay}
\label{sec:dalitzTFFs}

Similar in spirit to \cite{Korner:1976hv,Holmberg:2019ltw,Junker:2019vvy,Salone:2021bvx}, we introduce the transition form factors (TFFs) for the process $N^*(1520) \to N \gamma^*$ by
\begin{eqnarray}
  \label{eq:def-transFF}
  \langle N \vert j_\mu \vert N^* \rangle = 
  e \, \bar u_N(p_N,\lambda_N) \, \Gamma_{\mu\nu}(q) \, u^\nu_{N^*}(p_{N^{*}},\lambda_{N^*})
\end{eqnarray}
with
\begin{equation}
    \begin{split}
  \Gamma^{\mu\nu}(q) :=& i \left(\gamma^\mu q^\nu - \slashed{q} g^{\mu\nu} \right) m_N F_1(q^2) 
  + \sigma^{\mu\alpha} q_\alpha q^\nu F_2(q^2) + i \left(q^\mu q^\nu - q^2 g^{\mu\nu} \right) F_3(q^2)   
    \end{split}
          \label{eq:def-TFF-Gam}
\end{equation}
and $q^{\mu}:=p_{N^{*}}^{\mu}-p_{N}^{\mu}$. In \eqref{eq:def-transFF}, the $\lambda_B$ denotes the respective helicity ($B=N,N^*$). Spinors are defined in Appendix \ref{app:convent-spinors}.

The quark electromagnetic current $j_{\mu}$ is defined as
\begin{equation}
    j_{\mu}:=\sum_{q=u,d}Q_{q}\, e \, \bar{q}\gamma_{\mu}q
\end{equation}
and the $Q_q$ are the fractional charges of the quarks in units of the proton charge (i.e.\ $Q_u=2/3$,  $Q_d=-1/3$).  Note that the proton charge $e$ appears explicitly in \eqref{eq:def-transFF} and therefore does not appear in the unconstrained TFFs denoted by $F_1$, $F_2$, $F_3$. 

The $N^{*}(1520)$ transforms as a doublet  $(N^{*+},N^{*0})^{T}$ under isospin. We will refer to the FFs of the transition $N^{*+} \, (N^{*0}) \to p \, (n)$ as $F^{p}_{i}$ ($F^{n}_{i}$). We introduce also the isovector and isoscalar TFFs as the linear combination of $F^{p}$ and $F^{n}$ given as
\begin{equation}
\begin{split}
 F^{v}_{i}&:=(F^{p}_i-F^{n}_i)/2 \,,  \\ 
 F^{\text{s}}_{i}&:=(F^{p}_i+F^{n}_i)/2
\end{split}
\label{eq def isovector and isoscalar ffs}
\end{equation}
with $i=1,2,3$.
 In this paper we focus solely on the isovector TFFs. The two-pion intermediate state featured by our dispersive low-energy representation has isovector quantum numbers. To determine the isoscalar TFFs would require the treatment of the three-pion states. The mesonic aspects of the virtual photon coupled to three pions have been addressed in \cite{Hoferichter:2014vra}. But the coupling of the three-pion states to baryons is beyond the scope of the present work. We will see to which extent this focus on the isovector part limits our comparison to data (which are much better for protons than for neutrons).

The TFFs $F_{1,2,3}$, introduced in \eqref{eq:def-TFF-Gam}, have been constructed with the method introduced by Bardeen, Tung, and Tarrach (BTT construction method) \cite{Bardeen:1969aw,Tarrach:1975tu}. In that way it is ensured that these TFFs are free of kinematical constraints and therefore are perfectly suitable for a dispersive representation.
On the other hand, partial-wave decompositions \cite{Jacob:1959at} are better formulated with helicity amplitudes. Consequently we introduce next the (electromagnetic) helicity amplitudes \cite{Carlson:1985mm}
\begin{eqnarray}
  \label{eq:helampl-TFF}
  H_{-m} \sim \varepsilon^\mu(q,m) \, \langle N(p_z, +1/2) \vert j_\mu \vert N^*(p_z, m + 1/2) \rangle  \,,
\end{eqnarray}
where the incoming $N^{*}$ and outgoing $N$ fly in the $z$ direction with momentum $p_z$, the virtual photon $\gamma^*$ sits at rest. In this frame, the helicities ($+1/2$ and $m+1/2$, respectively) agree with the spin orientation measured along the positive $z$-axis. The photon spin orientation $s_\gamma$ is also measured along this direction. The polarization vectors are given in Appendix \ref{app:convent-spinors}.

Note that we have chosen $-m$ instead of $m$ for the index of $H$ in \eqref{eq:helampl-TFF} to obtain agreement with the labels\footnote{This leads to a slight mismatch with the labels used in \eqref{eq:finalMs2} below.} used in \cite{Carlson:1985mm}. The normalization will be chosen such that the relations to the constraint-free TFFs do not
contain non-analytic terms like square roots and that the kinematical constraints are simple. The normalization factors are calculated explicitly in Appendix \ref{sec:Normalization factors}.

We find that the helicity amplitudes $H_{\pm,0}$ can be related to the constraint-free TFFs $F_{1,2,3}$ as 
\begin{eqnarray}
  H_+(q^2) &:= & \frac{1}{m_{N^*}} 
  \left[ (m_{N^*} m_N - m_N^2 + q^2) m_N F_1(q^2) - (m_{N^*}+m_N) \left((m_{N^*}-m_N)^2-q^2 \right) F_2(q^2) 
    + m_{N^*} q^2 F_3(q^2) \right]   \,, \nonumber \\ 
  H_0(q^2) &:= & \frac{m_{N^*}-m_N}{2 m_{N^*}} 
  \left[ 2 m_{N^*} m_N F_1(q^2) - \left((m_{N^*}-m_N)^2-q^2 \right) F_2(q^2)  + (q^2+m_{N^*}^2-m_N^2) F_3(q^2) \right] \,, \nonumber \\ 
  H_-(q^2) &:= & (m_{N^*}-m_N) m_N F_1(q^2) + q^2 F_3(q^2) \,.
  \label{eq:defG0pm}
\end{eqnarray}
These helicity amplitudes satisfy the following kinematical constraints at the decay threshold $q^2=(m_{N^*}-m_N)^2$ (Siegert theorem \cite{Ramalho:2016zzo}): 
\begin{eqnarray}
  \label{eq:two-constr}
  H_+(q^2=(m_{N^*}-m_N)^2) =   H_0(q^2=(m_{N^*}-m_N)^2) =   H_-(q^2=(m_{N^*}-m_N)^2)  \,,
\end{eqnarray}
while there is an additional constraint at the production threshold: 
\begin{eqnarray}
  && H_0(q^2 =(m_{N^*}+m_N)^2) = \frac{m_{N^*}-m_N}{2 (m_{N^*}+m_N)} \left[ H_+(q^2 =(m_{N^*}+m_N)^2) + H_-(q^2 =(m_{N^*}+m_N)^2) \right] \,.
  \label{eq:one-constr}
\end{eqnarray}

The physical reason for these constraints is the following \cite{Salone:2021bvx}. The point $q^2=(m_{N^*}-m_N)^2$ 
is at the end of the phase space of the decay $N^* \to N \gamma^*$. 
Here the momenta of the baryons vanish in the rest frame of the virtual photon. Close to this point, one can apply 
non-relativistic arguments. In particular, spin and orbital angular momentum are separately conserved and can be used to 
characterize the system. The process $3/2^- \to 1^- + 1/2^+$ can take place via an s-wave or one of two d-waves. At the 
end of the phase space, the d-waves become irrelevant, which makes all three helicity amplitudes 
degenerate in \eqref{eq:two-constr}. 

The point $q^2 =(m_{N^*}+m_N)^2$ indicates the threshold of the production process $\gamma^* \to N^*(1520) \bar N$. Here the baryons 
are at rest in the center-of-mass frame, i.e.\ the rest frame of the virtual photon. The process $1^- \to 3/2^- + 1/2^-$ can take 
place via one of two p-waves or via an f-wave. The latter becomes suppressed close to the threshold, which leaves only two 
independent helicity amplitudes, i.e.\ constrains the third amplitude via \eqref{eq:one-constr}. 

The differential decay width of the reaction $N^*(1520) \to N \gamma^* \to N \, \ell^+ \ell^-$ is given by
\begin{equation}
 \begin{split}
&\frac{d \Gamma}{dq^2 \, d\cos\theta} = \frac{e^4 p_z }{(2\pi)^3 \, 96 m_{N^*}^3 q^2} \frac{\sqrt{q^2}}{2} \beta_\ell
  \left( (m_{N^*}+m_N)^2 - q^2 \right)   \\
  & \times \left[
    \left(1+\cos^2 \theta + \frac{4 m_\ell^2}{q^2}\sin^2 \theta \right) \left(3|H_-(q^2)|^2+|H_+(q^2)|^2  \right) \right. 
 \\
  & \phantom{m} \left. {}+ \left(\sin^2 \theta + \frac{4 m_\ell^2}{q^2}\cos^2 \theta \right) \frac{4 q^2}{(m_{N^*}-m_N)^2}
  \vert H_0(q^2) \vert^2  \right]  \,.
  \label{eq:diffdec}  
   \end{split}  
\end{equation}
Here $\theta$ denotes the angle between the lepton $\ell$ and the nucleon in the rest frame of the virtual photon. The velocity of the 
lepton is introduced by
\begin{eqnarray}
  \label{eq:defbetae}
  \beta_\ell := \sqrt{1-\frac{4 m_\ell^2}{q^2}}  
\end{eqnarray}
with the lepton mass $m_\ell$. The momentum of $N^*(1520)$ and $N$ in the rest frame of the virtual photon is given by
\begin{equation}
    p_z:=\frac{\lambda^{\frac{1}{2}}(m_{N^{*}}^2,m_N^2,q^2)}{2\sqrt{q^2}}\,,
\end{equation}
where we have introduced the K\"all\'en function 
\begin{eqnarray}
  \label{eq:kallenfunc}
  \lambda(a,b,c):=a^2+b^2+c^2-2(ab+bc+ac) \,.
\end{eqnarray}

The width of the real photon decay $N^{*} \to N\gamma$ is given as
\begin{equation}
   \Gamma_{N\gamma}=  \frac{e^2 (m_{N^{*}}-m_{N}) (m_{N}+m_{N^{*}})^3 (3 |H_{-}(0)|^2+|H_{+}(0)|^2)}{96 \pi  m_{N^{*}}^3}.
   \label{eq real photon}
\end{equation}

Along the lines of \cite{Junker:2019vvy}, one can also define a QED version of the differential decay width where one replaces $H_{\pm}(q^2) \to H_{\pm}(0)$ and $q^2 \vert H_{0}(q^2) \vert^2 \to 0$. This leads to 
\begin{equation}
 \begin{split}
&\frac{d \Gamma_\text{QED}}{dq^2 \, d\cos\theta} = \frac{e^4 p_z }{(2\pi)^3 \, 96 m_{N^*}^3 q^2} \frac{\sqrt{q^2}}{2} \beta_\ell
  \left( (m_{N^*}+m_N)^2 - q^2 \right)   \\
  & \times 
    \left(1+\cos^2 \theta + \frac{4 m_\ell^2}{q^2}\sin^2 \theta \right) \left(3|H_-(0)|^2+|H_+(0)|^2  \right) \,.
  \label{eq:diffdec QED}  
   \end{split}  
\end{equation}
In the result section, we will compare our calculations to this QED version to figure out if Dalitz decay data are capable to reveal the intrinsic structure of the baryons. Note, however, that this QED version is not based on a Lagrangian. A viable alternative would be the use of (constant) constraint-free TFFs to define another QED version \cite{Salone:2021bvx}.

\subsection{Electro-production}
\label{sec:electro-prodTFFs}

In \eqref{eq:def-transFF}, the nucleon is in the final state, as appropriate for the Dalitz decay process. 
For our description of electro-production data, i.e.\ for $e^- N \to e^- N^*$, and to facilitate the comparison to other conventions, it is worth writing down the TFFs for the case of a nucleon in the initial state. Crossing and charge conjugation symmetry translate \eqref{eq:def-transFF}, \eqref{eq:def-TFF-Gam} to 
\begin{eqnarray}
  \label{eq:def-transFF-cross}
  \langle N^* \vert j_\mu \vert N \rangle = 
  e \, \bar u^\nu_{N^*}(p_{N^*}, \lambda_{N^*}) \, \Gamma'_{\mu\nu}(q) \, u_N(p_N,\lambda_N) 
\end{eqnarray}
with
\begin{eqnarray}
  \Gamma'^{\mu\nu}(q) &:=& -i \left(\gamma^\mu q^\nu - \slashed{q} g^{\mu\nu} \right) m_N F_1(q^2) 
  \nonumber   \\
  && {} + \sigma^{\mu\alpha} q_\alpha q^\nu F_2(q^2)
  - i \left(q^\mu q^\nu - q^2 g^{\mu\nu} \right) F_3(q^2)   \,.
  \label{eq:def-TFF-Gam-cross}
\end{eqnarray}
We still use $q=p_{N^*}-p_N$. 

Primarily, we are interested in \eqref{eq:def-transFF-cross} for $q^2 = (p_{N^*}-p_N)^2 < 0$ (even though $N \, e^+ e^- \to N^*$
is in principle possible). In this kinematical regime, we {\em cannot} choose the frame $\vec q = 0$. However, one can choose the
Breit frame \cite{Carlson:1985mm} defined by $\vec p_N + \vec p_{N^*} = 0$. 
We choose furthermore that $\vec p_{N^*}$ points in the $z$-direction and introduce the positive quantity $Q^2:=-q^2$.
This gives us 
\begin{equation}
 \label{eq:Breit}
    \begin{split}
& p_N=(E_N,0,0,-p_z) \,, \quad p_{N^*} = (E_{N^*},0,0,p_z) \,, \quad q = (E_{N^*}-E_N,0,0,2p_z)  
    \end{split}
\end{equation}
with
\begin{equation}
   \label{eq:ENetc-breitQ2}   
   \begin{split}
E_N &= \frac{3 m_N^2 + m_{N^*}^2 + Q^2}{2 \sqrt{Q^2+2 m_{N^*}^2+2 m_N^2}} \,,  \quad  \quad \quad 
  E_{N^*}= \frac{3 m_{N^*}^2 + m_N^2 + Q^2}{2 \sqrt{Q^2+2 m_{N^*}^2+2 m_N^2}} \,, \\
  p_z& = \frac{\lambda^{1/2}(-Q^2,m_{N^*}^2,m_N^2)}{2 \sqrt{Q^2+2 m_{N^*}^2+2 m_N^2}} =\frac{q_z}{2} \,, \quad
 q^0 = \frac{m_{N^*}^2-m_N^2}{\sqrt{Q^2+2 m_{N^*}^2+2 m_N^2}}       \,.
   \end{split}
\end{equation}

It should be clear that for the nucleon $N$ its spin component along the $z$-axis is now
opposite to its helicity $\lambda_N$. Conservation of angular momentum implies $\lambda_{\gamma^*}-\lambda_N = \lambda_{N^*}$.
Note that the kinematics \eqref{eq:Breit} does not provide any direction perpendicular to the $z$-axis. Thus there cannot be any
orbital angular momentum. 

Following \cite{Carlson:1985mm} we define alternative helicity amplitudes 
\begin{equation}
    \label{eq:carlson-hel}
    \begin{split}
    G_m := \frac{i}{2 e m_N}\varepsilon^\mu(q,m)   \,
  \langle N^*(p_z,\lambda_{N^*}=m-1/2) \vert j_\mu \vert N(-p_z,\lambda_N=+1/2) \rangle   \,.
      \end{split}
\end{equation}
We can relate them to our TFFs $F_i$ and helicity amplitudes $H_m$. In \eqref{eq:carlson-hel}, we need the polarization
vectors for the space-like virtual photon. They are provided in Appendix \ref{app:convent-spinors}. 
Note the extra factor of $i$ appearing in \eqref{eq:carlson-hel}. It is caused by a phase freedom when choosing fields and states for fermions and relates in practice to the charge conjugation properties of the $N^*$ field. This aspect is discussed in Appendix \ref{sec:C-conj}.

We find the following relations between our and the alternative helicity amplitudes:
\begin{eqnarray}
  G_+ &=& \frac{1}{2 m_N} \, \frac{1}{\sqrt 3} \, \sqrt{Q^2+(m_{N^*}+m_N)^2} \, H_+(-Q^2) \,,
  \nonumber \\[1em]
  G_0 &=& -\frac{Q}{2 m_N} \, \sqrt\frac23 \, \frac{\sqrt{Q^2+(m_{N^*}+m_N)^2}}{m_{N^*}-m_N} \, H_0(-Q^2)  \,,
  \nonumber \\[1em]
  G_- &=& \frac{1}{2 m_N} \, \sqrt{Q^2+(m_{N^*}+m_N)^2} \, H_-(-Q^2) \,.
  \label{eq:G0H0}
\end{eqnarray}

Although our main focus is the low-energy structure of the TFFs, it is worth noting their high-energy behavior. In particular, one can justify in that way that the TFFs satisfy unsubtracted dispersion relations. In turn, this means that additional subtractions help to strengthen the influence of the low-energy region in the dispersive integral \cite{Niecknig:2012sj,Schneider:2012ez,Hoferichter:2014vra}. 

Using quark counting rules for $Q^2 \to + \infty$ one can show that the helicity amplitudes 
scale as \cite{Carlson:1985mm} 
\begin{eqnarray}
  \label{eq:scaling-helG}
  G_+ \sim \frac{1}{Q^3} \,, \quad G_0 \sim \frac{1}{Q^4}  \,, \quad G_- \sim \frac{1}{Q^5}  
\end{eqnarray}
and therefore
\begin{eqnarray}
  \label{eq:scaling-hel}
  H_+\sim \frac{1}{Q^4}, \, \quad H_0\sim \frac{1}{Q^6},  \, \quad H_- \sim \frac{1}{Q^6}.  \,
\end{eqnarray}
This leads to
\begin{eqnarray}
  \label{eq:scaling-TFFs}
    F_1 \sim \frac{1}{Q^6}, \, \quad F_2 \sim \frac{1}{Q^8}, \,\quad  F_3 \sim \frac{1}{Q^8}.
\end{eqnarray}
Interestingly, we deduce from \eqref{eq:def-TFF-Gam-cross} and \eqref{eq:scaling-TFFs} that 
$F_1$ is dominant at low and high energies. 

In other works, e.g.\ \cite{Tiator:2009mt,Tiator:2011pw,Aznauryan2008u0defhelicity,Ramalho:2013mxa,Ramalho:2016zgc,Eichmann:2018ytt}, one also uses helicity amplitudes defined in the rest frame of $N^{*}(1520)$: 
\begin{equation}
\begin{split}
        A_{\frac12} & = -\frac{1}{\sqrt{2\kappa_{r}}}\frac{i}{2\sqrt{m_{N^{*}}m_{N}}} \bra{N^{*}(s_z^*=+1/2)}j_{+}\ket{N(s_z=-1/2)} \,,  \\
       A_{\frac32} & = -\frac{1}{\sqrt{2\kappa_{r}}}\frac{i}{2\sqrt{m_{N^{*}}m_{N}}}\bra{N^{*}(s_z^* = +3/2)} j_{+} \ket{N(s_z = +1/2)}  \,, \\
         S_{\frac12}&=\frac{1}{\sqrt{2\kappa_{r}}}\frac{i}{2\sqrt{m_{N^{*}}m_{N}}}\bra{N^{*}(s_z^* = +1/2)} \rho \ket{N(s_z = +1/2)}  
         \end{split}
\label{eq:def helicity amplitudes}
\end{equation}
with $j_+:=-\frac{1}{\sqrt{2}}(j_{x}+ij_{y})$, $\rho:=j^0 = \epsilon^{\mu}(q,0) j_{\mu}\frac{|\vec{q}|}{Q}$ and  
$\kappa_r := \frac{m_{N^*}^2-m_N^2}{2m_{N^*}}$. In \eqref{eq:def helicity amplitudes}, 
the $z$-axis is defined by the direction of motion of the virtual photon. Note that this means that the nucleon moves against the $z$-direction. Its spin projection on the $z$-axis is opposite to the nucleon's helicity.

With the factor of $i$ in \eqref{eq:def helicity amplitudes}, we have taken care of the fact that our resonance states show a different behavior with respect to charge conjugation as compared to the states used in \cite{Tiator:2009mt,Tiator:2011pw,Aznauryan2008u0defhelicity,Ramalho:2013mxa,Ramalho:2016zgc,Eichmann:2018ytt}; see the discussion in Appendix \ref{sec:C-conj}. There could be an extra sign between experimentally and theoretically defined helicity amplitudes \cite{Aznauryan2008u0defhelicity}. This issue will be discussed further when we specify our hadronic input in Subsection \ref{sec nstar eff lag}.  

We find that the relation between $A_{\frac{1}{2},\frac{3}{2}}$, $S_\frac{1}{2}$ and our helicity amplitudes $H_{\pm,0}$ is given by
\begin{eqnarray}
A_{\frac{1}{2}} &=& 
\sqrt{\frac{(m_N + m_{N^*})^2 + Q^2}{12m_N (m_{N^*}^2-m_N^2 )}} \, e H_+  
\,, \nonumber \\
S_\frac{1}{2} &=& 
- \sqrt{\frac{(m_N - m_{N^*})^2 + Q^2}{24m_N (m_{N^*}^2 - m_N^2)}} \, 
\frac{(m_N + m_{N^*})^2 + Q^2}{m_{N^*} (m_{N^*}-m_N )} \, e H_0
\,, \nonumber \\
A_{\frac{3}{2}} &=& 
\sqrt{\frac{(m_N + m_{N^*})^2 + Q^2}{4m_N ( m_{N^*}^2-m_N^2 )}} \, e H_-   \,. 
\label{eq translation Hs and As}    
\end{eqnarray}
The assignment to the different $H_{m} \sim G_m$ can be easily understood. As already noted we have $s_z = - \lambda_N$. Starting with the previously introduced Breit frame, the rest frame of the $N^*$ can be obtained by a boost in the negative $z$-direction. This boost does not flip the helicity of $N^*$, i.e.\ $s_z^* = \lambda_{N^*}$. The label of the helicity amplitudes \eqref{eq:carlson-hel} is given by $\lambda_{N^*} + \lambda_N = s_z^* -s_z$ for the case $s_z = - \lambda_N = -1/2$. For $A_{\frac12}$ this is the case at hand and we find immediately $A_{\frac12} \sim H_+$. For the other two amplitudes in \eqref{eq:def helicity amplitudes}, one must rotate the system by $180^\circ$. The label of the helicity amplitudes \eqref{eq:carlson-hel} is then given by $\lambda_{N^*} + \lambda_N = -s_z^* + s_z$, leading to $A_{\frac32} \sim H_-$ and, of course, $S_{\frac12} \sim H_0$.

\section{Dispersive machinery}
\label{section:disp machinery}

\begin{figure}[h!]
    \centering
  $\text{Disc}\vcenter{\hbox{   \includegraphics{Figures/NTFF.pdf}}}=
\vcenter{\hbox{\includegraphics{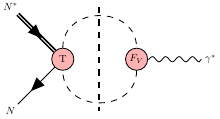}}}+ \ldots$
    \caption{Diagrammatic representation for the discontinuities along the unitarity cut. Dashed lines in the loop denote pions. The right-hand side of the equation displays explicitly the contribution from the two-pion cut. $F_V$ represents the pion vector FF and $T$ the pion-baryon reaction amplitude. The dots represent contributions from 
    $4\pi$, $K\bar{K}$, $K\bar{K}\pi$, $N\bar{N}$, $\ldots$.}
    \label{fig: disc representation}
\end{figure}

As has been explained previously, we calculate in this work the TFFs in a model independent way by resorting to dispersion theory, which is based on unitarity and analyticity \cite{Eden:1966dnq,Donoghue:1996kw,Donoghue:1990xh}. From the dispersive point of view, the unitarity cut can be understood diagrammatically in Fig.\ \ref{fig: disc representation}. 
Here the discontinuity of a TFF is approximated by the intermediate state of two pions in a p-wave. At low energies this saturates the discontinuity because the two-pion state is the lightest intermediate s-channel state. Correspondingly, it is expected that at low energies this lightest state has the most significant contribution to the whole TFF (and not only to its discontinuity). Contributions from heavier states to the discontinuity are neglected in the present work. They only appear indirectly via a subtraction constant that accounts for the impact of the high-energy region. Such an approximation is expected to be valid up to around the mass of the $\rho$ meson, $\vert q^ 2 \vert \approx m_\rho^2$ \cite{Leupold:2017ngs,Alvarado:2023loi}.

Fig.\ \ref{fig: disc representation} introduces also the pion vector FF (for an explicit definition see \cite{Leupold:2017ngs}) and a baryon-pion reaction amplitude. To preserve unitarity (Watson's theorem of final-state interactions \cite{Watson:1954uc}) this hadronic reaction amplitude must include the effect of pion re-scattering. For the energy regime that we want to explore, the re-scattering effect is non-perturbative, but essentially elastic. Therefore we utilize the parameterization of the p-wave phase shift provided in \cite{GarciaMartin:2011cn}, based on a dispersive analysis of two-pion scattering data. 
In this work, we use once-subtracted dispersion relations for the isovector TFFs $F_i^v(q^2)$, which are given as
\begin{eqnarray}
  &&  F_i^v(q^2) =  F_i^v(0) 
     +  \frac{q^2}{12\pi} \, \int\limits_{4 m_\pi^2}^{\Lambda^2} \frac{\text{d}s}{\pi} \, 
     \frac{T_i^F(s) \, p_{\rm cm}^3(s) \, F_{V}^*(s)}{s^{3/2} \, (s-q^2-i \epsilon)}
     + F^{\rm anom}_i(q^2)   \phantom{mm}
     \label{eq:dispbasic}  
\end{eqnarray}
for $i=1,2,3$. The pion momentum of a two-pion system with invariant mass $\sqrt{s}$ is called $p_{\rm cm}$. The pertinent pion-baryon reaction amplitudes are denoted by $T^ F_i$ and the pion vector FF by $F_V$. 
The last term on the right-hand side of \eqref{eq:dispbasic}, the ``anomalous'' piece, is caused by the anomalous singularity whose origin will be explained below in Section \ref{sec: Partial-wave projection}. There, also the corresponding analytic expressions will be specified.

The dispersive integral is cut off at an energy $\Lambda$ because we can only provide reliable input for the discontinuity at low energies where the two-pion state is dominant. Obviously, the contributions from the larger-$s$ regions of the integrand in \eqref{eq:dispbasic} are suppressed by several factors of $\sqrt{s}$ in the denominator (for small enough $\vert q^2 \vert$). Therefore a variation of $\Lambda$ in a reasonable range will not change the results very much. This has been confirmed in \cite{Granados:2017cib,Leupold:2017ngs,Junker:2019vvy,Aung:2024qmf} where the cutoff has been varied between 1 GeV (onset of two-kaon and four-pion physics) and 2 GeV (onset of two-baryon physics). 

The pion vector FF is introduced by
\begin{equation}
     \bra{0}j^{\mu}\ket{\pi^+ \pi^-} = e F_V(s) (p_{\pi^{+}}^{\mu}-p_{\pi^{-}}^{\mu})   
     \label{eq:def-pionVFF}  
\end{equation}
with $s=(p_{\pi^{+}} + p_{\pi^{-}})^2$. For our calculations, 
$F_V$ is taken from \cite{Leupold:2017ngs} (see also \cite{Hanhart:2012wi,Hoferichter:2016duk}):
\begin{eqnarray}
  \label{eq:FV-Omnes-alphaV}
  F_V(s) = (1+\alpha_V \, s) \, \Omega(s) 
\end{eqnarray}
with the Omn\`es function defined as
\begin{eqnarray}
  \Omega(s) = \exp\left\{ s \, \int\limits_{4m_\pi^2}^\infty \frac{\text{d}s'}{\pi} \, \frac{\delta(s')}{s' \, (s'-s-i \epsilon)} \right\}
  \label{eq:omnesele}  
\end{eqnarray}
where $\delta$ denotes the pion p-wave phase shift \cite{Colangelo:2001df,GarciaMartin:2011cn}. 
We utilize the pion phase shift from \cite{GarciaMartin:2011cn}. Then a value of 
\begin{eqnarray}
  \label{eq:alphaV}
  \alpha_V = 0.12 \, {\rm GeV}^{-2}
\end{eqnarray}
yields an excellent description of the data on the pion vector FF from tau decays \cite{Fujikawa:2008ma} for 
energies below 1 GeV.

Note that we follow always the same logic. Low-energy physics is provided explicitly, high-energy physics is encoded in polynomials (because high-energy physics cannot cause non-analytic structures at low energies but only contact terms). In \eqref{eq:dispbasic}, these are the subtraction constants. In \eqref{eq:FV-Omnes-alphaV}, this is the linear term that distinguishes the pion vector FF from the Omn\`es function. In general, the photon can couple to the pion via any intermediate state while the Omn\`es function accounts only for the coupling to intermediate two-pion states. 

In the dispersive framework, hadronic amplitudes $T_i(s)$ are calculated by solving a Muskhelishvili-Omn\`es equation \cite{zbMATH03081975,Omnes:1958hv}. This leads to \cite{Junker:2019vvy,Aung:2024qmf}
\begin{eqnarray}
  T_i(s) & = & K_i(s) + \Omega(s) \, {\rm Polyn}_i(s) + T_i^{\rm anom}(s) + \Omega(s) \, s \, 
  \int\limits_{4m_\pi^2}^{\Lambda^2}  \frac{\text{d}s'}{\pi} \, 
  \frac{K_i(s') \, \sin\delta(s')}{\vert\Omega(s')\vert \, (s'-s-i \epsilon) \, s'} \,.
  \label{eq:tmandel}
\end{eqnarray}

Before we explain all the ingredients of \eqref{eq:tmandel} in detail, we have to clarify the subtle difference between the hadronic amplitudes $T^F_i$ that appear in \eqref{eq:dispbasic} and the constraint-free hadronic amplitudes $T_i$ that appear in \eqref{eq:tmandel}. The relation is
\begin{eqnarray}
  T^F_{1,2}(s) := T_{1,2}(s) \,, \quad T^F_3(s) := \frac{m_N^2}{s} \, T_3(s) \,.
  \label{eq:rel-T-TF}
\end{eqnarray}
The basic idea is to construct constraint-free hadronic amplitudes $T_i$ that correspond to the constraint-free TFFs $F_i$. However, there is a subtlety related to gauge invariance (BTT construction) that applies to the FFs but not to the hadronic amplitudes. This will be further discussed in Subsection \ref{sec contact terms}. But the bottom line is that the constraint-free amplitudes $T_{1,2,3}$ correspond to $F_1$, $F_2$, and $q^2 \, F_3/m_N^2$ (and not just $F_3$). On the other hand, all FFs are reconstructed in \eqref{eq:dispbasic} from their two-pion inelasticity in the very same way. This leads to the identification in \eqref{eq:rel-T-TF}. 

In other words, the factor of $q^2$ that appears in \eqref{eq:def-TFF-Gam} and \eqref{eq:def-TFF-Gam-cross} in front of $F_3$ is caused by the demand of gauge invariance. But there is no such strict demand for pion-baryon amplitudes. On the other hand, $T_3^F(s)$ has a pole for $s=0$. This would not be a good starting point for a dispersive representation of type \eqref{eq:tmandel}. 

Now we turn to an explanation of all the building blocks of \eqref{eq:tmandel}. 
The already mentioned re-scattering effect is taken care of by the pion phase shift $\delta$ and the Omn\`es function $\Omega$. 
Also in this dispersion relation \eqref{eq:tmandel} an anomalous contribution appears which will be specified later. 

We have relegated the high-energy effects into subtraction polynomials ${\rm Polyn}_i$. We restrict their form by reasonable high-energy constraints. In QCD, the pion vector FF $F_V$ and the constraint-free baryon TFFs $F_i$ satisfy unsubtracted dispersion relations; see  \cite{Donoghue:1996bt} and \eqref{eq:scaling-TFFs}. This suggests that the hadronic amplitudes $T_i^F(s)$ should drop for large $s$. In turn, the rather conservative constraint is that $T_{1,2}$ should drop and $T_3$ might approach a constant. Given that the Omn\`es function $\Omega$ drops according to $\Omega(s) \sim 1/s$, we use 
\begin{eqnarray}
  {\rm Polyn}_{1,2}(s) \to P_{1,2} \,, \quad {\rm Polyn}_3(s) \to R_3 + P_3 \, \frac{s}{m_N^2} \,.
    \label{eq:polyn-replace}
\end{eqnarray}
The parameters $P_i$, $i=1,2,3$, will be fitted to data. As already stressed, they parameterize the high-energy aspects that we do not have under control in our low-energy dispersive framework. We will see below that $R_3$ is constrained by chiral symmetry.

\begin{figure}[h!]
  \centering
  \begin{subfigure}{0.45\textwidth}
  \centering
        \includegraphics[scale=1]{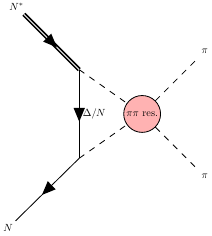}
        \caption{$\Delta$ and nucleon contribution to $T_i$}
        \label{fig:cross channal fey 1}
    \end{subfigure}
    \begin{subfigure}{0.45\textwidth}
     \centering
     \includegraphics[scale=1]{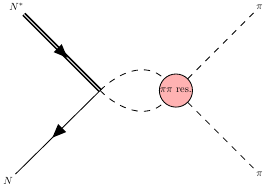}
        \caption{Contribution from the contact terms $\Omega(s) \, {\rm Polyn}_i$ }
        \label{fig:point channel fey 2}
    \end{subfigure}
  \caption{Diagrammatic representation for the reduced hadronic amplitudes $T_{i}$.}
  \label{fig: reduced hadronic amplitude}
\end{figure}

The remaining piece that is needed as input is constituted by the ``bare'' hadronic amplitudes $K_i$ that do not contain the pion-pion re-scattering. Consequently, they contain ``only'' the pion-baryon interactions, in particular the part that causes left-hand cuts from the exchange of states in the cross channels \cite{GarciaMartin:2010cw,Kang:2013jaa}. All parts that contain neither right- nor left-hand cuts must be polynomials. Those polynomials are explicitly provided by ${\rm Polyn}_i$. Our bare amplitudes $K_i$ are free of such polynomial parts, i.e.\ we remove them by hand when defining $K_i$. To obtain a low-energy approximation for these bare hadronic amplitudes $K_i$, we resort to a one-baryon exchange model. This is displayed in Fig.\ \ref{fig:cross channal fey 1}. 
To be more specific, the re-scattering integral in \eqref{eq:tmandel} (plus the anomalous part) is displayed in Fig.\ \ref{fig:cross channal fey 1}. If one drops the blob denoted by ``$\pi\pi$ res.'', one will obtain tree-level diagrams that correspond to $K_i$.  

What is not explicitly resolved by one-baryon exchange is covered by contact terms, expressed via the subtraction polynomials ${\rm Polyn}_i$. The corresponding diagram is shown in Fig.\ \ref{fig:point channel fey 2}. For FFs of ground-state baryons, this approximation of one-baryon exchange plus contact terms can be justified by ChPT \cite{Granados:2013moa,Leupold:2017ngs,Junker:2019vvy,Alvarado:2023loi,Aung:2024qmf}. The $N^*(1520)$ is beyond the applicability range of ChPT, but its hadronic decay properties are phenomenologically well described \cite{PDG} by the decay branches $N^*(1520) \to N \pi, \Delta \pi, N \rho$. We cover the physics of the $\rho$ meson in a model independent way by the re-scattering of the two pions where the $\rho$ meson appears in the p-wave phase shift. In principle, it is conceivable to cover the physics of the $\Delta$ (or more generally the $\pi$-$N$ interactions) by the corresponding pion-nucleon scattering phase shifts \cite{Hoferichter:2015hva}. However, this is far beyond the scope of the present work. In addition, it has been demonstrated in \cite{Leupold:2017ngs,Alvarado:2023loi} that nucleon and $\Delta$ exchange provide a very good approximation to the reduced hadronic amplitudes that are needed for the FFs of the nucleon. Therefore we approximate the pion-nucleon dynamics by its most prominent low-energy resonance, the $\Delta$ baryon. Heavier resonances such as the $N(1440)$ Roper resonance or the ever higher-lying $N(1520)$, $N(1535)$, \ldots\ are neglected as intermediate states due to the negligible couplings of the initial $N(1520)$ to these resonances \cite{CBELSATAPS:2015kka}. Hence, as is shown in Fig.\ \ref{fig: reduced hadronic amplitude}, we approximate the amplitudes in the pion-baryon channels by single-particle exchange diagrams denoted by $K_i$. Then the full amplitudes $T_i^F$ incorporating the pion re-scattering can be calculated via \eqref{eq:tmandel} and \eqref{eq:rel-T-TF}.  

 The details of the hadronic input such as the $N^*$-$\pi$-$N$ and $N^*$-$\pi$-$\Delta$ Lagrangians will be specified in detail in Section \ref{sec: hadronic input}.

\section{Hadronic Lagrangians as input}
\label{sec: hadronic input}

As has been explained in Section \ref{section:disp machinery}, we use amplitudes $\pi N^*(1520) \to N \to \pi N$ and $\pi N^*(1520) \to \Delta(1232) \to \pi N$ as the input for our hadronic amplitudes $T^F_i$. In turn, these amplitudes enter the dispersion relations \eqref{eq:dispbasic} for the TFFs. Thus we need the three-point interaction terms for $N^*(1520)$-$\pi$-$N$, $N$-$\pi$-$N$, $N^*(1520)$-$\pi$-$\Delta(1232)$, and $N$-$\pi$-$\Delta(1232)$.

The interaction terms for $N$-$\pi$-$N$ and $\Delta(1232)$-$\pi$-$N$ 
are fully specified in \cite{Granados:2017cib,Holmberg:2018dtv,Junker:2019vvy} in the framework of baryon ChPT. In Subsection \ref{sec:chiPT}, we will summarize the relevant aspects of the baryon ChPT Lagrangian. 
Subsection \ref{sec nstar eff lag} is devoted to the formulation of the pertinent interaction Lagrangians for the three-point interactions of 
$N^*(1520)$-$\pi$-$N$ and $N^*(1520)$-$\pi$-$\Delta(1232)$. Finally, a Lagrangian contributing to the subtraction constants in \eqref{eq:tmandel} will be formulated in Subsection \ref{sec contact terms}.

\subsection{Input from chiral perturbation theory}
\label{sec:chiPT}

In this work, the $N$-$N$-$\pi$  and $\Delta$-$N$-$\pi$ interactions are estimated using the leading-order (LO) baryonic chiral Lagrangian; see \cite{Jenkins:1991es,Pascalutsa:2006up,Ledwig:2014rfa,Holmberg:2018dtv,Junker:2019vvy}. Our conventions have been spelled out in \cite{Mommers:2022dgw}. Note that this is a three-flavor formalism that can easily be reduced to the two flavors that we need here. The advantage is that we do not need to specify the isospin quartet field for the $\Delta$ and the isospin transition operator 
that connects $\pi N$ to $\Delta$ (see, e.g., \cite{Pascalutsa:2006up}). 

The interactions relevant for our purposes are given by 
\begin{eqnarray}
  {\cal L}_{\rm baryon}^{(1)}  & \supseteq & \frac{g_A}{2} \, {\rm tr}(\bar B \, \gamma^\mu \, \gamma_5 \, u_\mu \, B ) 
    \nonumber \\ 
  && {} + \frac{h_A}{2\sqrt{2}} \, 
  \left(\epsilon^{ade} \, \bar T^\mu_{abc} \, (u_\mu)^b_d \, B^c_e
  + \epsilon_{ade} \, \bar B^e_c \, (u^\mu)^d_b \, T_\mu^{abc} \right) 
  \label{eq:baryonlagr}
\end{eqnarray}
with tr denoting a flavor trace. 
The octet baryons are collected in $B$. We need only the proton and neutron ($B^a_b$ is the entry in the $a$th row, $b$th column):
\begin{eqnarray}
  \label{eq:baroct}
  B \to   \left(
    \begin{array}{ccc}
      0 & 0 & p \\
      0 & 0 & n \\
      0 & 0 & 0 
    \end{array}   
  \right)  \,. 
\end{eqnarray}
The decuplet is expressed by a totally symmetric flavor tensor $T^{abc}$. We need the components 
\begin{eqnarray}
  && T^{111} = \Delta^{++} \,, \quad T^{112} = \frac{1}{\sqrt{3}} \, \Delta^+  \,, \nonumber \\
  && T^{122} = \frac{1}{\sqrt{3}} \, \Delta^0  \,, \quad T^{222} = \Delta^- \,. 
  \label{eq:tensorT}
\end{eqnarray}
The Goldstone bosons are encoded in $\Phi$ with the relevant pieces given by 
\begin{eqnarray}
  \Phi &\to &  \left(
    \begin{array}{ccc}
      \pi^0 & \sqrt{2}\, \pi^+ & 0 \\
      \sqrt{2}\, \pi^- & -\pi^0 & 0 \\
      0 & 0 & 0 
    \end{array}   
  \right)   \,.
  \end{eqnarray}
The non-linear representation manifests itself via the definitions   
\begin{eqnarray}
  u^2 & := & U := \exp(i\Phi/F_\pi) \,, \quad u_\mu := i \, u^\dagger \, (\nabla_\mu U) \, u^\dagger = u_\mu^\dagger  \phantom{mm}
  \label{eq:gold}
\end{eqnarray}
with the pion decay constant $F_\pi$. 

We use the following numbers for our coupling constants: $F_\pi = 92.28\,$MeV, $g_A = 1.26$, $h_A = 2.88$.

The chirally covariant derivatives acting on the baryon octet are defined by
\begin{eqnarray}
  \label{eq:devder}
  D_\mu B := \partial_\mu B + [\Gamma_\mu,B]   \,,
\end{eqnarray}
with the connection
\begin{eqnarray}
  \Gamma_\mu &:=&  \frac12 \, \left(
    u^\dagger \left( \partial_\mu - i (v_\mu + a_\mu) \right) u \right. \nonumber \\
    && \phantom{m} \left. {}+
    u \left( \partial_\mu - i (v_\mu - a_\mu) \right) u^\dagger
  \right) \,,
  \label{eq:defGammamu}
\end{eqnarray}
for a decuplet $T$ by
\begin{eqnarray}
  (D_\mu T)^{abc} &:=& \partial_\mu T^{abc} + (\Gamma_\mu)^a_{a'} T^{a' bc} + (\Gamma_\mu)^b_{b'} T^{a b' c} \nonumber \\
  && {} + (\Gamma_\mu)^c_{c'} T^{a bc'}   \,,
  \label{eq:devderdec}
\end{eqnarray}
for an anti-decuplet by 
\begin{eqnarray}
  (D_\mu \bar T)_{abc} &:=& \partial_\mu \bar T_{abc} - (\Gamma_\mu)_a^{a'} \bar T_{a' bc} - (\Gamma_\mu)_b^{b'} \bar T_{a b' c}  
  \nonumber \\
  && {} - (\Gamma_\mu)_c^{c'} \bar T_{a bc'}   \,,
  \label{eq:devderantidec}
\end{eqnarray}
and for the Goldstone boson fields by
\begin{eqnarray}
  \label{eq:devderU}
  \nabla_\mu U := \partial_\mu U -i(v_\mu + a_\mu) \, U + i U \, (v_\mu - a_\mu)
\end{eqnarray}
where $v$ and $a$ are external vectors and axial-vector sources, respectively. 
We introduce the coupling to photon fields $A_\mu$ via 
\begin{eqnarray}
    v_\mu \to e A_\mu \left(
\begin{array}{ccc}
    Q_u & 0 & 0 \\
    0 & Q_d & 0 \\
    0 & 0 & Q_s
\end{array}
\right)  \,.
\label{eq:subst-v-photon}
\end{eqnarray}

When confronted with the intricacies of working with relativistic spin-3/2 Rarita-Schwinger fields, a recurrent challenge emerges in the form of nonphysical spin-1/2 components. Our interaction term characterized by $h_A$ does not only provide a $\Delta$ exchange but introduces in addition a troublesome contact term. To mitigate this issue effectively, one can turn to the framework presented by Pascalutsa (as documented in \cite{Pascalutsa:1999zz,Pascalutsa:2005nd,Pascalutsa:2006up,Ledwig:2011cx}). The core concept behind this approach entails the substitution 
\begin{eqnarray}
  \label{eq:replace}
  T^\mu \to -\frac{1}{m_R} \, \epsilon^{\nu\mu\alpha\beta} \, \gamma_5 \, \gamma_\nu \, \partial_\alpha T_\beta   \,.
\end{eqnarray}
In this substitution, the symbol $m_R$ denotes the mass of the resonance. In practice we take here the mass of the $\Delta(1232)$. This substitution serves as an effective means of eliminating the unphysical spin 1/2 mode.

Given the usual sign ambiguity in the definition of $\epsilon^{0123}$ \cite{pesschr}, 
we note the following relations of practical relevance.
The antisymmetric combination of two and three gamma matrices are defined as \cite{Junker:2019vvy}
\begin{eqnarray}
  \label{eq:defgammunu}
  \gamma^{\mu\nu} := \frac12 [\gamma^\mu,\gamma^\nu]
\end{eqnarray}
and 
\begin{eqnarray}
  \label{eq:defgammunual}
  \gamma^{\mu\nu\alpha}&:=& \frac16 
  \left(\gamma^\mu \gamma^\nu \gamma^\alpha + \gamma^\nu \gamma^\alpha \gamma^\mu + \gamma^\alpha \gamma^\mu \gamma^\nu
  \right.  \nonumber \\ && \left. \phantom{m} {}
    - \gamma^\mu \gamma^\alpha \gamma^\nu - \gamma^\alpha \gamma^\nu \gamma^\mu - \gamma^\nu \gamma^\mu \gamma^\alpha \right)
  \nonumber \\ 
  & = & \frac12 \{\gamma^{\mu\nu},\gamma^\alpha\} = +i\epsilon^{\mu\nu\alpha\beta} \gamma_\beta \gamma_5  \,,
\end{eqnarray}
respectively.

\subsection{$N^*(1520)$ effective Lagrangian}
\label{sec nstar eff lag}

To formulate the interaction terms for $N^*(1520)$-$\pi$-$\Delta(1232)$ and $N^*(1520)$-$\pi$-$\Delta(1232)$, we start by figuring out how many independent three-point interactions exist. The $N^*(1520)$ state has quantum numbers 
$I=1/2$ and $J^P=3/2^-$. Concerning isospin there is always only one 
combination possible. Concerning partial waves, the system $\pi \, N$ couples with a d-wave to $N^*(1520)$; the system 
$\pi \, \Delta(1232)$ couples with an s- or a d-wave to $N^*(1520)$. Thus we have to write down three interaction terms. We use again the three-flavor formalism of \cite{Granados:2017cib,Holmberg:2018dtv,Junker:2019vvy,Mommers:2022dgw}.

A pertinent effective Lagrangian is given by 
\begin{equation}
\begin{split}
    {\cal L} =& \frac{ih}{m_N} \, \epsilon^{\mu\nu\alpha\beta} \, {\rm tr}\left( \bar B \gamma_\mu u_\nu D_\alpha B_\beta \right)\\
 & {}- \frac{ih}{m_N} \, \epsilon^{\mu\nu\alpha\beta} \, {\rm tr}\left( D_\alpha \bar B_\beta \gamma_\mu u_\nu B \right)
   \\ 
 & {}+ \frac{i H_1}{8 m_\Delta^2} \, \epsilon^{ade} \, (\bar T^{\mu\alpha})_{abc} 
  \left( \gamma_\mu \, (u_\nu)^b_d + \gamma_\nu \, (u_\mu)^b_d \right) (B^{\nu\beta})^c_e \; g_{\alpha\beta} 
  \\
 & {}- \frac{i H_1}{8 m_\Delta^2} \, \epsilon_{ade} \, (\bar B^{\nu\alpha})^e_c  
  \left( \gamma_\mu \, (u_\nu)_b^d + \gamma_\nu \, (u_\mu)_b^d \right) (T^{\mu\beta})^{abc} \; g_{\alpha\beta} 
   \\ 
 & {}+ \frac{i H_2}{8 m_\Delta^2} \, \epsilon^{ade} \, (\bar T^{\mu\alpha})_{abc} 
  \left( \gamma_\mu \, (u_\nu)^b_d - \gamma_\nu \, (u_\mu)^b_d \right) (B^{\nu\beta})^c_e \; g_{\alpha\beta} 
\\
&  {}- \frac{i H_2}{8 m_\Delta^2} \, \epsilon_{ade} \, (\bar B^{\nu\alpha})^e_c  
  \left( \gamma_\mu \, (u_\nu)_b^d - \gamma_\nu \, (u_\mu)_b^d \right) (T^{\mu\beta})^{abc} \; g_{\alpha\beta} 
  \end{split}
  \label{eq:lagrintNstar}
\end{equation}
where we have introduced field-strength type structures $T^{\mu\nu}:= D^\mu T^\nu - D^\nu T^\mu$ and 
$B^{\mu\nu}:= D^\mu B^\nu - D^\nu B^\mu$. 
The three dimensionless coupling constants $h$, $H_1$, and $H_2$ must be real. 
Then this Lagrangian is hermitian and symmetric with respect to charge conjugation; see the discussion in Appendix \ref{sec:C-conj}. The $N^*$ field is expressed via a (sparsely populated) $3 \times 3$ flavor matrix:
\begin{eqnarray}
  B_\mu \to \left(
    \begin{array}{ccc}
      0 & 0 & N^{*+}_\mu \\
      0 & 0 & N^{*0}_\mu \\
      0 & 0 & 0 
    \end{array}   
  \right)  \,. 
\end{eqnarray}

The following considerations went into the construction of \eqref{eq:lagrintNstar}. A meaningful formal low-energy limit is 
obtained for baryon masses much larger than pion masses and differences of baryon masses. In the rest frame of one of the 
baryons, temporal derivatives acting on baryons are as large as baryon masses, spatial derivatives acting on baryons and all 
derivatives acting on pions are small. If one uses the Pauli-Dirac representation \cite{bjorken-drell}
for the spinors and $\gamma$ matrices, the 
upper spinor components are large, the lower are small. $\gamma^0$ is diagonal, $\gamma^i$ is off-diagonal, i.e.\ connects upper 
and lower spinor components. The spatial components of the vector-spinors $T^\mu$ and $B^\nu$ are large, the temporal component 
is small. 

The dominant term in the $h$ interaction is provided for $\mu =0$ and all other indices spatial. One picks up two spatial 
derivatives, which fits to the d-wave. 
The dominant term in the $H_1$ interaction comes from $\mu=\nu=0$. Then $\alpha=\beta$ is a spatial index. There are no spatial
derivatives in the dominant term. This describes an s-wave interaction. By construction, the corresponding dominant term 
drops out for the $H_2$ interaction. The next terms come, e.g., from $\mu =0$ but spatial $\nu$. The indices $\alpha=\beta$ are 
then also spatial. In total, one picks up two momenta, the d-wave interaction. Note, however, that also the $H_2$ interaction has subleading parts that contribute to the s-wave. 
Thus a superposition of $H_1$ and $H_2$ describe the s-wave while $H_2$ describes the d-wave.

The coupling constant $h$ can be fitted to the decay width for $N^*(1520) \to \pi \, N$ \cite{PDG}. $H_2$ can be determined from the d-wave part of the decay $N^*(1520)\to \Delta \pi$. Finally, a superposition of $H_1$ and $H_2$ can be fitted to the corresponding s-wave part. 

At this point it is appropriate to discuss the signs of our coupling constants. Introducing a new bosonic quantum field allows to choose freely a convention about the sign of one interaction term where this field appears with an odd power. The reason is that the field can be replaced by its negative without changing any quantum physics (overall phases do not matter). Since fermions appear in pairs, the number of interaction terms where the sign is a pure convention is by 1 less than the number of fermion fields. The fields at hand are pions, photons, nucleons, $\Delta$ baryons, and $N^*$ baryons. Thus we can freely choose four signs. These conventions enter the substitution \eqref{eq:subst-v-photon} (related to the photon) and the choices $g_A > 0$ (related to the pion), and $h_A > 0$ (related to the $\Delta$). Finally we have a sign convention left that relates to the $N^*(1520)$. We can either choose the sign of $N^*$-$\pi$-$N$ or the sign of $N^*$-$\gamma$-$N$. One sign is a convention, but the other one contains the physics of interferences. 

We have decided to choose the convention for the $N^*$-$\gamma$-$N$ interactions such that our helicity amplitudes \eqref{eq translation Hs and As} agree with the experimental counter parts. This implies that we cannot freely choose the sign of $h$. But only its modulus is determined by 
the partial decay width of $N^* \to \pi N$. Fortunately, it has been specified in \cite{Aznauryan2008u0defhelicity,Aznauryan:2011qj} how the signs of the interactions are related. Note that this is possible because the experimental helicity amplitudes are determined from the single-pion electroproduction on the nucleon. A slight complication is caused by the fact that our conventions concerning charge conjugation (and time reversal) differ from the conventions used in \cite{Aznauryan2008u0defhelicity,Aznauryan:2011qj}. We have clarified this issue in our Appendix \ref{sec:C-conj}. The bottom line of these considerations is that $h < 0$. As a further cross-check we will also explore the consequences of the choice $h > 0$ in the result section. Indeed, the quality of our fit decreases for $h>0$. 

The previous considerations should have made clear that the signs of $H_1$ and $H_2$ are also not pure conventions. This can also be deduced from Fig.\ \ref{fig:cross channal fey 1}. The nucleon exchange is proportional to $h \, g_A$ while the $\Delta$ exchange is proportional to $H_{1,2} \, h_A$. These diagrams interfere at the amplitude level. Given that our conventions make $g_A$ and $h_A$ positive, it is obvious that the relative signs between $h$ and the two couplings $H_1$ and $H_2$ matter. The partial decay widths for $N^* \to \pi \Delta$ determine the modulus of $H_2$ and the modulus of a superposition of $H_1$ and $H_2$. This leaves four possibilities for the values of $H_1$ and $H_2$. However, we have checked that choosing the same sign for $H_1$ and $H_2$ leads to opposite signs of s- and d-wave in the $N^*(1520)\to\pi \Delta$ amplitudes, contradicting the partial-wave analysis \cite{Shrestha:2012ep} and quark model predictions \cite{Koniuk:1979vy,Korner:1976hv}. This leaves us with two possibilities, namely (1) $h < 0$, $H_1 > 0$, $H_2 < 0$; and (2) $h < 0$, $H_1 < 0$, $H_2 > 0$. In Section \ref{sec: results} we will explore these two scenarios.

\subsection{Contact terms}
\label{sec contact terms}

It is illuminating to construct a Lagrangian that gives tree-level contributions to the reaction 
matrix element \eqref{eq:def-transFF}. For the following considerations, isospin does not matter. Therefore we use just a single spin-1/2 field $\Psi$ for the nucleon and a single vector-spinor field $\Psi^\mu$ for the $N^*(1520)$. The photon field strength is denoted by $F_{\mu\nu} = \partial_\mu A_\nu - \partial_\nu A_\mu$. The Lagrangian 
\begin{eqnarray}
  {\cal L} &=& c_1 \left(\bar \Psi \gamma^\mu F_{\mu\nu} \Psi^\nu + \bar\Psi^\nu \gamma^\mu F_{\mu\nu} \Psi \right) \nonumber \\
  && {} + c_2 \left(\bar \Psi \sigma^{\mu\alpha} \partial_\nu F_{\mu\alpha} \Psi^\nu 
    + \bar\Psi^\nu \sigma^{\mu\alpha} \partial_\nu F_{\mu\alpha} \Psi \right) \nonumber \\
  && {} + i c_3 \left(\bar \Psi \partial^\mu F_{\mu\nu} \Psi^\nu - \bar\Psi^\nu \partial^\mu F_{\mu\nu} \Psi \right) 
  \label{eq:effL}
\end{eqnarray}
yields a contribution $\sim c_i$ to the TFF $F_i$. Charge conjugation (and hermiticity) demands that all three $c_i$ are real. We stress again that this statement is tied to the discussion of charge conjugation in Appendix \ref{sec:C-conj}. 

These simplest possible structures \eqref{eq:effL} for $\gamma^*$-$N$-$N^*(1520)$ interactions can also be used to argue why the 
constraint-free FFs $F_i$, $i=1,2,3$, are the starting point for dispersive representations.  
For these simplest 
Lagrangians the constraint-free FFs are constants, i.e.\ the most trivial analytic functions. 
In the presence of loops, these functions become more complicated. One resolves the intrinsic structure of composite objects and that the quantum vacuum is not empty. Starting from the constants, one avoids introducing spurious poles or extra cuts that could be caused by kinematical functions like momenta. 

One might also see the Lagrangian \eqref{eq:effL} as the provider of the subtraction constants $F^v_i(0)$ in \eqref{eq:dispbasic}. In the same spirit we aim at the construction of a Lagrangian that yields the contact terms in \eqref{eq:tmandel}, \eqref{eq:polyn-replace}. 
Mimicking (\ref{eq:effL}), one can replace the field strength $F_{\mu\nu}$ by $J_{\mu\nu}^\pi$
where
\begin{eqnarray}
  \label{eq:defjpi-fieldstr}
  J_{\mu\nu}^\pi := 
  i \left( \partial_\mu \pi^- \partial_\nu \pi^+ - \partial_\mu \pi^+ \partial_\nu \pi^- \right) \,.
\end{eqnarray}
One can then write down the following Lagrangian for four-point $N(1520)$-$N$-$\pi$-$\pi$ vertices:
\begin{eqnarray}
  {\cal L} &=&  \frac{p_1}{F_\pi^3}\left(\bar \Psi \gamma^\mu J^{\pi}_{\mu\nu} \Psi^\nu + \bar\Psi^\nu \gamma^\mu J^{\pi}_{\mu\nu} \Psi \right) \nonumber \\
  && {} + \frac{p_2}{F_\pi^4} \left(\bar \Psi \sigma^{\mu\alpha} \partial_\nu J^{\pi}_{\mu\alpha} \Psi^\nu 
    + \bar\Psi^\nu \sigma^{\mu\alpha} \partial_\nu J^{\pi}_{\mu\alpha} \Psi \right) \nonumber \\
  && {} + i \frac{p_3}{F_\pi^4} \left(\bar \Psi \partial^\mu J^{\pi}_{\mu\nu} \Psi^\nu - \bar\Psi^\nu \partial^\mu J^{\pi}_{\mu\nu} \Psi \right)  \,. 
  \label{eq:effL for pion current}
\end{eqnarray}

Note that \eqref{eq:effL for pion current} emerges from a chiral construction of interactions with pions. The commutator $i[u_\mu,u_\nu]$ yields the tensor $J_{\mu\nu}^\pi$. In principle, there is a simpler term that combines two pions to a p-wave, namely the scalar-QED construction of a charged current: 
\begin{eqnarray}
    j^\pi_\nu := i \left( \pi^- \partial_\nu \pi^+ - \pi^+ \partial_\nu \pi^- \right)  \,.
    \label{eq:scalarQED-current}
\end{eqnarray}
This suggests that instead of the $p_3$ term in \eqref{eq:effL for pion current} there could be a simpler interaction term $\sim \bar\Psi j^\pi_\nu \Psi^\nu$. A pertinent Lagrangian would then have the form\footnote{Note that $J^\pi_{\mu\nu} = \frac12 (\partial_\mu j_\nu^\pi - \partial_\nu j_\mu^\pi)$. }
\begin{eqnarray}
  {\cal L} &=&  \frac{p_1}{F_\pi^3}\left(\bar \Psi \gamma^\mu J^{\pi}_{\mu\nu} \Psi^\nu + \bar\Psi^\nu \gamma^\mu J^{\pi}_{\mu\nu} \Psi \right) \nonumber \\
  && {} + \frac{p_2}{F_\pi^4} \left(\bar \Psi \sigma^{\mu\alpha} \partial_\nu J^{\pi}_{\mu\alpha} \Psi^\nu 
    + \bar\Psi^\nu \sigma^{\mu\alpha} \partial_\nu J^{\pi}_{\mu\alpha} \Psi \right) \nonumber \\
  && {} + i \frac{r_3}{F_\pi^2} \left(\bar \Psi j^{\pi}_{\nu} \Psi^\nu - \bar\Psi^\nu j^{\pi}_{\nu} \Psi \right)  \,. 
  \label{eq:four-point-non-chiral}
\end{eqnarray}
However, such a structure $\bar\Psi j^\pi_\nu \Psi^\nu$ is at odds with chiral symmetry breaking. Indeed, the structure \eqref{eq:scalarQED-current} emerges from the chiral connection $\Gamma_\nu$ given in \eqref{eq:defGammamu}. But, of course, the chiral connection appears only together with a derivative. Yet, a term $\sim {\rm tr}(\bar B D_\nu B^\nu)$ would not make sense because it mixes the nucleon field with the unphysical spin-1/2 part of $B^\nu$. 

The whole discussion is completely equivalent concerning the photon field $A_\nu$ and the pion current $j^\pi_\nu$. Both terms are contained in the chiral connection $\Gamma_\nu$. Both interaction terms $\bar \Psi A_\nu \Psi^\nu$ and $\bar\Psi j^\pi_\nu \Psi^\nu$ are forbidden by symmetries. Instead, field strength structures $F_{\mu\nu}$ and $J_{\mu\nu}^\pi$ must be used (which emerge from the field strength $[D_\mu,D_\nu]$ related to the chiral connection \cite{Bijnens:1999sh}). 

On the other hand, gauge invariance, related to the use of the photon field strength in \eqref{eq:effL}, is an exact symmetry while chiral symmetry is only approximate. The appearance of the $q^2$ term in \eqref{eq:def-TFF-Gam} and \eqref{eq:def-TFF-Gam-cross} is mandatory. But in the projector formalism for pion-baryon amplitudes that we will introduce in the next section, there is no gauge invariance restriction that would induce a factor of $q^2$. Here emerges the subtle difference between standard partial-wave projected hadronic reaction amplitudes $T_i$ and the reduced amplitudes $T^F_i$ that are required in \eqref{eq:dispbasic} for the dispersive representation of the electromagnetic TFFs. 

The simplest structures that can appear for $T_i$ in \eqref{eq:tmandel} emerge from \eqref{eq:four-point-non-chiral}. In other words, the analogy to $F_i(0) \sim c_i$ is given by $P_{1,2} \sim p_{1,2}$ and $R_3 \sim r_3$. Yet this $R_3$ cannot come from a chiral Lagrangian. It can be caused, however, from the polynomial part of nucleon and $\Delta$ exchange diagrams. But then it is predicted from the low-energy exchange diagrams and not a free fit parameter (``low-energy constant'') that would correspond to the high-energy aspects. In \eqref{eq:polyn-replace}, we can calculate $R_3$, but will fit $P_{1,2,3}$.

\section{Projector formalism}
\label{sec:Projector formalism}

Let us spell out in detail the strategy how to obtain the pertinent ``bare'' hadronic reaction amplitudes. There are two general aspects: (1) We want to formulate dispersion relations for the TFFs $F_i$, which are free of kinematical constraints. (2) Feynman amplitudes for hadronic reactions $N^* \to N \, \pi \pi$ are most easily decomposed into hadronic helicity amplitudes $K_{\lambda_{N^*}-\lambda_N}$ \cite{Jacob:1959at}. How do we make contact between the $F_i$'s and the $K_m$'s? The discontinuities of the TFFs $F_i$ can be translated to discontinuities of the electromagnetic helicity amplitudes $H_{\lambda_N - \Lambda_{N^*}}$. Their two-pion inelasticities translate then directly to the hadronic helicity amplitudes $K_{\lambda_{N^*}-\lambda_N}$. Taken together, we obtain relations between the constraint-free TFFs $F_i$ and corresponding constraint-free hadronic amplitudes $K_i$. For the latter we can then formulate also dispersion relations --- provided that the $K_i$ do not contain spurious poles. To be completely precise, the relation between $F_1$, $F_2$, and $q^2 F_3/m_N^2$ on the one hand and $H_+$, $q^2 H_0/m_N^2$, and $H_-$ on the other translates to a relation between $K_1$, $K_2$, and $K_3$ on the one hand and $K_{-1}$, $K_0$, and $K_{+1}$ on the other. We find
\begin{eqnarray}
  \label{eq:introT-tilde5}
  \left(
    \begin{array}{c}
      H_+(q^2) \\ q^2 H_0(q^2)/m_N^2 \\ H_-(q^2)
    \end{array}
  \right) = \mathds{T}(q^2) \;
  \left(
    \begin{array}{c}
      F_1(q^2) \\ F_2(q^2) \\ q^2 F_3(q^2)/m_N^2 
    \end{array}
  \right) 
\end{eqnarray}
leading to 
\begin{eqnarray}
  \label{eq:introTprime}
  \left(
    \begin{array}{l}
      K_{-1}(s) \\ K_0(s) \\ K_{+1}(s)
    \end{array}
  \right) = \mathds{T}(s) \;
  \left(
    \begin{array}{l}
      K_1(s) \\ K_2(s) \\ K_3(s)
    \end{array}
  \right) 
\end{eqnarray}
with 
\begin{eqnarray}
  \label{eq:defT-tilde5}
  \mathds{T}(q^2) := \left(
    \begin{array}{ccc}
      \frac{(m_{N^*} m_N - m_N^2 + q^2) m_N }{m_{N^*}} & \frac{-(m_{N^*}+m_N) \left((m_{N^*}-m_N)^2-q^2 \right)}{m_{N^*}} & 
      {\scriptstyle m_N^2} \\
      \frac{q^2 (m_{N^*}-m_N)}{m_N} & \frac{-q^2 (m_{N^*}-m_N)\left((m_{N^*}-m_N)^2-q^2 \right)}{2 m_{N^*} m_N^2} & 
      \frac{(m_{N^*}-m_N)(q^2+m_{N^*}^2-m_N^2)}{2 m_{N^*}} \\
      {\scriptstyle (m_{N^*}-m_N) m_N} & {\scriptstyle 0} & {\scriptstyle m_N^2}
    \end{array}
  \right)  \,.
\end{eqnarray}
Of course, \eqref{eq:introT-tilde5} is a direct consequence of the original definitions \eqref{eq:defG0pm}.  
Had we used $H_0$ instead of the combination $q^2 H_0/m_N^2$, then the matrix $\mathds{T}$ would not just contain 
polynomials in $q^2$. The polynomials ensure that inelasticities for the helicity amplitudes can be translated directly to the inelasticities of the TFFs.  

After inverting \eqref{eq:introTprime}, the remaining tasks are then to obtain hadronic helicity amplitudes $K_m$ and to choose normalizations such that they fit to the electromagnetic helicity amplitudes $H_{-m}$. The first task is addressed in the present and the next section. The proper normalizations of the hadronic amplitudes is addressed in Appendix \ref{sec:Normalization factors}. 

\begin{figure}[h!]
    \centering
\includegraphics[scale=0.6]{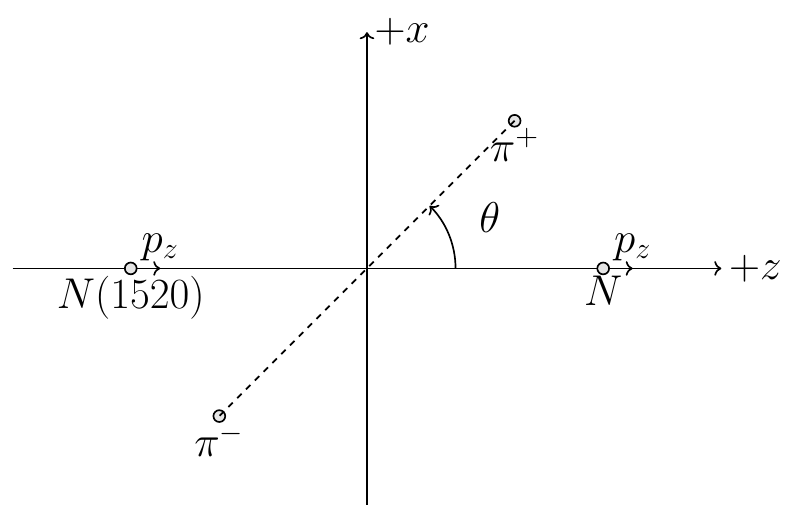}
    \caption{The $x$-$z$ plane of the center of mass frame of the 2-pion state. Both the $N^*(1520)$ and $N$ are chosen to fly in the $+z$ direction. }
    \label{fig: kinematic drawing}
\end{figure}

To describe the hadronic decay $N^* \to N \pi^+ \pi^-$, we choose a frame where the pion pair is at rest and the baryons move in the 
$+z$ direction. We choose the $x$-axis such that the whole reaction takes place in the $x$-$z$ plane and that the $\pi^+$ has a positive momentum component in the $x$-direction. A sketch for the kinematic configuration is shown in Fig.\ \ref{fig: kinematic drawing}.

It is convenient to introduce (see also \cite{Stoica:2011cy,Junker:2019vvy})
\begin{equation}
\label{eq: def kinematic variables}
    \begin{split}
    q &:= p_{N^*}-p_N = p_{\pi^+} + p_{\pi^-} \,,\\ k &:= p_{\pi^+} - p_{\pi^-}\,,\\  \bar k &:= p_{N^*} + p_N \,, \\
  r &:= \bar k - \frac{\bar k \cdot q}{q^2} \, q \,,\\  k_\perp &:= k - \frac{k \cdot r}{r^2} \, r \,.
    \end{split}
\end{equation}
If $\theta$ denotes the angle between the flight directions of the positive pion and the baryons (Fig.\ \ref{fig: kinematic drawing}), we find explicitly:
\begin{eqnarray}
  && p_{N^*}=(E_{N^*},0,0,p_z) \,, \quad p_N = (E_N,0,0,p_z) \,,    \nonumber \\
  && p_{\pi^+} = (E_\pi,p_{\rm cm} \sin\theta,0,p_{\rm cm} \cos\theta) \,, \quad   
  p_{\pi^-} = (E_\pi,-p_{\rm cm} \sin\theta,0,-p_{\rm cm} \cos\theta) \,,    \nonumber \\
  && q = (\sqrt{q^2},0,0,0) \,, \quad k_\perp = (0, 2 p_{\rm cm} \sin\theta ,0,0) \, \quad r = (0,0,0,2p_z)  
  \label{eq:all-fourvectors}
\end{eqnarray}
with
\begin{eqnarray}
  && E_{N^*} = \frac{m_{N^*}^2+q^2-m_N^2}{2 \sqrt{q^2}} \,, \quad E_N = \frac{m_{N^*}^2-q^2-m_N^2}{2 \sqrt{q^2}} \,,     
  \nonumber \\[1em]
  && p_z = \frac1{2 \sqrt{q^2}} \, \lambda^{1/2}(m_{N^*}^2,m_N^2,q^2) \,, \quad p_{\rm cm} = \frac{\sqrt{q^2}}2 \, \sigma(q^2)
  \label{eq:def-en-mom}
\end{eqnarray}
where we have used the K\"all\'en function \eqref{eq:kallenfunc}
and the velocity of the pions, 
\begin{eqnarray}
  \label{eq:velpions}
  \sigma(s) := \sqrt{1-\frac{4m_\pi^2}{s}}  \,.
\end{eqnarray}
Note that in the chosen frame, $q$ has only a zeroth component, $r$ has only a third component ($z$-component), 
and $k_\perp$ has only a positive first component ($x$-component). 

We want to relate the general matrix element $ \bar u_N \, M_\mu \, u^\mu_{N^*}$ to scalar quantities $a_i(q^2,\theta)$ and a pre-defined basis set of spinor-matrix Lorentz-vector objects such that only the scalar quantities depend on the explicit form of 
$M^\mu$, i.e. 
\begin{eqnarray}
  \label{eq:genM-Mi}
  \bar u_N \, M_\mu \, u^\mu_{N^*} = \sum\limits_i a_i(q^2,\theta) \; \bar u_N  \, M_i^\mu \, g_{\mu\nu} \, u^\nu_{N^*} \,.
\end{eqnarray}
One can show that there are only four independent spinor-matrix Lorentz-vector objects $M_i^\mu$, which we label by $i=+1$, $0$, $-1$, $-2$. The task of constructing the basis set of $M_i^\mu$ is similar to \cite{Stoica:2011cy,Junker:2019vvy} but technically 
significantly simpler since all our baryons have natural parity. We have found the following basis set of spinor-matrix Lorentz-vector objects:
\begin{eqnarray}
  M^\mu_{-1} &:=& i \slashed{k}_\perp \, p_N^\mu  \,, \nonumber \\
  M^\mu_{+1} &:=& i \left[(m_{N^*}-m_N)^2-q^2 \right] k_\perp^\mu + i m_{N^*} \, \slashed{k}_\perp \, p_N^\mu  \,, \nonumber\\
  M^\mu_{0} &:=& i p_N^\mu  \,, \nonumber \\
  M^\mu_{-2} &:=& i \left[(m_{N^*}+m_N)^2-q^2 \right]  \slashed{k}_\perp \, k_\perp^\mu + i m_{N^*} \, k_\perp^2 \, p_N^\mu  \,. 
  \label{eq:finalMs2}
\end{eqnarray}
We have constructed them 
such that each corresponds to one of the relevant helicity combinations, i.e.
\begin{eqnarray}
  \label{eq:orthoMi}
  \bar u_N(p_z,+1/2) \, M_i^\mu \, g_{\mu\nu} \, u^\nu_{N^*}(p_z,\lambda_{N^*}) \sim \delta_{i,\Delta\lambda}  
\end{eqnarray}
with the helicity change $\Delta \lambda := \lambda_{N^*}-\lambda_N$.

One can determine the scalar amplitudes $a_i$, $i=-2,-1,0,+1$, via a projector formalism along the lines of \cite{Junker:2019vvy}. 
The scalar amplitudes $a_i$, once properly normalized, can be further processed for partial-wave expansions and dispersive 
representations. In particular, they can be expanded as \cite{Jacob:1959at}
\begin{eqnarray}
  \label{eq:pwexp}
  a_i(s,\theta) \, \sin^{\vert i \vert} \theta 
  = \frac{1}{p_z} \sum\limits_J \left(J+\frac12 \right) a^J_i(s) \, d^J_{i,0}(\theta) 
\end{eqnarray}
where $J$ denotes the (total) angular momentum of the two-pion system and $d^J_{i,0}$ the Wigner rotation functions. 
Note that $k_\perp$ is the only quantity in \eqref{eq:finalMs2} that carries an angular dependence, which explains the appearance of the factor $\sin(\theta)^{|i|}$ in the partial-wave expansion in (\ref{eq:pwexp}).

\section{Partial-wave projection}
\label{sec: Partial-wave projection}

Next we present the partial-wave decomposition of the hadronic amplitude $ \bar u_N \, M_\mu \, u^\mu_{N^*}$ and the reduced amplitude $K_{\Delta \lambda}$.

Depending on the helicity configuration, one has for 
$\Delta\lambda = 0$ the partial-wave projection
\begin{equation}
 \label{eq:reduced amplitude k0}
\begin{split}
 K_0(s) =& 
  \frac32 \, \int\limits_0^\pi d\theta \, \sin\theta \, \cos\theta \\&\times
  \frac{\bar u_N(p_z,+1/2) \, M_\mu(q,p_N,k_\perp) \, u_{N^*}^\mu(p_z,+1/2)}{N_0(s) \, p_{\rm cm}} 
 \\ 
  =&   \frac32 \, \frac{\bar u_N(p_z,+1/2) \, M^\mu_0 \, g_{\mu\nu} u_{N^*}^\nu(p_z,+1/2)}{N_0(s) \, p_{\rm cm}} \, \\
  &\times\int\limits_0^\pi d\theta \, \sin\theta \,\cos\theta\, a_0(s,\theta) 
\end{split}
\end{equation}
while for the helicity-flip cases of $\vert \Delta\lambda \vert = 1$ one finds
\begin{equation}
\begin{split}
 K_{\pm 1}(s) =& 
  \frac34 \, \int\limits_0^\pi d\theta \, \sin^2\theta \,\\ &  \times  
  \frac{\bar u_N(p_z,+1/2) \, M_\mu(q,p_N,k_\perp) \, u_{N^*}^\mu(p_z,+1/2\pm 1)}{N_{\pm 1}(s) \, p_{\rm cm}} \, \\ 
  =&   \frac34 \, 
  \frac{\bar u_N(p_z,+1/2) \, \tilde M^\mu_{\pm 1} \, g_{\mu\nu} u_{N^*}^\nu(p_z,+1/2\pm 1)}{N_{\pm 1}(s) \, p_{\rm cm}} \, \\
  &\times \int\limits_0^\pi d\theta \, \sin\theta \, a_{\pm 1}(s,\theta) \, \sin^2\theta   \,.    
 \end{split}   
\label{eq:reduced amplitude kpm1}
\end{equation}
The normalization factors $N_{\pm1,0}$ which are defined in Appendix \ref{sec:Normalization factors} are chosen such that one gets the correct TFFs $F_{1,2,3}$, not multiples thereof. The $p_{\rm cm} $ factors in the denominator of \eqref{eq:reduced amplitude k0} and \eqref{eq:reduced amplitude kpm1} are purely conventional \cite{Junker:2019vvy}. 

We have checked explicitly that, for both the $N$ and $\Delta$ exchange, one gets the following analytic expression in each kinematical region:
\begin{equation}
  K_{\pm1,0}(s) = g_{\pm1,0}(s) - \frac{2 f_{\pm1,0}(s)}{Y(s) \, \kappa^2(s)}+  f_{\pm1,0}(s) \, \frac{1}{\kappa^3(s)} \, \mathscr{L}_{N/\Delta}
  \label{eq:Klog}
\end{equation}
where $\mathscr{L}_{N/\Delta}=\log\frac{Y(s)+\kappa(s)}{Y(s)-\kappa(s)}$ is the standard  logarithmic function for $(m_{N^*}+m_N)^2 \leq s$. Below this production threshold one has to choose the correct analytic continuation. We will come back to this point below.

The function $Y$ that appears in (\ref{eq:Klog}) and in the logarithm $\mathscr{L}_{N/\Delta}$ is defined by 
\begin{eqnarray}
  Y(s) := s + 2 m_{\rm exch}^2 - m_{N^*}^2 - m_N^2 - 2 m_\pi^2  \,.
  \label{eq:defYfunc}    
\end{eqnarray}
The mass $m_{\rm exch}$ of the exchanged state is either the nucleon or the $\Delta$ mass.
The function $\kappa(s)$ is defined as 
\begin{eqnarray}
  \label{eq:defKacser}
  \kappa(s) := \lambda^{1/2}(s,m_N^2,m_{N^*}^2) \, \sigma(s) \,,
\end{eqnarray}
with the K\"all\'en function \eqref{eq:kallenfunc} and the pion velocity $\sigma$ provided in \eqref{eq:velpions}.

The functions $f_{\pm1,0}(s)$, $g_{\pm1,0}(s)$ are functions without any cut but they have poles at kinematical thresholds. However, they conspire such that no poles show up for $K_{\pm1,0}(s)$. The same observation can be made for the corresponding expressions in \cite{Junker:2019vvy}. The analytic expressions for the nucleon exchange contribution to $f_{\pm1,0}(s)$, $g_{\pm1,0}(s)$ are explicitly given in Appendix \ref{sec f and g}. However, due to the lengthy analytic expressions of the $\Delta$ contributions, we refrain from explicitly presenting those in the paper. 

\begin{figure}[h!]
  {\centering
  \begin{subfigure}[c]{\textwidth}
  \centering
        \includegraphics[width=0.6\textwidth]{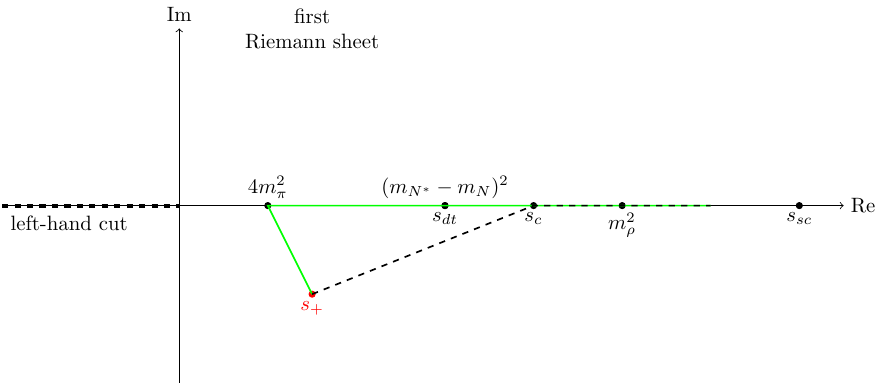}
        \caption{Analytic structure of the nucleon exchange diagram Fig.\ \ref{fig:cross channal fey 1} in the first Riemann sheet with $s_+$ being the anomalous branch point.  The first part of the green curve is defined in (\ref{eq:defsx}), the rest is the unitarity cut along the real axis as given in \eqref{eq:dispbasic} and \eqref{eq:tmandel}. An equivalent, numerically treatable path is shown by the dashed line.}
        \label{fig:nucleon integral contour}
    \end{subfigure}\\
    \begin{subfigure}[c]{\textwidth}
     \centering
        \includegraphics[width=0.6\textwidth]{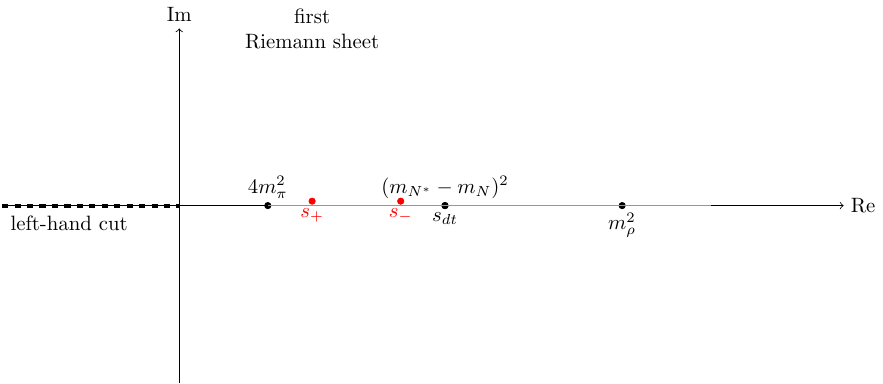}
        \caption{Analytic structure of the $\Delta$ exchange diagram Fig.\ \ref{fig:cross channal fey 1} in the first Riemann sheet. $s_+$ and $s_-$ are two singularities, both lying infinitesimally above the real axis, on the second Riemann sheet. The green curve shows the unitarity integration path. }
    \label{fig:delta integral contour}
    \end{subfigure}
}
  \caption{Analytic structure of the nucleon and $\Delta$ exchange diagrams in the first Riemann sheet}
  \label{fig:contour plot}
\end{figure}

We turn now to the analytic continuation of the logarithms. The right-hand cut is along the positive real axis starting at the two-pion threshold $4\mpi^2$, known as the unitarity cut. This cut defines two Riemann sheets. The logarithm of $\mathscr{L}_{N/\Delta}$ is caused by exchange diagrams in the cross channels. The corresponding branch point and attached cut are usually located in the second Riemann sheet \cite{Junker:2019vvy}. However, an additional branch point can appear on the physical sheet if the exchange particle in the $t$/$u$-channel is too light. The left-hand cut generated by $K_{\pm1,0}(s)$ might intersect with the right-hand cut and therefore produce a branch point in the first Riemann sheet. The criterion for the appearance of such a branch point on the first Riemann sheet is given by \cite{Karplus:1958zz,Junker:2019vvy}
\begin{eqnarray}
  \label{eq:anomthr}
  m_{\rm exch}^2 < \frac12 \, \left(m_{N^*}^2 + m_N^2 - 2 m_\pi^2 \right)  \,.
  \end{eqnarray}
For the $N^{*}(1520) \to  N \, \pi \pi$ amplitudes, the nucleon exchange satisfies the criterion (\ref{eq:anomthr}) and therefore there will be an anomalous singularity in the complex plane for the nucleon exchange diagram. 
One needs a proper analytic continuation for $\mathscr{L}_{N/\Delta}(s)$ down to the interval $4 \mpi^2\leq s<(m_{N^*}+m_N)^2$, which we now shall discuss. The following points along the real axis and in the complex plane, respectively, are of importance: 
\begin{itemize}
\item the two-pion threshold 
  \begin{eqnarray}
    \label{eq:def2pithr}
    s_{2\pi} := 4 m_\pi^2  \,; 
  \end{eqnarray}
\item the decay threshold below which the decay $N^*(1520) \to N  \pi \pi$ can happen,
  \begin{eqnarray}
    \label{eq:decthr}
    s_{\rm dt} := (m_{N^*}-m_N)^2  \,;    
  \end{eqnarray}
\item the point where the function $Y$ vanishes,
  \begin{eqnarray}
    \label{eq:yvan}
    s_Y := -2 m_{\rm exch}^2 + m_{N^*}^2 + m_N^2 + 2 m_\pi^2  \,;
  \end{eqnarray}
\item the scattering threshold for the scattering reaction $\pi \pi \to N^*(1520) \bar N$,
  \begin{eqnarray}
    \label{eq:scatthr}
    s_{\rm st} := (m_{N^*}+m_N)^2  \,;  
  \end{eqnarray}
\item the branch points of the logarithm that emerges from 
  the partial-wave projection \eqref{eq:reduced amplitude k0}, \eqref{eq:reduced amplitude kpm1} \cite{Junker:2019vvy},
  \begin{equation}
      \begin{split}
s_\pm &:=  -\frac{1}{4 m_{\rm exch}^2} \times \left[ \left( \lambda^{1/2}(m_{N^*}^2,m_{\rm exch}^2,m_\pi^2) \pm \lambda^{1/2}(m_N^2,m_{\rm exch}^2,m_\pi^2) \right)^2 - (m_{N^*}^2-m_N^2)^2\right].   
\end{split}
\label{eq:spm}
  \end{equation} 
\end{itemize}
Here $m_{\rm exch}$ denotes the mass of the exchanged baryon.
As long as $s_Y$ is below the two-pion threshold, no anomalous cuts appear \cite{Karplus:1958zz}. But for our case of 
nucleon and $\Delta$ exchange, $s_Y$ lies higher. What matters is the location of the points \eqref{eq:def2pithr}, 
\eqref{eq:decthr}, \eqref{eq:yvan}, \eqref{eq:scatthr}, and \eqref{eq:spm} relative to each other; see Fig.\ \ref{fig:contour plot}.

For the nucleon exchange diagram, the point $s_+$ lies on the first Riemann sheet in the fourth quadrant. The point $s_-$ lies 
on the second Riemann sheet and is of no concern. The ordering of the other points is 
\begin{eqnarray}
  \label{eq:ordering-points}
  s_{2\pi} < s_{\rm dt} < s_Y < s_{\rm st}  \,.
\end{eqnarray}
More explicitly, for the $N$ exchange, the logarithmic function $\mathscr{L}_{N}$ in (\ref{eq:Klog}) takes the following form:
\begin{equation}
\frac{\mathscr{L}_{N}}{\kappa(s)} = \begin{cases}
 \frac{1}{\kappa(s)}\left(\log\frac{Y(s)+\kappa(s)}{Y(s)-\kappa(s)}+2\pi i \right) \qquad &\mbox{for} \quad s_{2\pi} < s < s_{\rm{dt}}\,, \\[0.7em]
  \frac{2}{|\kappa(s)|}\left(\arctan\frac{|\kappa(s)|}{Y(s)}  + \pi \right) \qquad &\mbox{for} \quad s_{\rm{dt}}< s < s_Y \,, \\[0.7em]
\frac{2}{|\kappa(s)|}\arctan \frac{|\kappa(s)|}{Y(s)} \qquad &\mbox{for} \quad s_Y < s < s_{\rm{st}} \,, \\[0.7em]
 \frac{1}{\kappa(s)}\log\frac{Y(s)+\kappa(s)}{Y(s)-\kappa(s)}  \qquad &\mbox{for} \quad s_{\rm{st}} < s 
\end{cases}
\label{eq nucleon analytic}
\end{equation}
where $\log$ is the logarithmic function that has a cut along the negative axis. The arctan function maps the set of real numbers onto the interval $(-\pi/2,+\pi/2)$. 

For the $\Delta$ exchange, both $s_{\pm}$ lie on the second Riemann sheet, and therefore there is no anomalous singularity. Both $s_{\pm}$ lie infinitely close to, but slightly above the real axis, if one uses the prescription $m_{N^*}^2 \to m_{N^*}^2+i\epsilon$. 
The ordering of points is 
\begin{eqnarray}
  \label{eq:ordering-points2}
  s_{2\pi} < s_+ < s_Y < s_- < s_{\rm dt} < s_{\rm st}  \,.
\end{eqnarray}
Numerically, $s_+$ is rather close to the two-pion threshold while $s_-$ is close to the decay threshold. 
We find that the correct analytic continuation of the logarithm for the $\Delta$ exchange is
\begin{equation}
\frac{\mathscr{L}_{\Delta}}{\kappa(s)} = \begin{cases}
\frac{1}{\kappa(s)}\log\frac{Y(s)+\kappa(s)}{Y(s)-\kappa(s)}\qquad &\mbox{for} \quad s_{2\pi} < s <  s_+\,, \\[0.7em]
\frac{1}{\kappa(s)}\left(\log \left\vert \frac{Y(s)+\kappa(s)}{Y(s)-\kappa(s)} \right\vert +\pi i \right) \qquad &\mbox{for} \quad  s_+< s <  s_-\,, \\[0.7em]
\frac{1}{\kappa(s)}\log\frac{Y(s)+\kappa(s)}{Y(s)-\kappa(s)}\qquad &\mbox{for} \quad  s_-< s <  s_{\rm dt}\,, \\[0.7em]
\frac{2}{|\kappa(s)|} \, \arctan\frac{|\kappa(s)|}{Y(s)}  \qquad &\mbox{for} \quad  s_{\rm dt}< s < s_{\rm st} \,, \\[0.7em]
\frac{1}{\kappa(s)}\log\frac{Y(s)+\kappa(s)}{Y(s)-\kappa(s)}\qquad &\mbox{for} \quad  s_{\rm st}< s \,. 
\end{cases}
\label{eq delta analytic conti}
\end{equation}
Note that the $\log$ function has led to an imaginary part between the two singularities $s_+$ and $s_-$.
We have checked that (\ref{eq nucleon analytic}) and (\ref{eq delta analytic conti}) are exactly the correct analytic expressions along the unitarity cut by a numerical comparison of our dispersive calculations with the respective one-loop result for the scalar case \cite{tHooft:1978jhc}.

From the tree-level hadronic helicity amplitudes $K_{\pm1,0}$ given in (\ref{eq:Klog}) and logarithmic functions $\mathscr{L}_{N/\Delta}$ properly analytically continued, it is straightforward to invert \eqref{eq:introTprime} and to calculate the pion re-scattering loop diagrams in (\ref{eq:tmandel}). 

So far, we have described the situation as if one-baryon exchange diagrams yielded only left-hand cut structures $K_m$. However, also polynomial contributions emerge from the corresponding Feynman diagrams \cite{Gasser:1987rb,Granados:2017cib,Leupold:2017ngs,Junker:2019vvy,Alvarado:2023loi}. In principle, one can just drop these contributions as long as they can be merged with the fit parameters $P_i$ that appear in \eqref{eq:polyn-replace}. Yet this is not the case for $R_3$ as pointed out in Subsection \ref{sec contact terms}. Therefore we present explicitly those constants emerging from the baryon exchange diagrams. In addition, we match the subtraction constants $P_i$ to the contact interaction terms of \eqref{eq:effL for pion current}. This is a further cross-check for the procedure on the hadronic side that starts with Feynman amplitudes, translates them to helicity amplitudes and finally combines those to unconstrained reduced hadronic amplitudes. Carried out for the contact terms of \eqref{eq:effL for pion current} leads indeed to the correct results. We find
\begin{eqnarray}
P_1 &=& \frac{p_1}{F_\pi^3 m_N}   \,, \nonumber \\
P_2 &=& -\frac{2 p_2}{F_\pi^4}   \,, \nonumber \\
P_3 &=& -\frac{p_3}{F_\pi^4}    \,, \nonumber \\
R_3 &=& R_3^N + R_3^\Delta   
    \label{eq:poly-conn-contactL}
\end{eqnarray}
with $R_3^{N,\Delta}$ presented in \eqref{eq:polyn-N-Delta}.

The anomalous contributions that enter (\ref{eq:dispbasic}) and \eqref{eq:tmandel} concern the nucleon exchange. They are given by \cite{Junker:2019vvy}
\begin{eqnarray}
  F^{\rm anom}_i(q^2) &=& \frac{q^2}{12\pi} \, \int\limits_{0}^1 \text{d}x \, \frac{ds'(x)}{dx} \,
                          \frac{1}{s'(x)-q^2}   \nonumber \\ && \times
  \frac{f_i^F(s'(x)) \, F_{V}(s'(x))}{-4\, (-\lambda(s'(x),m_{N}^2,m_{N^*}^2))^{3/2}}  \,,
  \label{eq:Fanom}
\end{eqnarray}
and
\begin{eqnarray}
  T_{i}^{\rm anom}(s) &=& \Omega(s) \, s \, \int\limits_0^1 \text{d}x \, \frac{\text{d}s'(x)}{\text{d}x}  \,
                      \frac{1}{s'(x)-s}  \nonumber \\
                  && \times 
                     \frac{2f_i(s'(x))}{(-\lambda(s'(x),m_{N^*}^2,m_N^2))^{1/2} \, \kappa^2(s'(x))}  \nonumber \\
                  && \times  \frac{t_{\rm{IAM}}(s'(x))}{\Omega(s'(x)) \, s'(x)} \,.
  \label{eq:tanom}
\end{eqnarray}
Further details are provided in Appendix \ref{append:Technical aspects on the anomalous singularity}. 
Note that in \eqref{eq:Fanom} and \eqref{eq:tanom}, the quantities $f^F_i$ and $f_i$ are obtained from $f_{-1,0,1}$ in the very same way how $T^F_i$ are obtained from $T_i$ in \eqref{eq:rel-T-TF} and how $K_{1,2,3}$ are obtained from $K_{-1,0,1}$ in \eqref{eq:introTprime}. 

In \eqref{eq:Fanom} and \eqref{eq:tanom} we have chosen the anomalous cut to lie along the straight-line path
\begin{eqnarray}
  \label{eq:defsx}
  s'(x) := (1-x) s_+ + x\, 4m_\pi^2 
\end{eqnarray}
which connects the anomalous singularity $s_+$ to $ 4m_\pi^2$; see Fig. \ref{fig:nucleon integral contour}. 
However, for the nucleon exchange, one faces an integrand that is singular at the decay 
threshold $s_{\rm dt}$, as can be deduced from \eqref{eq nucleon analytic}. Even though this constitutes an inverse-square-root singularity, which is integrable and well-defined with the epsilon prescription $m_{N^*}^2 \to m_{N^*}^2+i\epsilon$ \cite{Bronzan:1963mby}, it is 
numerically easier to avoid this problem \cite{Aung:2024qmf}. We can continuously deform the integration contour from the green path to the black-dashed path in Fig.\ \ref{fig:nucleon integral contour} because the integrand is analytic. In practice, we connect the points $s_+$ and $s_c$ in our calculation where we have chosen $s_{dt}<s_c=0.56 \, \text{GeV}^2<m_\rho^2$. A detailed treatment of the deformation of the contour can be found in Appendix \ref{append:Technical aspects on the anomalous singularity}.


\section{Results}
\label{sec: results}

How well does our formalism describe the available (low-energy) data on the (isovector) TFFs in the space-like region? How do our predictions look like for the time-like (low-energy) region? To answer these questions we will discuss first to which extent the isovector dominates over the isoscalar channel, based on the existing electro- and photo-production data (Subsection \ref{sec: extracting radii}). As a next step, we need to fix our input parameters, namely the three-point coupling constants $h$, $H_{1,2}$ together with the subtraction constants $P_{1,2,3}$, $F^v_{1,2,3}(0)$, a total of 9 parameters. Subsection \ref{sec: fix parameters} is devoted to this task. Finally, the results for the space-like and the predicted time-like TFFs will be presented in Subsection \ref{sec: dispersive predictions}.

\subsection{Extracting the isovector TFFs}
\label{sec: extracting radii}

Based on \eqref{eq def isovector and isoscalar ffs}, we need data for protons and neutrons to extract the isovector TFFs. Naturally, the data situation is better for the proton \cite{CLAS:2009ces,Mokeev:2015lda,jlabdata3} than for the neutron \cite{Tiator:2009mt,Tiator:2011pw}. Our low-energy formalism should definitely not be used for photon virtualities above $1 \,$GeV. But for the discussion of the data situation we use the full range of available data. 

We start with the photon point and use the results from the recent compilation of the Particle Data Group (PDG) \cite{PDG}. Results are shown in Tab.\ \ref{tab:F0 and derivetives}. 
Since we will also use the results of the MAID analysis \cite{Tiator:2009mt,Tiator:2011pw} for the space-like data, it is illuminating to compare also the photon-point results. The deviation between PDG and MAID it explained by the poorer data situation at the time when the MAID analysis has been carried out.
\begin{table}[H]
  \centering
  \begin{tabular}{|c|c|c|}
    \hline
    TFF & PDG & MAID \\  \hline \hline 
    $F_1^v(0)$ & $0.73\pm 0.03 $  & 0.89\\  \hline 
     $F_2^v(0)$ & $0.32\pm 0.06$ &0.31\\ \hline $F_3^v(0)$ & - & $-1.03$ \\  \hline 

  \end{tabular}
  \caption{The isovector TFFs at the photon point, $F_i^v(q^2=0)$, based on the estimates from PDG \cite{PDG} and MAID \cite{Tiator:2009mt,Tiator:2011pw}. All quantities in GeV$^{-2}$. The uncertainties are based on the PDG results for $A_{\frac32}$ and $A_{\frac12}$, assuming no correlations between the uncertainties of these helicity amplitudes.}
  \label{tab:F0 and derivetives}
\end{table}

\begin{figure}[h!]
   \begin{subfigure}{0.45\textwidth}
\includegraphics[width=\textwidth]{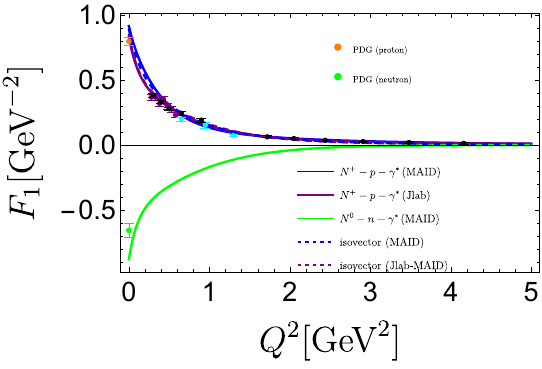}
   \end{subfigure}  \hfill
   \begin{subfigure}{0.45\textwidth}
\includegraphics[width=\textwidth]{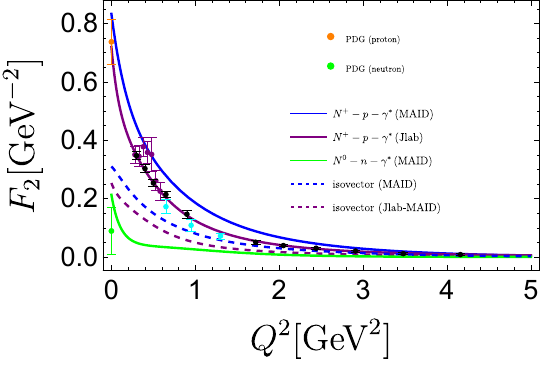}
\end{subfigure}
   \begin{subfigure}{0.45\textwidth}
   \centering
\includegraphics[width=\textwidth]{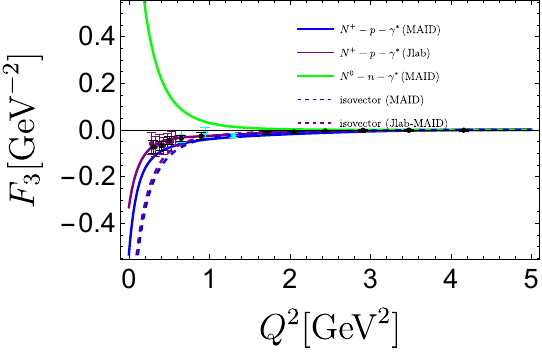}
\end{subfigure}
\caption{The data situation for the TFFs in the space-like region, $-q^2= Q^2 \ge 0$. $F_i^p$ data are shown in black \cite{CLAS:2009ces}, cyan \cite{Mokeev:2015lda} and purple \cite{jlabdata3}. Data at $Q^2=0$ (orange for proton and green for neutron) are taken from the PDG \cite{PDG}. The solid purple (``Jlab'') \cite{couplings} and green (``MAID'') \cite{Tiator:2009mt} lines are the respective parameterization for the TFFs of $p \gamma^* \to N^{*+}(1520)$ and $n \gamma^* \to N^{*0}(1520)$. Subtracting the Jlab and the MAID proton TFFs by the MAID neutron TFFs, the dashed purple and the dashed blue lines are obtained, respectively. The former is denoted as ``Jlab-MAID''. (Color online. The different colors refer to lines from top to bottom, except for the $F_3$ panel where they refer to lines from bottom to top.)}
\label{fig. tff data}
\end{figure}
In Fig.\ \ref{fig. tff data}, we show the space-like TFFs (electro-production) and also the photon point (photo-production). The first thing to notice is that there are data with error bars for the proton and data parametrizations without uncertainties for the neutron. This is because there are no systematic error estimates, as pointed out by the MAID group \cite{Tiator:2009mt,Tiator:2011pw}. This problem makes it also a bit unclear down to which $Q^2$ values one should trust the parameterizations; see also the corresponding discussion in \cite{Ramalho:2016zzo,Ramalho:2019ocp} where it has been pointed out that the MAID parameterization violates the Siegert theorem \eqref{eq:two-constr}. Though this constraint resides in the time-like region, it likely influences also the photon point and the low-energy space-like region. 

Our framework is suited for the isovector part of the TFFs, but not for the isoscalar. Therefore we cannot directly compare to the high-quality proton data. Instead we need to subtract proton and neutron parameterizations from each other according to \eqref{eq def isovector and isoscalar ffs}. Therefore the uncertainties for the isovector TFFs are hard to assess. Clearly, the situation would drastically improve with better neutron data. 

For the present paper, our strategy is to identify TFFs that are isovector dominated. This means that we check if an isovector TFF is close to the corresponding proton data (in the low-energy regime). In such a case, we use the proton error bars as a rough indication of the data uncertainties. Already the photon point shows that $F_1$ might be isovector dominated, but $F_2$ is clearly not. This finding is further supported by comparing in Fig.\ \ref{fig. tff data} the dashed isovector lines with the proton data points. In view of these observations, we will spend a larger degree of effort to reproduce $F_1^v$ with our framework. For $F_2^v$ we will stay more schematic in lack of any reasonable estimate of the data uncertainties.

Concerning $F_3$, the situation is further complicated by the fact that it cannot be extracted at the photon point. Note that $H_0$ is related to the longitudinal polarization, which does not exist for real photons. For $H_\pm$, the contribution of $F_3$ is multiplied in \eqref{eq:defG0pm} by the photon virtuality $q^2$. The absence of photon-point data increases the complications associated with the Siegert theorem \eqref{eq:two-constr} and its violation by data parameterizations. We observe in Fig.\ \ref{fig. tff data} that the dashed isovector lines coincide with the proton data for $Q^2\gtrsim 0.40\,\text{GeV}^2$. Given that the situation is unclear for low $Q^2$, we boldly assume isovector dominance also for $F_3$. Of course, whether $F_3$ is really isovector dominated at small $Q^2<0.4 \,\text{GeV}^2$ remains to be seen in future experiments addressing neutron TFFs. 

In short, we assume that
\begin{equation}
       F_{i,\rm{data}}^v(-Q^2) = F_{i,\rm{data}}^p(-Q^2) \quad \mbox{for} \quad i=1,3 \,.
\label{eq my estimate for spacelike TFFs}
 \end{equation}
Concerning $F_2$, we have a value for the photon point; see Tab.\ \ref{tab:F0 and derivetives}. We will fit our parameters such that we reproduce the photon point and get a slope that agrees reasonably well with the isovector parameterizations shown in Fig.\ \ref{fig. tff data}.

\subsection{Fixing hadronic and electronic parameters}
\label{sec: fix parameters}

Our strategy is to fit our parameters to space-like TFFs and to hadronic two-body decays $N^* \to \pi N, \pi \Delta, \rho N$ \cite{PDG,HADES:2020kce,Burkert:2022bqo}. It is worth to point out the conceptual difference between those two-body decays. The $\Delta$ baryon is an explicit degree of freedom in our framework. Therefore the decays $N^* \to \pi N, \pi \Delta$ are just described by tree-level calculations based on the Lagrangian \eqref{eq:lagrintNstar}. On the other hand, the $\rho$ meson is covered via the pion phase shift and not as an explicit quantum field.\footnote{As already pointed out, the use of the $\Delta$ instead of the corresponding pion-nucleon phase shift is an approximation.} To make contact with the experimental determination of the decays $N^* \to \rho N$ (with its three different partial waves) we use from our amplitudes \eqref{eq:tmandel} those parts that contain the pion phase shift, i.e.\ we use $T_i-K_i$. For the details of the calculation of $\Gamma_{N\rho}$, we refer the reader to Appendix \ref{sec decay width rho}. 

We start with the $N^{*}(1520)$-$N$-$\pi$ coupling $h$ and the $N^{*}(1520)$-$\Delta$-$\pi$ couplings $H_{1,2}$ that appear in the Lagrangian \eqref{eq:lagrintNstar}. 
The s-wave $N^{*}(1520)$-$\Delta$-$\pi$ amplitude contains a linear combination of $H_1$ and $H_2$ while the d-wave $N^{*}(1520)$-$\Delta$-$\pi$ amplitude depends on $H_2$, only. We take the values for the $N^*(1520) \to N \pi, \Delta \pi$ decay widths from the PDG \cite{PDG}. The corresponding solutions for $h$, $H_{1,2}$ are shown in Tab.\ \ref{tab. hadronic h H1 H2}.
 \begin{table}[H]
 \centering
\begin{tabular}{|c|c|c|}
\hline 
decay width & PDG value [GeV] & parameter  \\
\hline \hline  
$\Gamma_{\pi N}$ & $0.066\pm0.0055$& $h=-0.64\pm0.025$ \\
$\Gamma_{\pi\Delta,L=0}$& $(2.09\pm0.6)\times 10^{-2}$& $H_1=\pm 0.28\pm0.06$\\
$\Gamma_{\pi\Delta,L=2}$& $(9.9\pm2.2)\times 10^{-3}$&  $H_2=\mp 5.57\pm0.6$\\
\hline
\end{tabular}
\caption{The result for $h$, $H_{1,2}$ from the fit. The experimental values and uncertainties are taken from \cite{PDG}.}
\label{tab. hadronic h H1 H2}
\end{table}
As already pointed out in Subsection \ref{sec nstar eff lag}, we choose a negative sign for $h$ and signs for $H_1$ and $H_2$ that are opposite to each other. We have also explored a positive sign for $h$, but found that we cannot describe simultaneously the space-like isovector TFFs and the $\Gamma_{N\rho,L=0}$ decay width. 

As one can see in Tab.\ \ref{tab. hadronic h H1 H2}, we are left with one sign ambiguity for $H_1$ (and the respective opposite sign for $H_2$). We refer to the choice of $H_1=+0.28, H_2=-5.57$ and $H_1=-0.28, H_2=+5.57$ as parameter choices 1 and 2, respectively. In principle, one could distinguish the two signs by studying the photo-production process near the $N^*$ pole \cite{GomezTejedor:1995kj}, i.e.\ performing an amplitude analysis for $\gamma N \to N^*(1520) \to \pi \Delta$ and a background term, e.g.\ $\gamma N \to N \to \pi \Delta$. In the current work, we are focusing on the decay widths and therefore we are not able to distinguish the two choices. Consequently we will perform fits and present predictions based on both parameter choices.  

$F_2^v(0)$, together with its uncertainty, is fixed to the PDG value in Tab.\ \ref{tab:F0 and derivetives} while $p_2$ is adjusted such that the slope of the MAID and Jlab-MAID isovector TFF $dF_2^v(Q^2)/dQ^2$ at $Q^2=0$ is reproduced reasonably well. As already stressed, we do not attempt more than a schematic reproduction. For parameter choice 1, we find $p_2=-1.7\times10^{-3}$, while for choice 2, we obtain $p_2=-1.5\times10^{-3}$. It turns out that the impact of $F_2^v$ on $\Gamma_{N\rho}$ is rather small. Hence, the only way to get a better description and understanding of $F_2^v$ requires improved data on the neutron.

Now we turn to the determination of the parameters $F_{1,3}^v(0)$ and $p_{1,3}$. To this end, we employ a global fit to the TFF (proton) data and to the main $N^* \to N \rho$ decay width \cite{PDG,HADES:2020kce}. The best-fit parameters are found by minimizing 
\begin{equation}
    \chi^2 = \left(\frac{\Gamma_{N\rho,L=0}-\Gamma_{N\rho,L=0}^{\rm data}}{\sigma_{N\rho}}\right)^2+\sum_{i=1,3} \left(\frac{F_i^v-F_{i,\rm{data}}^v}{\sigma_{i} }\right)^2 
    \label{eq chisure}
\end{equation}
where  $\sigma_{\ldots}$ are the respective error estimates.
For the space-like TFFs, we take $F_{i,\rm{data}}^v$ from the range $0\leq Q^2< 0.5\,\text{GeV}^2$, where $i=1,3$. There are 8 data points for $F_{1,\rm{data}}^v$ and 7 data points for $F_{3,\rm{data}}^v$ in this region.\footnote{$F_{3}^v(0)$ is not measurable.}

\begin{table}[H]
  \centering
  \begin{tabular}{|c|c|c|c|c|c|c|c|c|c|}
    \hline
    &  $p_1$  & $p_3$&$F_1^v(0)\,[\text{GeV}^{-2}]$& $F_3^v(0) \,[\text{GeV}^{-2}]$ & $\chi^2$/dof\\\hline \hline  
parameter choice 1&$0.075_{-0.005}^{+0.005}$&$-0.0181_{-0.0008}^{+0.0008}$& $0.70_{-0.024}^{+0.024}$&$-0.35_{-0.04}^{+0.04}$&7.1/12\\
parameter choice 2&$0.076_{-0.005}^{+0.005}$&$-0.0166_{-0.0008}^{+0.0008}$&$0.697_{-0.024}^{+0.024}$&$-0.150_{-0.04}^{+0.04}$&8.4/12\\
\hline 
  \end{tabular}
  \caption{Predictions for our parameters, i.e.\ the TFFs $F_{1,3}^v(0)$ at the photon point and the parameters $p_{1,3}$ entering the hadronic subtraction constants via \eqref{eq:poly-conn-contactL}.}
  \label{tab:couplings}
\end{table}
\begin{table}[h]
\centering

\begin{tabular}{|c|c|c|c|c|c|}
\hline 
  decay width [GeV]& experimental results & parameter choice 1& parameter choice 2\\\hline \hline  

$\Gamma_{N\rho,L=0,S=\frac{3}{2}}$ & $(1.43\pm0.33) \times10^{-2}$ \cite{PDG,HADES:2020kce}& $(1.60\pm 0.26)\times 10^{-2}$ & $(1.48\pm 0.25)\times 10^{-2}$ \\
$\Gamma_{N\rho,L=2,S=\frac{1}{2}}$& $(3.3\pm2.2) \times10^{-4}$ \cite{PDG,HADES:2020kce} & $(1.00\pm 0.11)\times 10^{-4}$ & $(0.78\pm 0.10)\times 10^{-4}$ \\
$\Gamma_{N\rho,L=2,S=\frac{3}{2}}$ & $\approx0$ \cite{Burkert:2022bqo}& $(1.07\pm 0.20)\times 10^{-5}$ & $(1.35\pm 0.20)\times 10^{-5}$ \\
\hline
\end{tabular}
\caption{Results for the $N\rho$ decay widths (different partial waves) from the fit in (\ref{eq chisure}). For each parameter choice, the input is $\Gamma_{N\rho,L=0,S=\frac{3}{2}}$ while the two $L=2$ widths are pure predictions.}
\label{tab. fit width}
\end{table}
The results for our parameters and the $N\rho$ decay widths are shown in Tab.\ \ref{tab:couplings} and \ref{tab. fit width}, respectively. We estimate the uncertainty of the parameters by reading off the tangent values of each parameter in the hyper-ellipse of $\chi^2$ \cite{Behnke:2013pga}. 

In this way, we are very conservative in estimating our statistical uncertainty with the purpose to compensate for the lack of isovector data at very low energies $Q^2<0.3 \,\text{GeV}^2$. Clearly the errors must be taken with a grain of salt. Note that we only include the s-wave hadronic decay width $\Gamma_{N\rho, L=0}$ in the fit (\ref{eq chisure}). The two experimentally much smaller \cite{PDG,HADES:2020kce,Burkert:2022bqo} d-wave decay widths are not used in the fit (\ref{eq chisure}). Thus they are a prediction of our formalism and it is reassuring that these subleading decay widths come out as small.

This way of fixing the subtraction constants $p_{1,3}$ maximally takes advantage of the experimental data available, including hadronic, photo-, and electro-production data. 
We explore now a second option by not including the decay width $\Gamma_{N\rho,L=0}$ in the $\chi^2$ fit in (\ref{eq chisure}). Instead we include only the electro-production data by using $\chi^2=\sum_{i=1,3}\left(\frac{F_i^v-F_{i,\rm{data}}^v}{\sigma_{i}}\right)^2$. Interestingly, we find that even the s-wave $\Gamma_{N\rho,L=0}$ decay width can be reproduced. Thus we can predict all three hadronic decay widths $\Gamma_{N\rho}$. The results are presented in Tab.\ \ref{tab. fit width method2}.
\begin{table}[h]
\centering
\begin{tabular}{|c|c|c|c|c|c|}
\hline 
  decay width [GeV]& experimental results & parameter choice 1& parameter choice 2 \\ \hline \hline  

$\Gamma_{N\rho,L=0,S=\frac{3}{2}}$ & $(1.43\pm0.33) \times10^{-2}$ \cite{PDG,HADES:2020kce}& $(1.89\pm 0.29)\times 10^{-2}$ & $(1.55\pm 0.26)\times 10^{-2}$ \\
$\Gamma_{N\rho,L=2,S=\frac{1}{2}}$& $(3.3\pm2.2) \times10^{-4}$ \cite{PDG,HADES:2020kce} & $(0.98\pm 0.14)\times 10^{-4}$ & $(0.75\pm 0.10)\times 10^{-4}$ \\
$\Gamma_{N\rho,L=2,S=\frac{3}{2}}$ & $\approx0$ \cite{Burkert:2022bqo}& $(1.25\pm 0.10)\times 10^{-5}$ & $(1.30 \pm 0.20)\times 10^{-5}$ \\
\hline
\end{tabular}
\caption{Prediction of all $N\rho$ decay widths for the case where only TFF data are used in the fit.}
\label{tab. fit width method2}
\end{table}

Those encouraging results provide evidence that the dispersive formalism is compatible with both the TFFs and the hadronic decays. Nevertheless, to better make use of all the available data including all the hadronic, photo- and electro-production data, we still adopt \eqref{eq chisure} for the results presented in the following. For all figures we adopt the scheme
\begin{itemize}
    \item parameter choice 1: $H_1 > 0$, $H_2 < 0$, orange bands/lines in figures;
    \item parameter choice 2: $H_1 < 0$, $H_2 > 0$, black bands/lines in figures.
\end{itemize}

\begin{figure}[h!]
	\centering
	\begin{subfigure}{0.45\textwidth}
		\includegraphics[width=\textwidth]{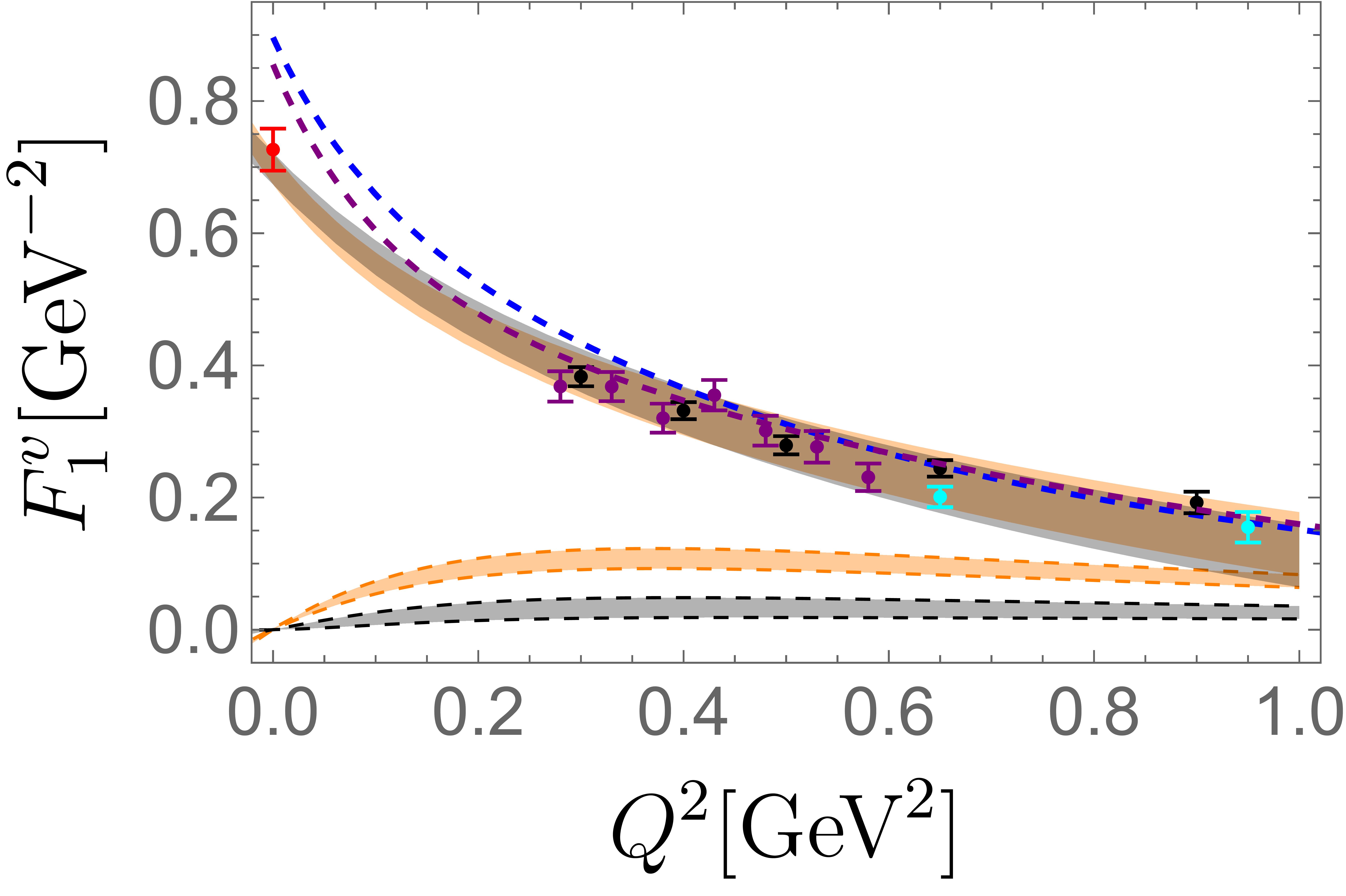}
	\end{subfigure} \hfill
	\begin{subfigure}{0.45\textwidth}
		\includegraphics[width=\textwidth]{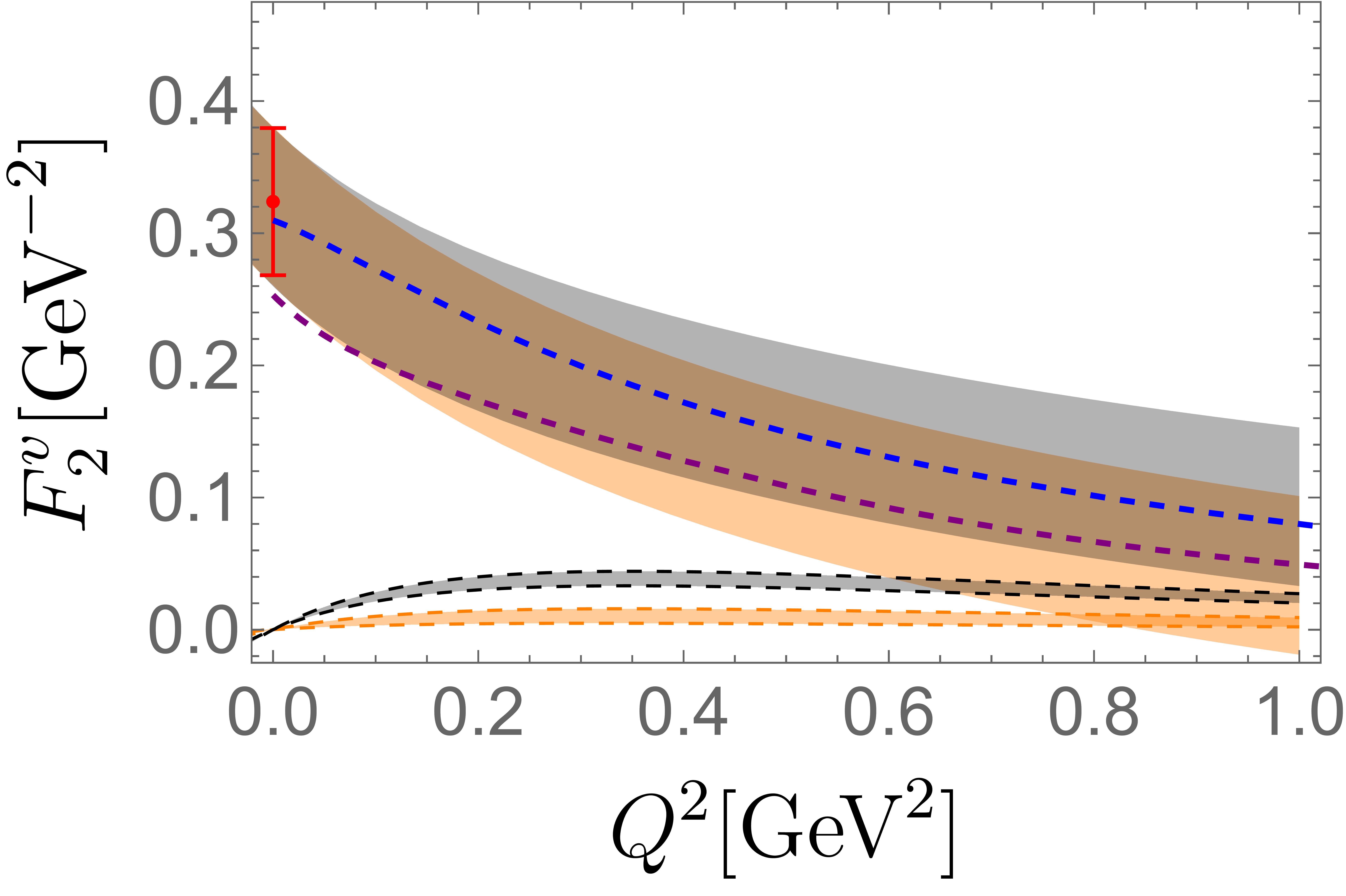}
	\end{subfigure}
	\begin{subfigure}{0.45\textwidth}
		\centering
		\includegraphics[width=\textwidth]{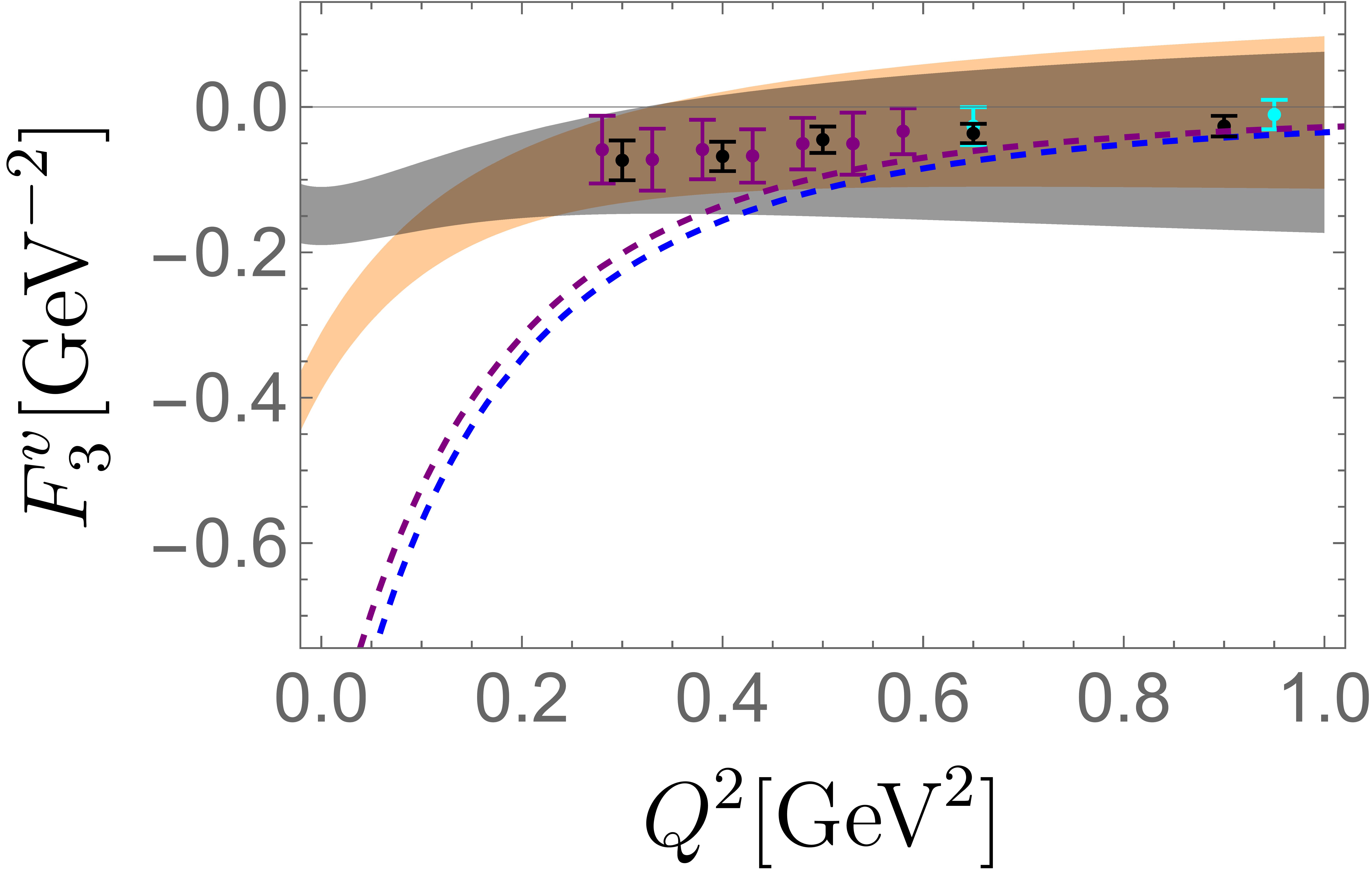}
	\end{subfigure}
	\caption{Dispersive results (bands) for the space-like isovector TFFs in comparison to proton data and Jlab and MAID parameterizations. Bands limited by full (dashed) lines display the real (imaginary) part. Orange and black refers to parameter choices 1 and 2, respectively. The dashed blue and purple curves show the MAID and Jlab-MAID parameterizations, respectively. They are the same as in Fig.\ \ref{fig. tff data}. For the $F_3^v(Q^2)$ panel, only the real parts are displayed to prevent the figure from becoming overly crowded. (Both real and imaginary parts are shown in Fig.\ \ref{fig timelike ff} below.) (Color online. Curves and bands show from top to bottom (to the left of the respective figure): $F_1^v$ panel: MAID isovector parameterization, Jlab-MAID isovector parameterization, real part for parameter choice 2, real part for parameter choice 1, imaginary part for parameter choice 2, imaginary part for parameter choice 1. $F_2^v$ panel: real part for parameter choice 2, real part for parameter choice 1, MAID isovector parameterization, Jlab-MAID isovector parameterization, imaginary part for parameter choice 1, imaginary part for parameter choice 2. $F_3^v$ panel: real part for parameter choice 2, real part for parameter choice 1, Jlab-MAID isovector parameterization, MAID isovector parameterization.)}
	\label{fig spacelike ff}
\end{figure}
The resulting space-like TFFs $F_{1,2,3}^v$ are shown in Fig.\ \ref{fig spacelike ff}, calculated from the dispersion relations \eqref{eq:dispbasic} with the input parameters presented in Tab.\ \ref{tab:couplings}. 
For all three TFFs in the space-like region, the dispersive method provides an accurate description of the data not only in the fit region $Q^2< 0.5 \, \text{GeV}^2$ but also up to $Q^2=1 \, \text{GeV}^2$. For $F_1^v$, both parameter choices 1 and 2 yield similar predictions (for the real part), resulting in comparable radii $r_1^2$ (as defined below in \eqref{eq:radii} and listed in Tab.\ \ref{tab:Predictions}) and curvatures. One can also see in Tab.\ \ref{tab:couplings} that the parameters $F_1^v(0)$ and $p_1$ are essentially the same for both parameter choices. Thus the impact of the $\Delta$ triangle diagrams of Fig.\ \ref{fig:cross channal fey 1} is of minor importance for $F_1^v$. 

Both parameter choices also produce similar predictions for the space-like $F_2^v(-Q^2)$. In particular, our results align with the MAID parametrization (blue dashed) up to $Q^2 =1 \, \text{GeV}^2$, even though only the radius from MAID is used to fix $p_2$. 

Regarding $F_3^v$, the two parameter choices provide similar predictions for $Q^2 \geq 0.3 \,\text{GeV}^2$ but differ at lower $Q^2$. As illustrated in Fig.\ \ref{fig spacelike ff}, parameter choice 2 (black) predicts a smaller $F_3^v(0)$ at the photon point as compared to choice 1 (orange). Additionally, parameter choice 2 has a smaller radius and curvature due to the opposite sign of the $\Delta$ triangles and a lower value of $p_3$, as indicated in Tab.\ \ref{tab:couplings}. Note that our bands agree with the (proton) data but disagree at low values of $Q^2$ with the isovector parameterizations. As already pointed in out in the discussion preceding \eqref{eq my estimate for spacelike TFFs}, there are severe reasons to distrust these parameterizations at low $Q^2$. 

So far we have focused on the discussion of the respective real parts of the TFFs.
Some words of clarification are in order. TFFs of unstable particles are complex, even in the space-like region \cite{Unal:2021byi,Junker:2019vvy,Aung:2024qmf}. The experimental results to which we compare correspond to the real parts of the TFFs, if the results have been extracted from the peak position of the resonance seen e.g.\ in the partial-wave analysis of the pion-nucleon final state \cite{Aung:2024qmf}. In principle, our fit parameters $F_i^v(0)$ are also complex, but we obtain only their real parts from the data. As a self-consistency check we have determined the TFFs also from an unsubtracted dispersion relation using the previously determined parameter values. We found rather small imaginary parts at the photon point. In Fig.\ \ref{fig spacelike ff} we have used the better convergent subtracted dispersion relations \eqref{eq:dispbasic} and dropped any possible (presumably small) imaginary part at the photon point.

\begin{figure}[h!]
\centering\includegraphics[width=0.5\textwidth]{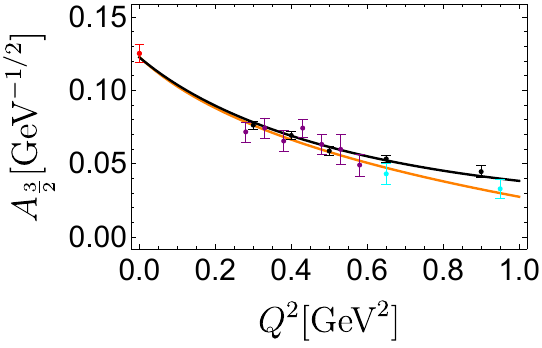}
    \caption{Results for the helicity amplitude $A_{\frac32}$ as compared to proton data. Color code of the data is the same as in Fig.\ \ref{fig. tff data}. Color code of the theory lines is the same as in Fig.\ \ref{fig spacelike ff}. (Color online. The bottom (top) line shows the results from parameter choice 1 (2).)}
    \label{fig: A32 spacelike}
\end{figure}

It is worth to discuss also the helicity amplitudes and not only the constraint-free TFFs. But here the problem with the poor neutron data and the corresponding uncertainties related to $F_2$ comes back. Fortunately, there is one helicity amplitude that does not depend on $F_2$. 
For the helicity amplitudes (\ref{eq:defG0pm}), the quantity that only depends on $F_{1,3}$ is $H_-$, which is essentially $A_\frac{3}{2}$ (see (\ref{eq translation Hs and As})). In Fig.\ \ref{fig: A32 spacelike}, our best-fit results for the $A_\frac{3}{2}$ helicity amplitude, assumed to be isovector dominated, is compared to proton data. We observe an excellent description of the data.\footnote{Of course, we compare here the real part of our helicity amplitude to the data.}


\subsection{Dispersive predictions}
\label{sec: dispersive predictions}

\begin{figure}[h!]
	\centering
	\begin{subfigure}{0.45\textwidth}
		\includegraphics[width=\textwidth]{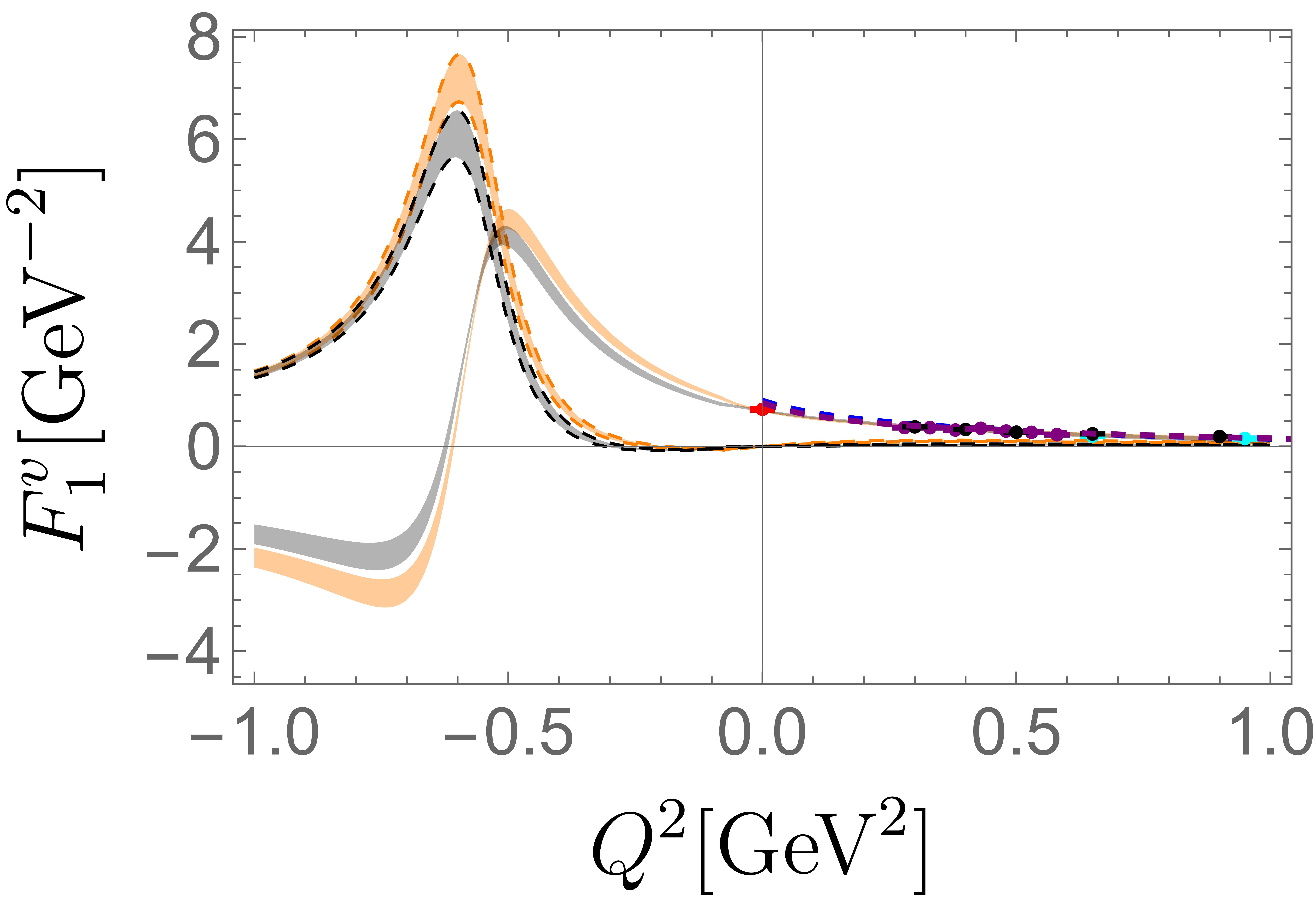}
	\end{subfigure} \hfill
	\begin{subfigure}{0.45\textwidth}
		\includegraphics[width=\textwidth]{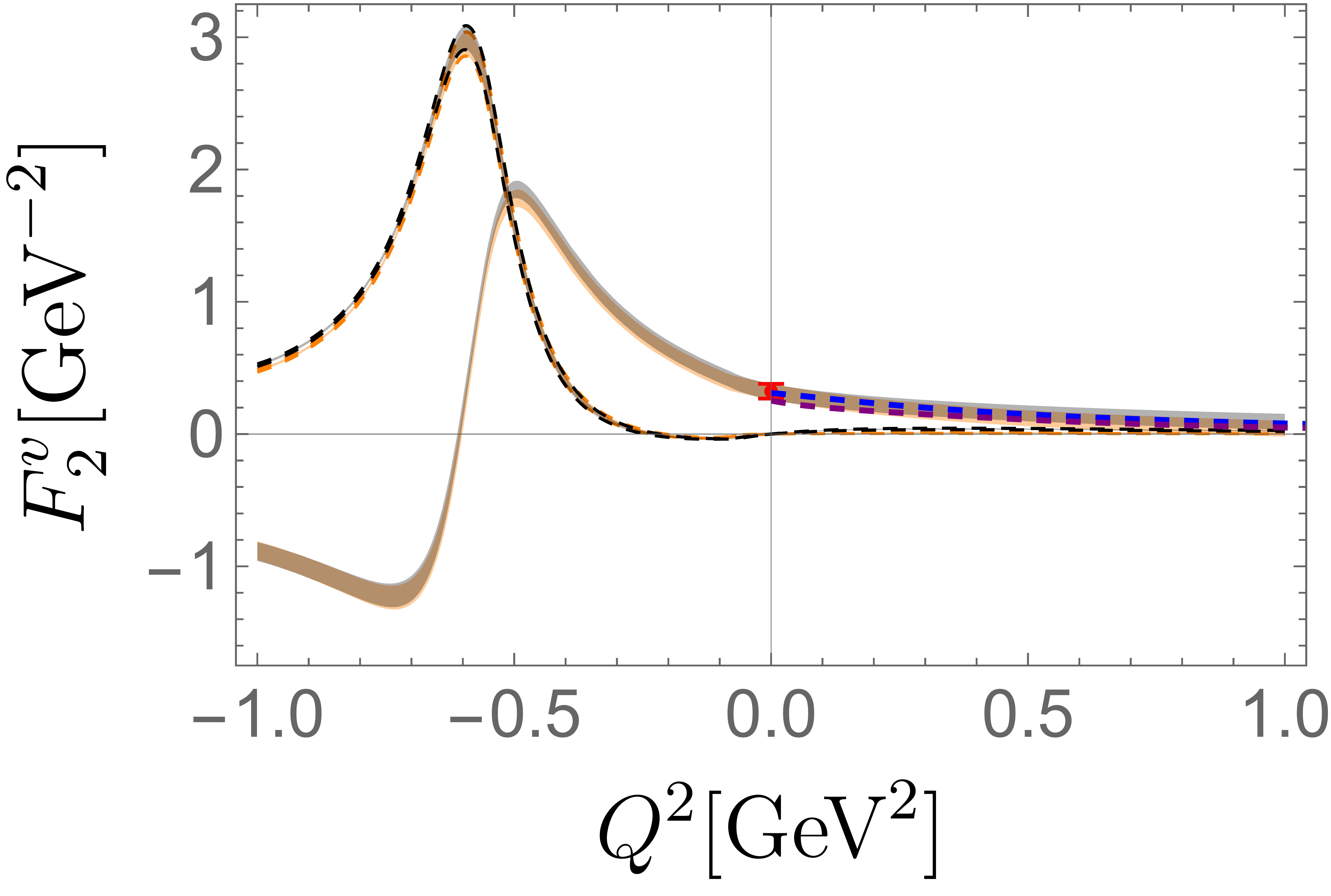}
	\end{subfigure} 
	\begin{subfigure}{0.45\textwidth}
		\centering
		\includegraphics[width=\textwidth]{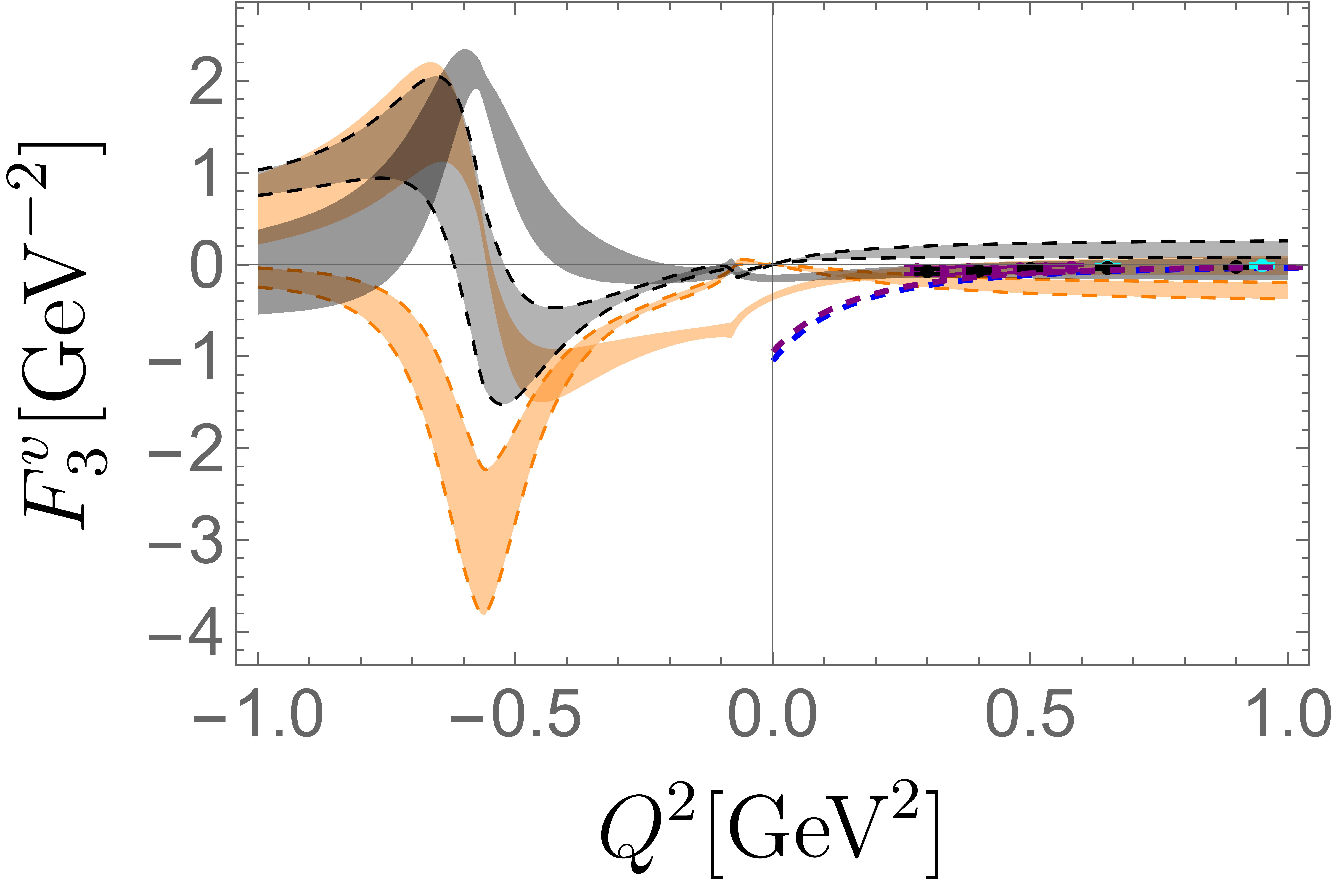}
	\end{subfigure}
	\caption{Time-like ($Q^2 < 0$) and space-like ($Q^2 > 0$) isovector TFFs. Same color codes as in Fig.\ \ref{fig spacelike ff}.  }
	\label{fig timelike ff}
\end{figure}
Dispersion theory naturally relates the space-like and time-like regions. Once the free parameters are fixed from space-like data, dispersion theory gains predictive power in the time-like region. The TFFs in the full space- and time-like low-energy region are depicted in Fig.\ \ref{fig timelike ff}. The first thing to observe is that for $F_{1}^v$ as well as for $F_2^v$ both parameter choices yield similar predictions in the time-like region. However, for $F_3^v$, the predictions for the two parameter choices are markedly different in the time-like region. 

Real and imaginary parts of $F_1^v$ and $F_2^v$ show the usual pattern \cite{Granados:2013moa} in the time-like region: a peak of the imaginary part and a zero crossing of the real part around the mass of the $\rho$ meson, i.e.\ for 
$q^2 = -Q^2 \approx m_\rho^2 \approx 0.6 \,\text{GeV}^2$. This pattern fits to the behavior of $F_3^v$ for parameter choice 1 (orange bands), but is inverted for parameter choice 2 (black bands). This discrepancy arises from the dominance of triangle diagrams (Fig.\ \ref{fig:cross channal fey 1}) for parameter choice 2, in contrast to the normal dominance of the contact diagram (Fig.\ \ref{fig:point channel fey 2}). We observe here a drastic qualitative difference to vector-meson dominance models where baryonic triangles are not considered. Such models lead always to a peak in the imaginary part. At present, we have no means to decide which parameter choice is more realistic. Consequently, we do not know whether the scenario with the high importance of the baryon triangles (parameter choice 2) is realized for $F_3^v$. Exploring real and imaginary parts of amplitudes in more detail requires polarization data; see, e.g.\ \cite{Salone:2021bvx} for the corresponding case of hyperon decays.

\begin{figure}[h!]
    \centering
\begin{subfigure}{0.45\textwidth}
		\centering
		\includegraphics[width=\textwidth]{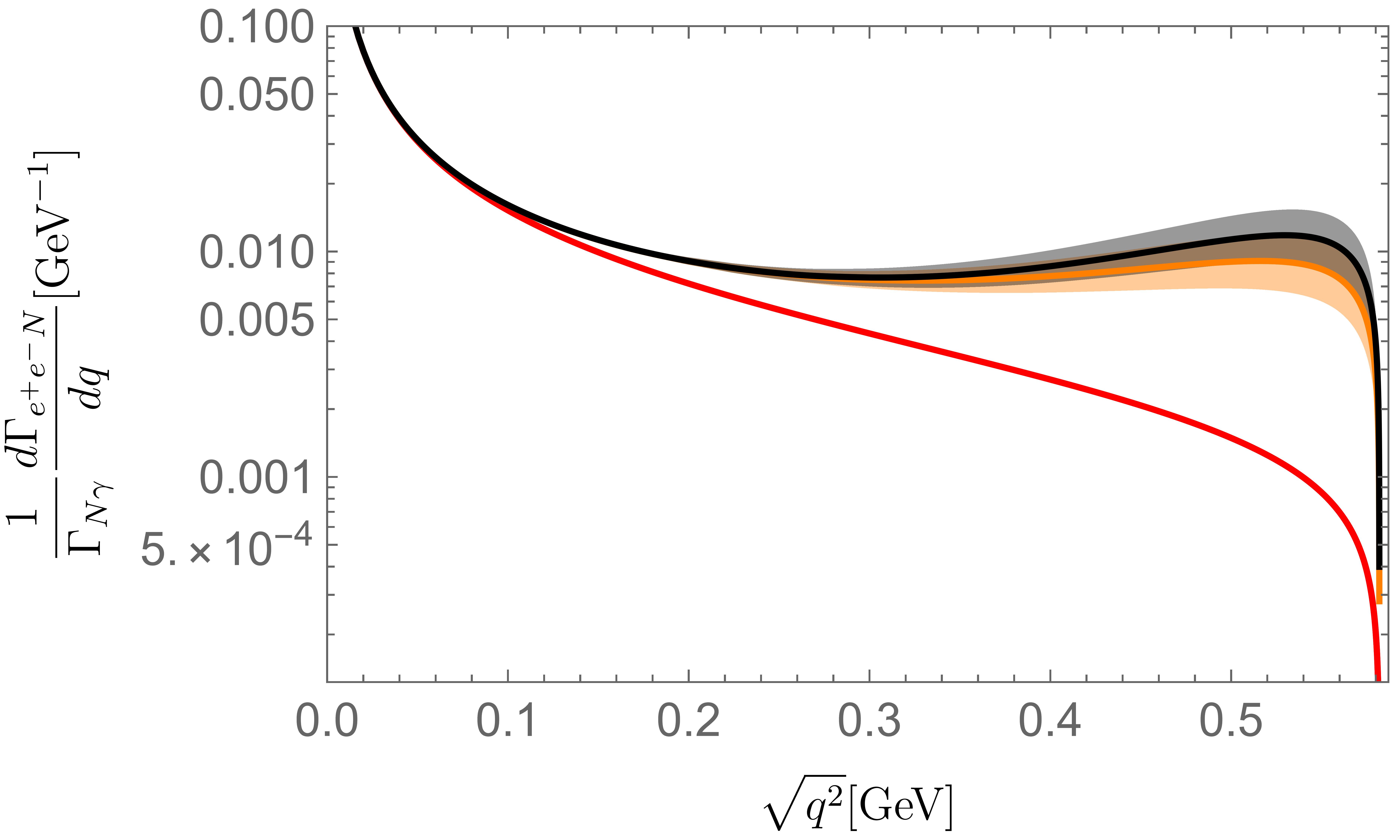}
  \caption{invariant-mass distribution for the electronic case}
  \label{fig:Dalitz-invmass-elec}
	\end{subfigure} \hfill 
 \begin{subfigure}{0.45\textwidth}
		\centering
		\includegraphics[width=\textwidth]{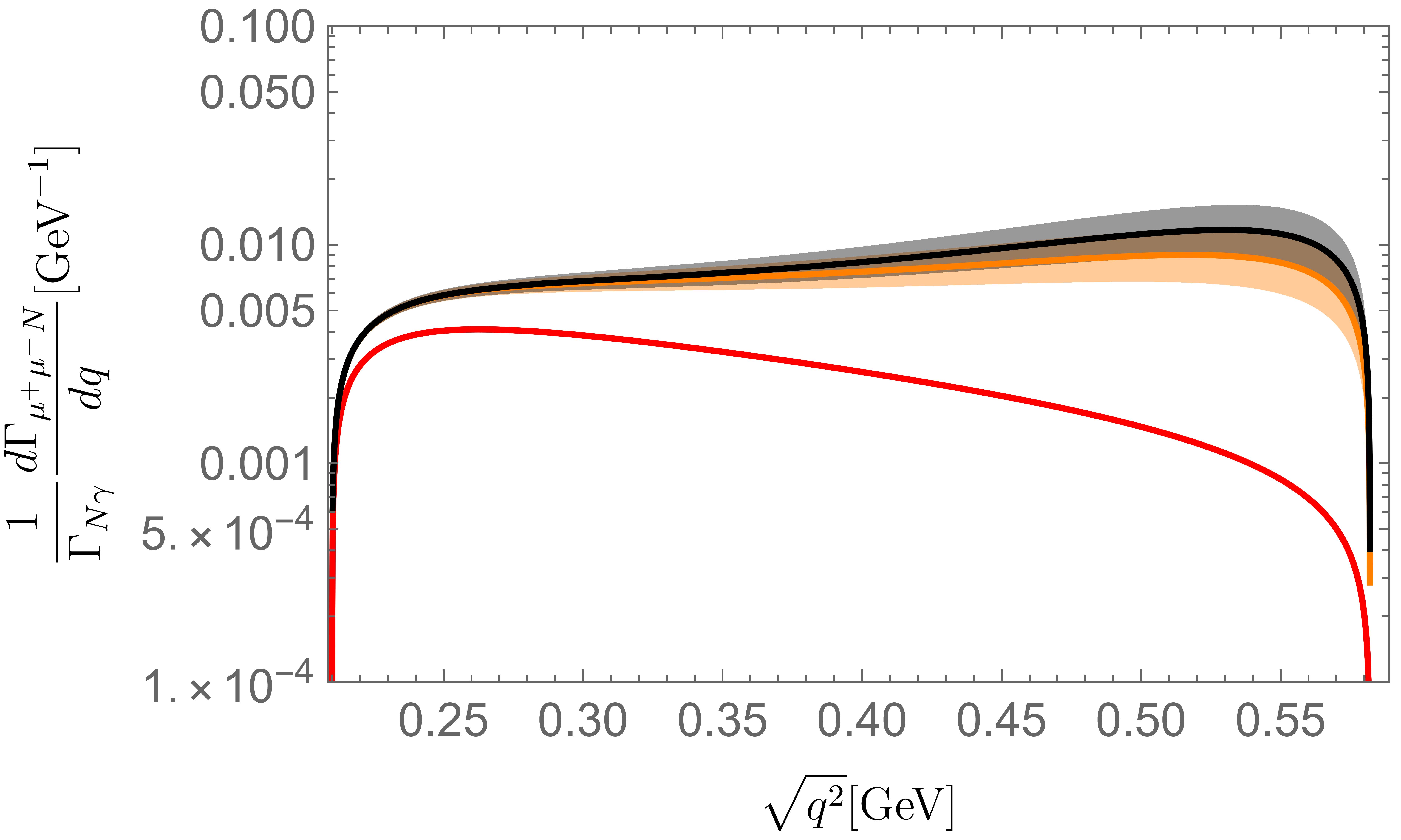}
  \caption{invariant-mass distribution for the muonic case}
  \label{fig:Dalitz-invmass-muno}
	\end{subfigure}
  \begin{subfigure}{0.45\textwidth}
		\centering
		\includegraphics[width=\textwidth]{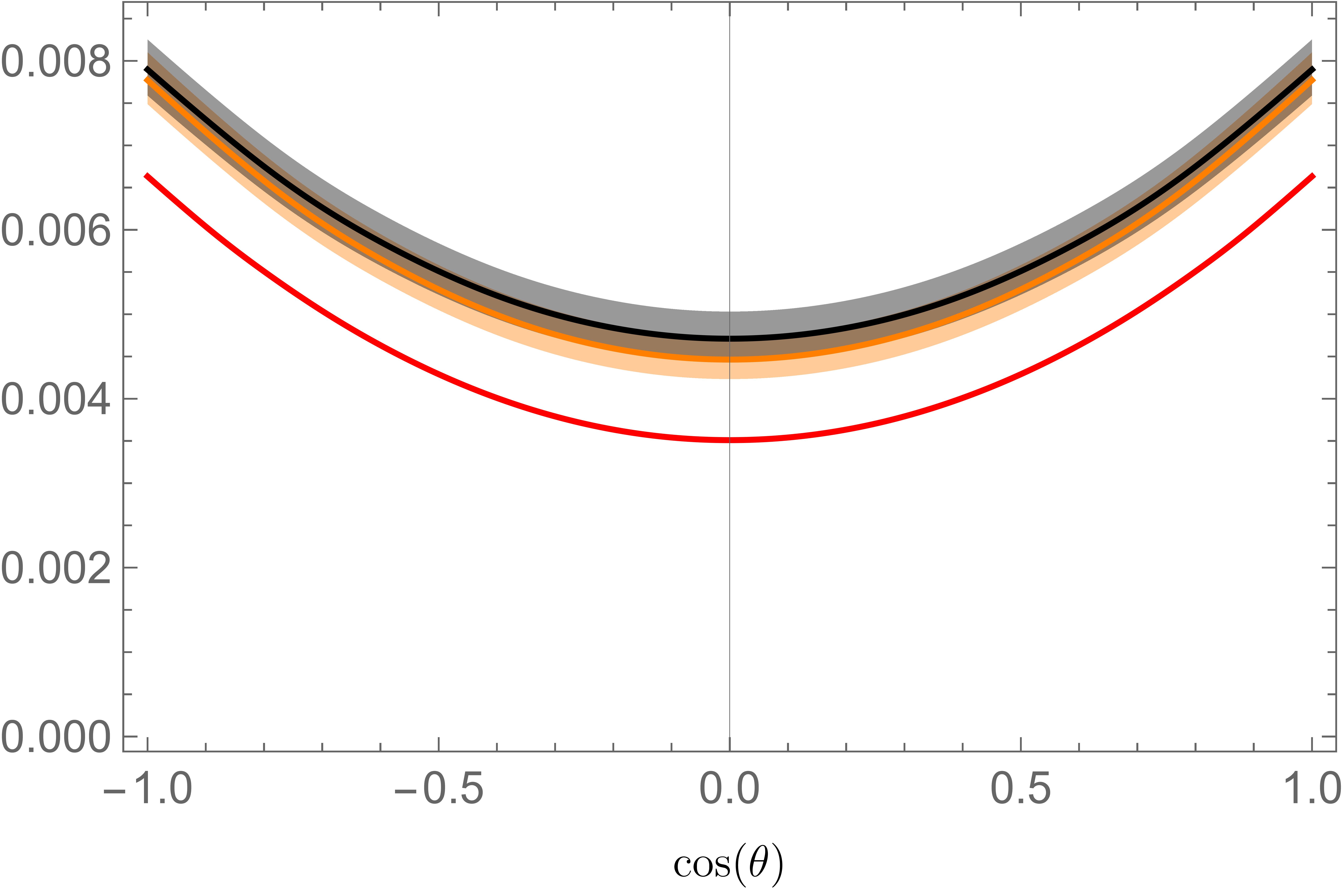}
  \caption{angular distribution $\frac{1}{\Gamma_{N\gamma}} \frac{d\Gamma_{Ne^+e^-}}{d\cos{\theta}}$}
  \label{fig:Dalitz-angle-elec}
	\end{subfigure}  \hfill
 \begin{subfigure}{0.45\textwidth}
		\centering
		\includegraphics[width=\textwidth]{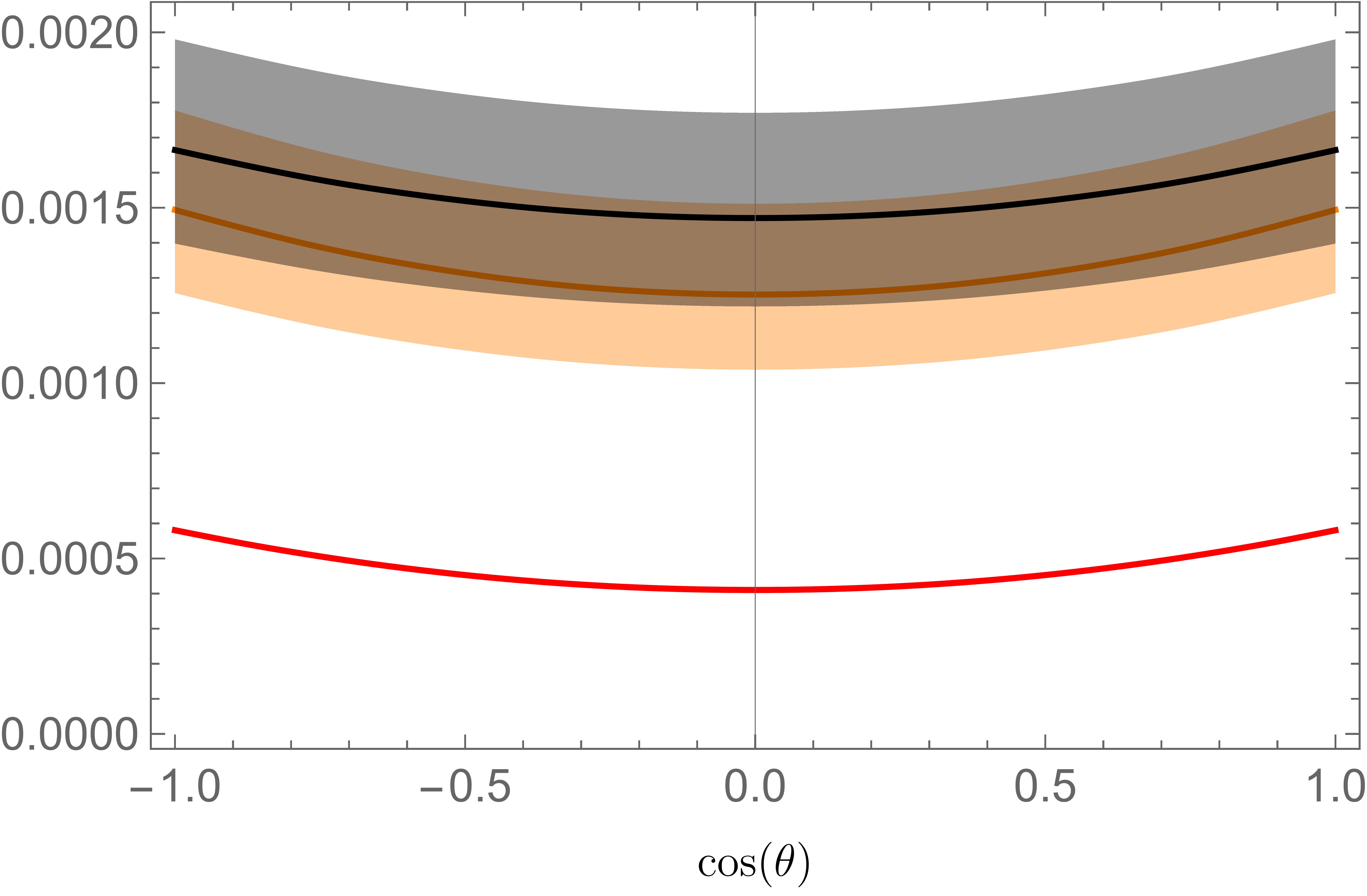}
    \caption{angular distribution $\frac{1}{\Gamma_{N\gamma}} \frac{d\Gamma_{N\mu^+\mu^-}}{d\cos{\theta}}$}
    \label{fig:Dalitz-angle-muon}
	\end{subfigure}
 \caption{The predicted electronic and muonic differential decay widths. $\theta$ is the angle between $l^+$ and $N$ in the virtual photon's rest frame. Orange and black band/lines are plotted using parameter choice 1 and 2, respectively. The QED curve is in red. (Color online. From top to bottom: black, orange, red.)}
 \label{fig diff decay width}
\end{figure}

Concerning observables in the time-like region, we focus in the present work on electronic and muonic Dalitz decays $N^* \to N \, \ell^+ \ell^-$ without accounting for a polarization of the initial resonance or the final nucleon. Instead of Dalitz plots of \eqref{eq:diffdec} we show in Fig.\ \ref{fig diff decay width} the single-differential quantities $\frac{1}{\Gamma_{N\gamma}}\frac{d\Gamma}{d q}=\frac{2\sqrt{q^2}}{\Gamma_{N\gamma}}\frac{d\Gamma}{d q^2}$ and $\frac{1}{\Gamma_{N\gamma}}\frac{d\Gamma}{d \cos \theta}$. 

Before discussing the results we stress again that our prediction is based on the assumption that the isovector part of the TFFs dominates over the isoscalar part. This is supported by the radiative decay $N^*(1520) \to N \gamma$, which has equal strength for proton and neutron \cite{PDG}, and by the observation that $F_1$ as the most important TFF is isovector dominated in the space-like region; see Fig.\ \ref{fig. tff data}.

The first thing to notice from Fig.\ \ref{fig diff decay width} is the fact that black and orange bands lie relatively close together and are significantly different from the red line which shows the QED version based on \eqref{eq:diffdec QED}. Thus we see structure effects both for the electronic and for the muonic channel. And our formalism makes a clear quantitative prediction. 

Having normalized to the decay width for real photons, there cannot be any difference between the QED curve and our results for low $q^2$ in Fig.\ \ref{fig:Dalitz-invmass-elec}. However, at higher $q^2$, the dispersive predictions display a more pronounced shoulder compared to the QED prediction, which is attributed to the significant impact of the $\rho$ meson. This interesting part of the $q^2$ region is already at so high $q^2$ that there is not much difference between the muonic case of Fig.\ \ref{fig:Dalitz-invmass-muno} and the electronic case displayed in Fig.\ \ref{fig:Dalitz-invmass-elec}.

The noticeable difference to the QED case is also evident in the predicted angular distribution depicted in Figs.\ \ref{fig:Dalitz-angle-elec} and \ref{fig:Dalitz-angle-muon}. For the electron case the difference is not so pronounced because of the large influence from the photon peak at small $q^2$. The QCD structure effect is more visible in the muonic Dalitz decay where the photon peak is cut away by the limited phase space related to the higher muon mass. 

The angular distribution shown in Fig.\ \ref{fig diff decay width} takes the form 
 \begin{equation}
    \frac{1}{\Gamma_{N\gamma}} \frac{d\Gamma}{d \cos \theta} =A^\ell(1+B^\ell \cos^2\theta)\,,
 \end{equation}
 where $\ell=e,\mu$.
Our predictions for those parameters $A^\ell$, $B^\ell$ are provided in Tab.\ \ref{tab:Predictions}, together with the integrated electronic and muonic decay widths and the radii \cite{KubisPhD,Kubis:2000aa} associated with the respective TFF, 
\begin{equation}
  \langle r^2_i \rangle := \frac{6}{F_i(0)} \left. \frac{\text{d} F_i(q^2)}{\text{d}q^2} \right\vert_{q^2=0}  
  \label{eq:radii}
\end{equation}
for $i=1,2,3$. 

The typical radius of a hadron is about $1\,$fm. Examples are the electric radius of the proton with $\langle r_E^2 \rangle = 0.705 \,\text{fm}^2$ or its magnetic radius, $\langle r_M^2 \rangle = 0.724 \,\text{fm}^2$ \cite{PDG}. As seen in Table \ref{tab:Predictions}, the radii $\langle r_1^2 \rangle$ and $\langle r_2^2 \rangle$ are similar to the proton radii and also comparable to radii from other transitions as e.g.\ $\Delta \to N$ \cite{Aung:2024qmf} and $\Sigma^{(*)} \to \Lambda$ \cite{Junker:2019vvy,Granados:2017cib}. In addition, the respective real part is larger than the imaginary part for our predictions for $\langle r_1^2 \rangle$ and $\langle r_2^2 \rangle$, though the size of the imaginary part depends on the parameter choice 1 or 2. This is even more pronounced for $\langle r_3^2 \rangle$. We obtain very different results depending on the parameter choice. For choice 1, the real part of our $\langle r_3^2 \rangle$ is significantly larger than the nucleon radius, whereas for choice 2 the value is tiny. But there, the imaginary part is particularly large. Only combinations of these quantities enter the Dalitz decay widths. Determining the TFF radii separately would require polarization data. More generally, Fig.\ \ref{fig diff decay width} shows which experimental accuracy is required to distinguish parameter choices 1 and 2.

Our dispersive result for $\langle r_3^2 \rangle$ can also be compared with the corresponding result for the Jlab-MAID parameterization, $\langle r_3^2 \rangle_{\text{Jlab-MAID}} = 1.58 \,\text{fm}^2$, and the MAID isovector parameterization, $\langle r_3^2 \rangle_{\text{MAID}} = 1.53 \,\text{fm}^2$. Both the Jlab-MAID and MAID parameterization agree with our parameter choice 1. But we stress once more that should be sceptical about the extrapolation of those parameterizations to the not-measured low-$\vert q^2 \vert$ region. 

Our predictions for the integrated electronic and muonic Dalitz decay widths for parameter choice 1 and 2 are compatible with each other within their uncertainties. For the electronic case the prediction is a bit larger than for the structureless QED case. But for the muonic decay there is a large difference of about a factor of 3 between QCD and QED. 

Tab.\ \ref{tab:Predictions}  also shows that in order to distinguish the composite baryon structure of QCD from pointlike QED, one can use $A^l/A^l_{\text{QED}}$ as a measure. We find that $A^e/A^e_{\text{QED}} = 1.29$ whereas $A^\mu/A^\mu_{\text{QED}} = 3$. Therefore, we conclude that the best way to distinguish QCD from QED is by measuring the muonic Dalitz decay.

\begin{table}[H]
  \centering
  \begin{tabular}{|c|c|c|}
    \hline
    quantity & parameter choice 1  & parameter choice 2\\  \hline \hline
  
    $\langle r_1^2 \rangle$[fm$^2$] & $0.69-0.28i$ & $0.53-0.05i$ \\  \hline
   
    $\langle r_{2}^2 \rangle$[fm$^2$] & $0.55 -0.07\,i$   & $0.55 -0.27\,i$\\  \hline
  
    $\langle r_{3}^2 \rangle$[fm$^2$] & $1.59 - 0.49 \,i$&$0.02 + 2.69 \,i$\\ \hline \hline
    $\Gamma_{N^*\rightarrow N e^+ e^-}$/$\Gamma_{N^*\rightarrow N \gamma}$   & $(1.12\pm0.06)\times 10^{-2}$ &$(1.16\pm0.06)\times 10^{-2}$\\  \hline
    $\Gamma_{N^*\rightarrow N e^+ e^-}^{\rm{QED}}$/$\Gamma_{N^*\rightarrow N \gamma}$   &    \multicolumn{2}{c|}{$9.10\times 10^{-3}$}  \\ \hline
     $\Gamma_{N^*\rightarrow N \mu^+ \mu^-}$/$\Gamma_{N^*\rightarrow N \gamma}$   & $(2.71\pm0.50)\times 10^{-3}$ &$(3.12\pm0.55)\times 10^{-3}$\\  \hline
  $\Gamma_{N^*\rightarrow N \mu^+ \mu^-}^{\rm{QED}}$/$\Gamma_{N^*\rightarrow N \gamma}$ & \multicolumn{2}{c|}{$9.34\times 10^{-4}$} \\ \hline \hline
  $A^e,B^e$& $A=4.5\times10^{-3},B=0.74$ & $A=4.7\times10^{-3},B=0.67$ \\ \hline
$A^e_{\text{QED}},B^e_{\text{QED}}$& \multicolumn{2}{c|}{$A=3.5\times 10^{-3}$, $B=0.89$}  \\ \hline
    $A^\mu,B^\mu$& $A=1.2\times 10^{-3},B=0.19$  & $A=1.5\times 10^{-3},B=0.13$ \\
 \hline 
$A^\mu_{\text{QED}},B^\mu_{\text{QED}}$& \multicolumn{2}{c|}{$A=0.41\times 10^{-3}$, $B=0.41$}\\\hline
  \end{tabular}
  \caption{Predictions for radii and Dalitz decay widths. The real photon decay width is calculated using (\ref{eq real photon}) where we have used the isovector TFFs at the photon point.}
  \label{tab:Predictions}
\end{table}

\section{Summary and outlook} 
\label{sec:summary}

We have established a model-independent determination of the isovector TFFs for $N^*(1520)\longleftrightarrow N \gamma^*$ using dispersion theory. The formalism is appropriate for the low-energy region where the two-pion intermediate state dominates the structure aspect. All other asymptotic intermediate states are too heavy to transport information over long distances of about $1\,$fm. What we do not resolve is encoded in subtraction constants of the dispersive integrals for the TFFs and for the pion-baryon reaction amplitudes. Our parameters have been fixed from fits to hadronic two-body decays ($N^*(1520)\to N\pi$, $N^*(1520)\to \Delta\pi$ and s-wave $N^*(1520)\to N\rho$) and to space-like TFF data in the energy range $0.28\leq Q^2<0.5\,{\rm GeV^2}$. We obtain a decent description of the TFFs in the space-like region with the main limitation coming from the lack of high-quality neutron data. 

Predictions for the differential decay widths of Dalitz decays $N^*(1520) \to N \, e^+ e^-$ and $N^*(1520) \to N \, \mu^+ \mu^-$ have been made, which can be tested by present and future experiments such as HADES and CBM at GSI/FAIR \cite{fair-homepage}. 
We would like to stress again, however, that we are facing a situation where we lack neutron data, and even for the proton, low-energy data do not exist. Experimentally, a better measurement of both the proton and neutron space-like TFFs at low energies can certainly help to reduce the uncertainties on the time-like TFF predictions \cite{An:2023dqx}. 

From the theoretical point of view, it is conceivable to replace the Delta-exchange diagram by full-fledged pion-nucleon scattering phase shifts. Even more advanced would be a generalization of the optical theorem to the complex plane such that one can treat the $N^*(1520)$ as a pole on the second Riemann sheet instead of using its peak mass on the real axis. A further extension would be the three-pion ($\omega$) physics of the isoscalar part of the TFFs with the corresponding coupling of three pions (or $\pi$-$\rho$) to baryons. 

The purpose of our framework is to isolate the universal long-range pion physics from the process specific short-distance physics. The information about the latter is contained in the subtraction constants and it would be appealing to calculate them from QCD instead of fitting them to data.

\begin{acknowledgments}
This work has been supported by the Swedish Research Council (Vetenskapsr\aa det) (grant number 2019-04303). The authors acknowledge stimulating discussions with B.\ Ramstein, P.\ Salabura, A.\ Sarantsev, and K.\ Sch\"onning. The author D. An also acknowledges the Liljewalch travel scholarship for supporting him in traveling to international conferences. 
\end{acknowledgments}

\appendix

\section{Conventions for spinors and polarization vectors}
\label{app:convent-spinors}

We have to deal with objects describing states with spin 1/2, spin 1, and spin 3/2. For the spin-1/2 spinors $u(\vec p,\lambda)$ we use the conventions of \cite{pesschr}. But note that typically we use helicity and not the spin orientation relative to a fixed axis. We need spin-1 polarization vectors in various frames both for virtual photons and for the construction of spin-3/2 vector-spinors. For Dalitz decays, we work often in the frame where the virtual photon is at rest. Its spin orientation is then measured along the axis of the moving baryons, typically the $+z$-direction. We use
\begin{eqnarray}
  \label{eq:expl1pol-timelike}
  \varepsilon^\mu(\vec q = 0,s_\gamma= \pm 1) = \frac{\mp 1}{\sqrt{2}}(0,1,\pm i,0) \,\qquad 
  \varepsilon^\mu(\vec q = 0,s_\gamma=0) = (0,0,0,1) \,.
\end{eqnarray}

For electro-production, we utilize the Breit frame and choose the space-like photon with viruality $Q^2 = -q^2$ to move along the positive $z$-axis. There we have 
\begin{eqnarray}
  \label{eq:expl1pol-spacel}
  \varepsilon^\mu(q_z,\pm 1) = \frac{\mp 1}{\sqrt{2}}(0,1,\pm i,0) \,\qquad 
  \varepsilon^\mu(q_z,0) = (q_z/Q,0,0,q_0/Q) \,.
\end{eqnarray}

Spin-3/2 states of the $N^*(1520)$, moving in the positive $z$-direction, are constructed by \cite{Rarita:1941mf,deJong:1992wm,Penner:1999jia}
\begin{eqnarray}
  u^\mu_{N^*}(p_z,\pm 3/2) &=& u_{N^*}(p_z,\pm 1/2) \, \varepsilon_{N^*}^\mu(p_z,\pm 1)  \,, \nonumber \\
  u^\mu_{N^*}(p_z,\pm 1/2) &=& \frac{1}{\sqrt{3}} \, u_{N^*}(p_z,\mp 1/2) \, \varepsilon^\mu_{N^*}(p_z,\pm 1) \nonumber \\
  && {} +\frac{\sqrt{2}}{\sqrt{3}} \, u_{N^*}(p_z,\pm 1/2) \, \varepsilon^\mu_{N^*}(p_z,0)   \,.
  \label{eq:expl32spinor}
\end{eqnarray}
For the spin-1 polarization vectors for massive (time-like) states we use 
\begin{eqnarray}
  \label{eq:expl1pol}
  \varepsilon^\mu_{N^*}(p_z,\pm 1) = \frac{\mp 1}{\sqrt{2}} \, (0,1,\pm i,0) \,, \nonumber \\
  \varepsilon^\mu_{N^*}(p_z,0) = (p_z/m_{N^*},0,0,E_{N^*}/m_{N^*}) \,.
\end{eqnarray}

\section{Normalization factors}
\label{sec:Normalization factors}

In this appendix, we present the normalization factors $N_m$ for the hadronic reaction amplitudes, which shall be chosen such that one obtains correctly the helicity amplitudes and not multiples thereof. 
We make heavily use of the projector formalism introduced in Section \ref{sec:Projector formalism}.

We need to evaluate our matrix elements of the electromagnetic current \eqref{eq:def-transFF} for specific helicity configurations \cite{Granados:2017cib,Junker:2019vvy}.  
The normalization factor $N_m$ is essentially the ratio of the matrix element 
and the helicity amplitude $H_{-m}$, possibly modified by the $q^2/m_N^2$ factor. 
We recall that we work in the frame where $\vec q$ vanishes. Then current conservation implies that the matrix element vanishes for $\mu = 0$. 

We start with the non-flip case and calculate 
\begin{eqnarray}
  && \frac{1}{e} \, \langle N(p_z,+1/2) \vert j^3 \vert N^*(p_z,+1/2) \rangle =
  - \frac{1}{e p_z} \, p_{N^*}^\mu \langle N(p_z,+1/2) \vert j_\mu \vert N^*(p_z,+1/2) \rangle 
  \nonumber \\
  &&=  - \frac{i}{p_z} \, \bar u_N(p_z,+1/2) q_\nu u^\nu_{N^*}(p_z,+1/2)  
  \nonumber \\ && \phantom{m} \times 
  \left[ m_{N^*} \, (m_N F_1 + (m_{N^*}+m_N) F_2) -2 m_{N^*}^2 F_2 + p_{N^*} \cdot q \; (F_2+F_3) \right]
  \nonumber \\ 
  &&= \frac{1}{p_z} \, \bar u_N(p_z,+1/2) M_0^\mu \, g_{\mu\nu} u^\nu_{N^*}(p_z,+1/2) \, \frac{m_{N^*}}{m_{N^*}-m_N} \, H_0 \,.
\end{eqnarray}
The quantity $M_0^\mu$ is defined in \eqref{eq:finalMs2}.
Since we need the extra $q^2/m_N^2$ factor from \eqref{eq:introT-tilde5}, this leads to 
\begin{eqnarray}
  \label{eq:N0fixed}
  N_0 = \frac{1}{p_z} \, \bar u_N(p_z,+1/2) M_0^\mu \, g_{\mu\nu} u^\nu_{N^*}(p_z,+1/2) \, \frac{m_{N^*}}{m_{N^*}-m_N} \, 
  \frac{m_N^2}{q^2}
\end{eqnarray}

Next, we determine the spin-flipping case
\begin{eqnarray}
  && \frac{1}{e} \, \langle N(p_z,+1/2) \vert j^1 \vert N^*(p_z,+3/2) \rangle = 
  - \frac{1}{2e p_{\rm cm}\sin\theta} \, k_\perp^\mu \langle N(p_z,+1/2) \vert j_\mu \vert N^*(p_z,+3/2) \rangle 
  \nonumber \\ 
  &&= \frac{1}{2p_{\rm cm}\sin\theta} \, \bar u_N(p_z,+1/2) i k_\perp^\mu \, g_{\mu\nu} u^\nu_{N^*}(p_z,+3/2) \,  H_-
  \nonumber \\ 
  &&= \frac{1}{2p_{\rm cm} \, ((m_{N^*}-m_N)^2-q^2)} \, \bar u_N(p_z,+1/2) \tilde M_{+1}^\mu \, g_{\mu\nu} u^\nu_{N^*}(p_z,+3/2)
      \,  H_- \,.
\end{eqnarray}
Thus,
\begin{eqnarray}
  \label{eq:N+1fixed}
  N_{+1} = \frac{1}{2p_{\rm cm} \, ((m_{N^*}-m_N)^2-s)} \, \bar u_N(p_z,+1/2) \tilde M_{+1}^\mu \, g_{\mu\nu} u^\nu_{N^*}(p_z,+3/2) \,.
\end{eqnarray}

Finally, 
\begin{eqnarray}
  && \frac{1}{e} \, \langle N(p_z,+1/2) \vert j^1 \vert N^*(p_z,-1/2) \rangle =
     - \frac{1}{e 2p_{\rm cm}\sin\theta} \, k_\perp^\mu \langle N(p_z,+1/2) \vert j_\mu \vert N^*(p_z,-1/2) \rangle 
  \nonumber \\ 
  &&= \frac{1}{2p_{\rm cm}\sin\theta} \left[ \bar u_N(p_z,+1/2) i \slashed{k}_\perp p_N^\mu g_{\mu\nu} u^\nu_{N^*}(p_z,-1/2)
      \left( m_N F_1 + (m_{N^*}+m_N) F_2 \right) \right.
      \nonumber \\
  && \phantom{mmmmmmn} {}+  \left. \bar u_N(p_z,+1/2) i k_\perp^\mu g_{\mu\nu} u^\nu_{N^*}(p_z,-1/2)
     \left( m_N \, (m_{N^*}-m_N) F_1 + q^2 F_3 \right) \right]
     \nonumber \\
  &&= \frac{1}{2p_{\rm cm}\sin\theta} \, \bar u_N(p_z,+1/2) M_{-1}^\mu \, g_{\mu\nu} u^\nu_{N^*}(p_z,-1/2)
      \frac{-m_{N^*}}{(m_{N^*}-m_N)^2-q^2} \, H_+  \,.
\end{eqnarray}
This brings us to
\begin{eqnarray}
  \label{eq:N-1fixed}
  N_{-1} =
  \frac{-m_{N^*}}{2p_{\rm cm} \, ((m_{N^*}-m_N)^2-s)} \, \bar u_N(p_z,+1/2) \tilde M_{-1}^\mu \, g_{\mu\nu} u^\nu_{N^*}(p_z,-1/2)  \,.
\end{eqnarray}

\section{Charge conjugation}
\label{sec:C-conj}

Charge conjugation maps particles on antiparticles and correspondingly fields on their conjugates. Charge-conjugation symmetry 
relates an interaction that contains a particular field to an interaction term that contains the conjugate field. 
On the other hand, a Lagrangian needs to be hermitian. This is a second relation that connects a field and its conjugate. 
Taken together, one has to demand 
\begin{eqnarray}
  \label{eq:C-hermit-cond}
  C^{-1} \, {\cal L}^\dagger \, C = {\cal L}
\end{eqnarray}
for every single interaction term. This condition leads in the end to real- instead of complex-valued coupling constants. \eqref{eq:lagrintNstar} is constructed by requiring \eqref{eq:C-hermit-cond}. For an elementary spin-1/2 fermion field $\psi$, charge conjugation implies the following relations for the 
bilinears \cite{pesschr}:
\begin{eqnarray}
  C^{-1} \, \bar\psi \psi \, C &=&  \bar\psi \psi  \,,  \nonumber \\
  C^{-1} \, \bar\psi i \gamma_5 \psi \, C &=&  \bar\psi i \gamma_5 \psi  \,,  \nonumber \\
  C^{-1} \, \bar\psi \gamma^\mu \psi \, C &=&  -\bar\psi \gamma^\mu \psi  \,,  \nonumber \\
  C^{-1} \, \bar\psi \gamma^\mu \gamma_5 \psi \, C &=&  \bar\psi \gamma^\mu \gamma_5 \psi  \,,  \nonumber \\
  C^{-1} \, \bar\psi \sigma^{\mu\nu} \psi \, C &=&  - \bar\psi \sigma^{\mu\nu} \psi  \,,  \nonumber \\
  C^{-1} \, \bar\psi (\partial^\mu \psi) \, C &=&  (\partial^\mu \bar\psi) \psi  \,.
  \label{eq:list-from-Peskin}
\end{eqnarray}

When a new fermion field is introduced, it would be useful to spell out how its bilinears built with previously introduced fields 
transform. However, this is often not done in the literature, in spite of the fact that there is a freedom of choice here, i.e.\ 
an implicit convention. The point is that one could introduce a new field $\psi_{\rm new}$ or instead the field 
$\psi'_{\rm new} := i \psi_{\rm new}$. It would not matter for the free-field parts. But clearly these two fields behave 
differently in \eqref{eq:C-hermit-cond}. 

What is done in practice? When coupling nucleon and $\Delta$ fields\footnote{ground-state octet and decuplet fields, but we suppress
the flavor indices here} one uses a convention that leads to relations that are in formal agreement 
with \eqref{eq:list-from-Peskin}, namely
\begin{eqnarray}
  C^{-1} \, \bar\Delta^\nu N \, C &=&  \bar N \Delta^\nu  \,,  \nonumber \\
  C^{-1} \, \bar\Delta^\nu i \gamma_5 N \, C &=&  \bar N i \gamma_5 \Delta^\nu  \,,  \nonumber \\
  C^{-1} \, \bar\Delta^\nu \gamma^\mu N \, C &=&  -\bar N \gamma^\mu \Delta^\nu  \,,  \nonumber \\
  C^{-1} \, \bar\Delta^\nu \gamma^\mu \gamma_5 N \, C &=&  \bar N \gamma^\mu \gamma_5 \Delta^\nu  \,,  \nonumber \\
  C^{-1} \, \bar\Delta^\nu \sigma^{\alpha\beta} N \, C &=&  - \bar N \sigma^{\alpha\beta} \Delta^\nu  \,,  \nonumber \\
  C^{-1} \, \bar\Delta^\nu (\partial^\mu N) \, C &=&  (\partial^\mu \bar N) \Delta^\nu  \,.
  \label{eq:list-Delta-N}
\end{eqnarray}
This leads to real-valued coupling constants $h_A$, $c_M$ etc.\ in \cite{Holmberg:2018dtv} and consequently in \eqref{eq:baryonlagr}. It is then practical to use 
the same relations for every $J^P = {\frac 32}^+$ state. 
But for states with $J^P = {\frac 32}^-$, one has the same freedom again. In \cite{Penner:1999jia}, parity is flipped by 
introducing an extra factor of $i \gamma_5$ into all interaction terms. References \cite{Penner:2002md,Salone:2021bvx} follow 
this prescription. So does the present paper. As a result, one finds for the $N^*$ field 
\begin{eqnarray}
  C^{-1} \, \bar N^\nu N \, C &=&  \bar N  N^\nu  \,,  \nonumber \\
  C^{-1} \, \bar N^\nu i \gamma_5 N \, C &=&  \bar N i \gamma_5  N^\nu  \,,  \nonumber \\
  C^{-1} \, \bar N^\nu \gamma^\mu N \, C &=&  -\bar N \gamma^\mu  N^\nu  \,,  \nonumber \\
  C^{-1} \, \bar N^\nu \gamma^\mu \gamma_5 N \, C &=&  \bar N \gamma^\mu \gamma_5  N^\nu  \,,  \nonumber \\
  C^{-1} \, \bar N^\nu \sigma^{\alpha\beta} N \, C &=&  - \bar N \sigma^{\alpha\beta}  N^\nu  \,,  \nonumber \\
  C^{-1} \, \bar N^\nu (\partial^\mu N) \, C &=&  (\partial^\mu \bar N)  N^\nu  \,.
  \label{eq:list-Nstar-N}
\end{eqnarray}

But a possible alternative would be to add just $\gamma_5$ instead of $i \gamma_5$ when changing from positive- to 
negative-parity resonances. This is the convention chosen in \cite{Aznauryan2008u0defhelicity,Aznauryan:2011qj}. Let us call the corresponding field $N^\nu_{\rm alt}$. In practice this means $N^\nu_{\rm alt} = i N^\nu$.
With this modified convention, one obtains 
\begin{eqnarray}
  C^{-1} \, \bar N^\nu_{\rm alt} N \, C &=&  -\bar N  N^\nu_{\rm alt}  \,,  \nonumber \\
  C^{-1} \, \bar N^\nu_{\rm alt} i \gamma_5 N \, C &=&  -\bar N i \gamma_5  N^\nu_{\rm alt}  \,,  \nonumber \\
  C^{-1} \, \bar N^\nu_{\rm alt} \gamma^\mu N \, C &=&  \bar N \gamma^\mu  N^\nu_{\rm alt}  \,,  \nonumber \\
  C^{-1} \, \bar N^\nu_{\rm alt} \gamma^\mu \gamma_5 N \, C &=& - \bar N \gamma^\mu \gamma_5  N^\nu_{\rm alt}  \,,  \nonumber \\
  C^{-1} \, \bar N^\nu_{\rm alt} \sigma^{\alpha\beta} N \, C &=&   \bar N \sigma^{\alpha\beta}  N^\nu_{\rm alt}  \,,  \nonumber \\
  C^{-1} \, \bar N^\nu_{\rm alt} (\partial^\mu N) \, C &=&  -(\partial^\mu \bar N)  N^\nu_{\rm alt}  \,.
  \label{eq:list-Nstar-N-new}
\end{eqnarray}
In some sense the signs appearing in \eqref{eq:list-Nstar-N-new} are quite natural. If one builds a spin 3/2 state from 
spin 1/2 (spinor structure) and spin 1 (Lorentz index), the natural parity is opposite to the original spin 1/2 state. 
Essentially one ``adds'' a vector. Thus for a $J^P= {\frac 32}^-$ state, the negative sign in the first equation of 
\eqref{eq:list-Nstar-N-new} is what one naturally expects (vectors are negative under charge conjugation). What would happen, if we adopt the convention that leads to \eqref{eq:list-Nstar-N-new}? We would introduce new fields and new 
states via 
\begin{eqnarray}
  N^\nu(x) = -i N^\nu_{\rm alt}(x) \,, \qquad \vert N^* \rangle = i \, \vert N^*_{\rm alt} \rangle  
  \,, \qquad \langle N^* \vert = -i \, \langle N^*_{\rm alt} \vert  \,.  
  \label{eq:oldNstar-to-new}  
\end{eqnarray}
As a consequence, there would be no $i$'s in \eqref{eq:lagrintNstar} (and relative signs turn around).

\section{Explicit expressions for $f$ and $g$}
\label{sec f and g}

The functions $g_{\pm1,0}=g_{\pm1,0}^N + g_{\pm1,0}^\Delta$, $f_{\pm1,0}=f_{\pm1,0}^N + f_{\pm1,0}^\Delta$ have both contributions from the nucleon and the $\Delta$ diagrams.  The nucleon contributions $f_{\pm1,0}^N$ are given as
\begin{equation}
   f_{-1}^N= -\frac{3g_Ah m_{N^*} (m_N^2 - m_{N^*}^2 - 2m_{\pi}^2 + s)^2 ((m_N + m_{N^*}) m_{\pi}^2 - m_N s)}{F_\pi^2((m_N + m_{N^*})^2 - s)}\,,
\end{equation}
\begin{equation}
\begin{split}
f_0^N&=-\frac{1}{F_\pi^2 m_N^2 ((m_N + m_{N^*})^2 - s)} 6g_Ah (m_N - m_{N^*}) (m_N^2 - m_{N^*}^2 - 2m_{\pi}^2 + s) \\
&\times \Bigl((m_N - m_{N^*}) (m_N + m_{N^*})^2 m_{\pi}^2 (m_N^2 + m_N m_{N^*} - 2m_{N^*}^2 - m_{\pi}^2) \\
&- m_N (m_N^2 - m_{N^*}^2) s - 2m_{N^*} (2m_N^2 + m_N m_{N^*} + m_{N^*}^2) m_{\pi}^2 s + (m_N + 2m_{N^*}) m_{\pi}^4 s \\
&+ m_N (m_N^2 + m_N m_{N^*} + m_{N^*}^2 - m_{\pi}^2) s^2 \Bigr)\,,
\end{split}
\end{equation}
and
\begin{equation}
 \begin{split}
f_{+1}^N&=-\frac{3g_A h (m_N^2 - m_{N^*}^2 - 2m_{\pi}^2 + s)^2}{F_\pi^2((m_N + m_{N^*})^2 - s)} \\
&\times \Bigl((m_N + m_{N^*})((m_N - m_{N^*})^2(m_N + m_{N^*}) - (m_N - 2m_{N^*})m_{\pi}^2) \\
&- (m_N^2 + m_N m_{N^*} + m_{N^*}^2 - m_{\pi}^2)s\Bigr)\,.
\end{split}
\end{equation}

The functions $g_{\pm1,0}^N$ are given as
\begin{equation}
    g_{-1}^N=-\frac{g_A h m_{N^*} (m_N^2 - m_{N^*}^2 + s)}{F_\pi^2 m_N ((m_N + m_{N^*})^2 - s)}\,,
\end{equation}
\begin{equation}
    g_{0}^N=-\frac{g_A h (m_N - m_{N^*}) ((m_N^2 - m_{N^*}^2)^2 - 2 m_N m_{N^*} s - s^2)}{2 F_\pi^2 m_N^3 ((m_N + m_{N^*})^2 - s)}\,,
\end{equation}
and
\begin{equation}
g_{+1}^N=\frac{g_A h (2 m_N^3 - m_N^2 m_{N^*} + m_{N^*}^3 - m_{N^*} s - 2 m_N (m_{N^*}^2 + s))}{F_\pi^2 m_N ((m_N + m_{N^*})^2 - s)}\,.
\end{equation}
Due to the extensive nature of the $\Delta$ contributions, we have opted not to display them explicitly within this paper.

The relevant constant contributions from the nucleon and $\Delta$ diagrams are given by
\begin{eqnarray}
R_3^N &=& \frac{g_A h m_{N^*}}{F_\pi^2 m_N^3}  \,, \nonumber \\[0.7em]
R_3^\Delta &=& \frac{h_A}{720 \sqrt{2} F^2 m_\Delta^3 m_N^2} \Bigg[ H_1 \Big(110 m_\Delta^3 + 20 m_\Delta^2 (m_N - 2 m_{N^*}) - 10 m_\Delta \left(4 m_N m_{N^*} - 5 m_{N^*}^2 + 2 m_\pi^2\right) 
\nonumber \\&& \phantom{mmmmmmmmmmm}
- m_N^3 + m_N^2 m_{N^*} + 21 m_N m_{N^*}^2 - 6 m_N m_\pi^2 - m_{N^*}^3\Big) 
\nonumber \\&& \phantom{mmmmmmmmm}
- H_2 \Big(70 m_\Delta^3 + 20 m_\Delta^2 (2 m_N - 5 m_{N^*}) + 10 m_\Delta \left(-4 m_N m_{N^*} + 3 m_{N^*}^2 + 2 m_\pi^2\right)  \nonumber \\
&& \phantom{mmmmmmmmmmmm}  +(m_N - m_{N^*})^2 (m_N + m_{N^*}) + 6 m_N m_\pi^2\Big)\Bigg]  \,.
\label{eq:polyn-N-Delta}
\end{eqnarray}

\section{Technical aspects about the anomalous singularity}
\label{append:Technical aspects on the anomalous singularity}

The calculation of the TFFs $F_{1,2,3}$ (\ref{eq:Fanom}) and the hadronic amplitudes (\ref{eq:tanom}) involves an anomalous cut. 
This cut does not lie on the real axis and therefore one needs also the pion scattering amplitude in the complex plane. The $\pi\pi$ scattering amplitude in the complex plane cannot be directly obtained from experimental data. In principle, one might resort to ChPT. However, our formalism makes massive use of exact unitarity. On the other hand, ChPT only satisfies unitarity order by order (perturbative unitarity) \cite{Dobado:1996ps}. One way to enhance the application range of ChPT is the resummation of higher-order terms by combining ChPT with unitarity. 
In this paper we take the inverse amplitude method \cite{Dobado:1996ps} to unitarize ChPT at the chiral order $p^{4}$ \cite{Dax:2018rvs,Junker:2019vvy}.

In ChPT, the p-wave pion scattering amplitude is given as
\begin{equation}
\label{tChPT1}
t_{\chi\rm PT}(s) \approx t_2(s)+t_4(s)
\end{equation}
and its unitarized version is
\begin{equation}
\label{tIAM1}
t_{\text{IAM}}(s)=\frac{t_2^2(s)}{t_2(s)-t_4(s)} 
\end{equation}
with
\begin{eqnarray}
  t_2(s) &=& \frac{s\sigma^2}{96\pi F_0^2}  \,, 
\end{eqnarray}

\begin{eqnarray}
  t_4(s) &=& \frac{t_{2}(s)}{48\pi^2F_0^2}\bigg[s\left(\bar{l}+\frac{1}{3}\right)-\frac{15}{2} m_\pi^2
             -\frac{m_\pi^4}{2s}\Big(41-2L_\sigma\big(73-25\sigma^2\big)
             +3L_\sigma^2\big(5-32\sigma^2+3\sigma^4\big) \Big)\bigg] -\hat\sigma(s) \,t_2^2(s)   \,,
            \label{t2t4}
\end{eqnarray}

\begin{equation}
  \label{Abbrev_Lsig_sig}
  L_\sigma=\frac{1}{\sigma^2}\left(\frac{1}{2\sigma}\log\frac{1+\sigma}{1-\sigma}-1\right)   \,.
\end{equation}
The function $\sigma(s)$ is  defined in \eqref{eq:velpions} and 
$ \hat\sigma$ is defined as
\begin{eqnarray}
  \label{eq:defsigmahat}
  \hat\sigma(z) := \sqrt{\frac{4 m_\pi^2}{z}-1}  \,. 
\end{eqnarray}

The value for the pion decay constant in the chiral limit $F_0$ is taken from the ratio $F_\pi/F_0=1.064(7)$, where
$F_\pi = 92.28(9)\,$MeV is the pion decay constant at the physical point. 

In the original paper \cite{Dax:2018rvs}, the low-energy constant $\bar{l}=5.73(8)$ has been
adjusted such as to reproduce the position of the pole of the $\rho$-meson resonance on the second Riemann sheet. In our work instead, we use $\bar{l}=6.59$ which is obtained by requiring agreement between the pion p-wave phase shifts from \eqref{tIAM1} and from \cite{GarciaMartin:2011cn} at the point $s_c$. This point $s_c$ has been introduced at the end of Section \ref{sec: Partial-wave projection}; see also Fig.\ \ref{fig:nucleon integral contour}. 
It is chosen to lie on the positive real axis above the decay threshold $(m_{N^*}-m_N)^2$. Its purpose is to avoid a numerical singularity in the dispersive integration that appears at the decay threshold.

We will next explain how to deform the dispersive integrals in  (\ref{eq:tanom}), (\ref{eq:Fanom}).
Let the original integral that we want to calculate be given by 
\begin{eqnarray}
  \label{eq:original}
  F(s) := \int\limits_{{\cal C}_{+,2\pi}} \! dz \, \frac{J_1(z)}{z-s-i\epsilon'} 
  + \int\limits_{4m_\pi^2}^\infty \! ds' \, \frac{J_2(s')}{s'-s-i\epsilon'} 
\end{eqnarray}
with the path ${\cal C}_{+,2\pi}$ connecting $s_+$ to the two-pion threshold; see Fig.\ \ref{fig:nucleon integral contour}. 
The function $J_1$ is of the type of the integrands in (\ref{eq:tanom}), (\ref{eq:Fanom}). The most important part for our discussion is the K\"all\'en function in the denominator. The function $J_2$ is the type of the integrands in \eqref{eq:tmandel}, \eqref{eq:dispbasic}. Most important there is the input amplitude $K_i$ which in turn includes the logarithm as shown in \eqref{eq:Klog} and further specified in \eqref{eq nucleon analytic}. 

Consider now a path along the closed triangle formed by $s_+$, 
the two-pion threshold $4m_\pi^2$ and an arbitrary point $s_{\rm c}$ on the real axis beyond the decay threshold $(m_{N^*}-m_N)^2$. 
An integral over a function along this closed path vanishes if this function is analytic inside of this triangle. 
This is the case for integrands $I(z)$ of the type
\begin{eqnarray}
  \label{eq:type}
  I(z) = \frac{J_1(z)}{z-s-i\epsilon'} \sim \frac{1}{[-\lambda(z,m_{N^*}^2+i\epsilon,m_N^2)]^{3/2}} \, \frac{1}{z-s-i\epsilon'} \,.
\end{eqnarray}
Here $s$ lies on the real axis and the square root function is defined with a cut along the negative real axis. With the 
$\epsilon$ prescription for the mass of the unstable ${N^*}$, the function $-\lambda(z,m_{N^*}^2+i\epsilon,m_N^2)$ adopts 
negative real values slightly above the real axis (with real parts below $(m_{N^*}-m_N)^2$ or above $(m_{N^*}+m_N)^2$). 

Thus instead of \eqref{eq:original} we can write 
\begin{eqnarray}
  \label{eq:better}
  F(s) = \int\limits_{{\cal C}_{+,{\rm c}}} \! dz \, \frac{J_1(z)}{z-s-i\epsilon'} 
  + \int\limits_{4m_\pi^2}^{s_{\rm c}} \! ds' \, \frac{J_2(s')-J_1(s')}{s'-s-i\epsilon'} 
  + \int\limits_{s_{\rm c}}^\infty \! ds' \, \frac{J_2(s')}{s'-s-i\epsilon'} 
\end{eqnarray}
with the path ${\cal C}_{+,{\rm c}}$ connecting $s_+$ to $s_{\rm c}$. What we have used to obtain \eqref{eq:better} is 
\begin{eqnarray}
  \label{eq:zeroclosedpath}
  0 = \int\limits_{{\cal C}_{+,2\pi}} \! dz \, \frac{J_1(z)}{z-s-i\epsilon'} 
  + \int\limits_{4m_\pi^2}^{s_{\rm c}} \! ds' \, \frac{J_1(s')}{s'-s-i\epsilon'} 
  - \int\limits_{{\cal C}_{+,{\rm c}}} \! dz \, \frac{J_1(z)}{z-s-i\epsilon'} \,.
\end{eqnarray}
The important point is that the difference $J_2(s')-J_1(s')$ in \eqref{eq:better} involves just the standard 
logarithm/arctan without the extra term $\sim 2\pi i$ from \eqref{eq nucleon analytic}. This difference diverges neither at 
the decay threshold $(m_{N^*}-m_N)^2$ nor at the two-pion threshold. To be slightly more specific: 
\begin{eqnarray}
  J_2(s')-J_1(s') &\sim& \log \qquad \mbox{for} \quad 4 m_\pi^2 < s' < (m_{N^*}-m_N)^2 \,, \nonumber \\
  J_2(s')-J_1(s') &\sim& \arctan \qquad \mbox{for} \quad (m_{N^*}-m_N)^2 < s' < s_{\rm c} \,, \nonumber \\
  J_2(s') &\sim& \arctan + \pi \qquad \mbox{for} \quad s_{\rm c} < s' < s_Y \,, \nonumber \\
  J_2(s') &\sim& \arctan  \qquad \mbox{for} \quad s_Y < s' < (m_{N^*}+m_N)^2 \,, \nonumber \\
  J_2(s') &\sim& \log  \qquad \mbox{for} \quad (m_{N^*}+m_N)^2 < s' \,.
  \label{eq:morespec}
\end{eqnarray}
We recall that $s_Y$ denotes the point where $Y(s)$ vanishes. Obviously, the simplest choice would be $s_{\rm c}=s_Y$. 
Then we could 
use in \eqref{eq:better} the standard log or standard arctan along the whole real axis. But we will argue 
below that this is {\em not} a good choice for $s_{\rm c}$. 

For the calculation of \eqref{eq:better}, the only numerically problematic point is at $s=s_{\rm c}$. 
Since this point is arbitrary, the resulting function $F(s)$ must be 
smooth at this point. Schematically we can rewrite each of the integrals of \eqref{eq:better} into 
\begin{eqnarray}
  \label{eq:rewritesmooth}
  \int \! dz \, \frac{J_{\ldots}(z)}{z-s-i\epsilon'} = \int \! dz \, \frac{J_{\ldots}(z)-J_{\ldots}(s)}{z-s-i\epsilon'} 
  + J_{\ldots}(s) \int \! dz \, \frac{1}{z-s-i\epsilon'} \,.
\end{eqnarray}
Here $J_{\ldots}$ denotes $J_1$, $J_2$ or $J_1-J_2$. The first term on the right-hand side of \eqref{eq:rewritesmooth} 
is smooth for any value of $s$ if 
\begin{eqnarray}
  \label{eq:smoothness}
  J_{\ldots}(z)-J_{\ldots}(s) \sim z-s \qquad \mbox{for} \quad z \to s \,.
\end{eqnarray}
The second term is proportional to $J_{\ldots}(s) \, \log(s_{\rm c}-s)$. Such a term diverges logarithmically for $s \to s_{\rm c}$. 
If one takes all integrals of \eqref{eq:better} together, one obtains for the potentially divergent terms a sum proportional to 
\begin{eqnarray}
  \label{eq:sumlogs}
  J_1(s) \, \log(s_{\rm c}-s) + \left(J_2(s)-J_1(s) \right) \, \log(s_{\rm c}-s) - J_2(s) \, \log(s_{\rm c}-s) = 0 \,.
\end{eqnarray}
Thus there is no numerical problem with \eqref{eq:better} if something like \eqref{eq:rewritesmooth} and \eqref{eq:sumlogs} 
is numerically implemented and {\em if} \eqref{eq:smoothness} is satisfied. 

To be more specific one needs in particular
\begin{eqnarray}
  \label{eq:smoothness2}
  J_1(z)-J_1(s_{\rm c}) \sim z-s_{\rm c} \qquad \mbox{for} \quad z \to s_{\rm c} \,.
\end{eqnarray}
Here $z$ is a complex number on the line that connects $s_+$ with $s_{\rm c}$. 
All this resembles to some extent the discussion for the two-pion threshold in \cite{Junker:2019vvy}. We use 
in practice two versions for the two-pion scattering amplitude. One along the real axis based on the measured phase shift, 
$t_{\rm ps}$, and the other, $t_{\rm IAM}$, employed in the complex 
plane for the definition of $J_1$. Those two versions must agree at the connection $s=s_{\rm c}$ to make $F(s)$ smooth, i.e.\ to 
ensure that \eqref{eq:smoothness2} is satisfied. The crucial point is that $J_1(z)$ is obtained from $t_{\rm IAM}$ 
while $J_1(s_{\rm c})$ is obtained from $t_{\rm ps}$. The latter is necessary, otherwise, the cancellation \eqref{eq:sumlogs} does 
not happen.  
What needs to be done in practice is to readjust the low-energy constant that appears in $t_{\rm IAM}$ such that 
\begin{eqnarray}
  \label{eq:tpsIAM}
  t_{\rm ps}(s_{\rm c}) = t_{\rm IAM}(s_{\rm c}) 
\end{eqnarray}
holds.

As spelled out in \cite{Junker:2019vvy}, we trust the complex-plane two-pion amplitude $t_{\rm IAM}$ in the low-energy region. 
Thus $s_{\rm c}$ should not be chosen too large. More generally, the whole path ${\cal C}_{+,{\rm c}}$ should lie in the 
low-energy region. Therefore we prefer our choice of $s_c = 0.56\,$GeV$^2$ over the point $s_Y \approx 1.47\,$GeV$^2$.

\section{Partial wave decomposed decay widths for $N^{*} \to N \rho$}
\label{sec decay width rho}

In this appendix, we introduce the formalism to calculate the partial wave amplitudes $N^\star \to N \rho$ and the corresponding partial widths. Our ``$\rho$'' is defined as the two pions in a p-wave state.

Our hadronic reaction amplitudes \eqref{eq:tmandel} include the rescattering of pions via the pion phase shift $\delta$, which also enters $\Omega$. In turn, the pion p-wave phase shift contains the information about the $\rho$ meson. Thus we use $T_i - K_i$ to calculate from our hadronic reaction amplitudes the decay width for the ``$N$-$\rho$'' channel, i.e.\ for the process $N^*(1520) \to N \, (2 \pi)_{L=1}$.

Essentially we invert the procedure that brought us from the ``bare'' (tree-level, without pion rescattering) Feynman reaction amplitudes to the reduced amplitudes $K_i$. We just use now $T_i - K_i$, $i=1,2,3$, instead of $K_i$ as starting point. We can then obtain reduced helicity amplitudes $T_m - K_m$, $m=0,\pm 1$, like in \eqref{eq:introTprime}. Inverting \eqref{eq:reduced amplitude k0}, \eqref{eq:reduced amplitude kpm1}, we obtain functions $a_m(s)$. Those are the $J=1$ objects entering \eqref{eq:pwexp} and we can obtain in this way the quantities $a_m(s,\theta)$. This is the required input to reconstruct the ``$N$-$\rho$'' part of the Feynman matrix elements \eqref{eq:genM-Mi}. 

To compare to experimental results \cite{PDG,HADES:2020kce}, we need to translate our amplitudes with fixed helicities to partial-wave amplitudes in the $LS$ basis \cite{Jacob:1959at,Chung:1971ri,Chung:1102240}. This must be done in the rest frame of the decaying $N^*$. Of course, the Feynman matrix elements are Lorentz invariant, but the meaning of helicities depends on the frame of reference. We have used so far the frame where the two-pion system (the $\rho$) is at rest and the baryons move in the positive $z$-direction. The label $m$ of our amplitudes $a_m$ is given by $m=\lambda_{N^*} - \lambda_N$. If we boost to the rest frame of the $N^*$, the nucleon moves still in the positive $z$-direction, while the $\rho$ moves in the negative $z$-direction. This means that we find for the helicities $\lambda_\rho = \lambda_N - \lambda_{N^*} = -m$. 

We define helicity amplitudes $F_{\lambda_{N},\lambda_{\rho}}$ for the $N$-$\rho$ system via
\begin{eqnarray}
  && 2 p_{\rm cm} F_{\lambda_N=+\frac12,\lambda_\rho=-m}(s)  \, d^{J=1}_{-m,0}(\theta) \nonumber \\
  &&:= a_m(s,\theta) \, \bar u_N(p_z,\lambda_N=+1/2) \, M^\mu_m \, u_{N^*}^\nu(p_z,\lambda_{N^*}=\lambda_N+m) \, g_{\mu \nu}
  \nonumber \\
  &&= {\cal M}(s,\theta;\lambda_N,\lambda_{N^*})  \,. 
  \label{eq:def-my-B}  
\end{eqnarray}

One can expand the helicity decay amplitude $ F_{\lambda_1,\lambda_2}$ in terms of the partial-wave amplitudes $a_{L,S}$ \cite{Jacob:1959at}
\begin{equation}
    F_{\lambda_1,\lambda_2}=\sum_{S,L}\left(\frac{2L+1}{2J_{N^*}+1}\right)^{\frac{1}{2}}a_{S,L} C(L,S,J_{N^*};0,\lambda_1-\lambda_2) C(s_1,s_2,S;\lambda_1,-\lambda_2)  \,,
    \label{eq helicity real}
\end{equation}
where the $C$ are the Clebsch-Gordan coefficients. 
We find the following relations between the helicity amplitudes in the $N^*$ rest frame and the partial-wave amplitude $a_{S,L}$:
\begin{equation}
    \begin{array}{l}
         F_{\frac{1}{2},1}=\frac{a_{\frac{3}{2},0}}{2 \sqrt{3}}+\frac{a_{\frac{1}{2},2}}{\sqrt{3}}-\frac{a_{\frac{3}{2},2}}{2 \sqrt{3}}\,,\\[1em]
           F_{\frac{1}{2},0}= \frac{a_{\frac{3}{2},0}}{\sqrt{6}}-\frac{a_{\frac{1}{2},2}}{\sqrt{6}}-\frac{a_{\frac{3}{2},2}}{\sqrt{6}}\,,\\[1em]
           F_{\frac{1}{2},-1}= \frac{a_{\frac{3}{2},0}}2 + \frac{a_{\frac{3}{2},2}}2\,,
    \end{array}
\end{equation}
Conversely,
\begin{equation}
\begin{split}
     a_{\frac{3}{2},0}&= \frac{  F_{\frac{1}{2},1}}{\sqrt{3}}+\sqrt{\frac{2}{3}}   F_{\frac{1}{2},0}+  F_{\frac{1}{2},-1}\,,\\a_{\frac{1}{2},2}&= \frac{2   F_{\frac{1}{2},1}-\sqrt{2}   F_{\frac{1}{2},0}}{\sqrt{3}}\,,\\a_{\frac{3}{2},2} &=
   -\frac{  F_{\frac{1}{2},1}}{\sqrt{3}}-\sqrt{\frac{2}{3}}   F_{\frac{1}{2},0}+  F_{\frac{1}{2},-1}.
\end{split}
 \label{eq pwa}
\end{equation}
It is straightforward to check that
\begin{equation}
\begin{split}
\sum_{s,l}|a_{s,l}|^2=\sum_{\lambda_N,\lambda_{\rho}}|F_{\lambda_N,\lambda_{\rho}}|^2
   \end{split}
   \label{eq helicity sum}
\end{equation}
where the sum of $\lambda_N$ goes from $-1/2$ to $1/2$ and $\lambda_\rho$ is from $-1$ to $1$. 
Moreover, by explicit calculation, we find that 
\begin{equation}
\begin{split}
   \langle|{\cal{M}}|^2\rangle (s,\theta) &= \frac{1}{4} \times 2 \left( |2 p_{\text{cm}} F_{\lambda_{N}=\frac{1}{2},\lambda_{\rho}=1} d_{1,0}^{J=1}|^2 + |2 p_{\text{cm}} F_{\lambda_{N}=\frac{1}{2},\lambda_{\rho}=0} d_{0,0}^{J=1}|^2 + |2 p_{\text{cm}} F_{\lambda_{N}=\frac{1}{2},\lambda_{\rho}=-1} d_{-1,0}^{J=1}|^2 \right) \\
   &= p_{\text{cm}}^2 \left( 2 F_{\lambda_{N}=\frac{1}{2},\lambda_{\rho}=0}^2 z^2 - F_{\lambda_{N}=\frac{1}{2},\lambda_{\rho}=-1}^2 (-1 + z^2) - F_{\lambda_{N}=\frac{1}{2},\lambda_{\rho}=1}^2 (-1 + z^2) \right) \\
   &= p_{\text{cm}}^2 (\frac{1}{3}|a_{\frac{1}{2},2}|^2  + \frac{1}{3}|a_{\frac{3}{2},2}|^2+ \frac{1}{3}|a_{\frac{3}{2},0}|^2 + \text{cross terms}\times(-1 + 3 z^2))
\end{split}
\end{equation}
with $z=\cos\theta$. Note that the interference term appear with a factor $\times(-1 + 3 z^2)$ factor which will drop out in \eqref{eq width def} below. The factor $\frac{1}{4}$ comes from the average over the initial state and the factor of 2 accounts for both $\lambda_N=\frac{1}{2}$ and $-\frac{1}{2}$.  The total decay width $N^*\to N \rho$ is given as
\begin{equation}
\begin{split}
\Gamma_{N^*\to N \rho}= &\int\limits_{4m_{\pi}^2}^{(m_{N^{*}}-m_N)^2} ds \int\limits_{-1}^{1} d\cos\theta \, J \, \frac{\langle|{\cal{M}}|^2\rangle (s,\theta)}{(2\pi)^332m_{N^{*}}^3} \\
 =&\int\limits_{4m_{\pi}^2}^{(m_{N^{*}}-m_N)^2} ds \, \frac23 p_{\text{cm}}^2 J \, \frac{|a_{\frac{1}{2},2}|^2+|a_{\frac{3}{2},2}|^2+|a_{\frac{3}{2},0}|^2}{(2\pi)^3 32m_{N^{*}}^3}  \\
\end{split}
    \label{eq width def}
\end{equation}
where the Jacobian is given by $J:=p_\text{cm}(s)\lambda^{\frac{1}{2}}(s,m_{N^*}^2,m_N^2)/\sqrt{s}$. Based on this rewriting, a partial-wave decay width can be un-ambiguously defined. As an example, the s-wave decay width can be calculated as 
\begin{equation}
    \Gamma_{N\rho,L=0} = \int\limits_{4m_{\pi}^2}^{(m_{N^{*}}-m_N)^2} ds \, \frac23 p_{\text{cm}}^2 J \, \frac{|a_{\frac{3}{2},0}|^2}{(2\pi)^3 32m_{N^{*}}^3}   \,.
\end{equation}

\bibliography{lit}{}

\begin{thebibliography}{130}%
\makeatletter
\providecommand \@ifxundefined [1]{%
 \@ifx{#1\undefined}
}%
\providecommand \@ifnum [1]{%
 \ifnum #1\expandafter \@firstoftwo
 \else \expandafter \@secondoftwo
 \fi
}%
\providecommand \@ifx [1]{%
 \ifx #1\expandafter \@firstoftwo
 \else \expandafter \@secondoftwo
 \fi
}%
\providecommand \natexlab [1]{#1}%
\providecommand \enquote  [1]{``#1''}%
\providecommand \bibnamefont  [1]{#1}%
\providecommand \bibfnamefont [1]{#1}%
\providecommand \citenamefont [1]{#1}%
\providecommand \href@noop [0]{\@secondoftwo}%
\providecommand \href [0]{\begingroup \@sanitize@url \@href}%
\providecommand \@href[1]{\@@startlink{#1}\@@href}%
\providecommand \@@href[1]{\endgroup#1\@@endlink}%
\providecommand \@sanitize@url [0]{\catcode `\\12\catcode `\$12\catcode `\&12\catcode `\#12\catcode `\^12\catcode `\_12\catcode `\%12\relax}%
\providecommand \@@startlink[1]{}%
\providecommand \@@endlink[0]{}%
\providecommand \url  [0]{\begingroup\@sanitize@url \@url }%
\providecommand \@url [1]{\endgroup\@href {#1}{\urlprefix }}%
\providecommand \urlprefix  [0]{URL }%
\providecommand \Eprint [0]{\href }%
\providecommand \doibase [0]{http://dx.doi.org/}%
\providecommand \selectlanguage [0]{\@gobble}%
\providecommand \bibinfo  [0]{\@secondoftwo}%
\providecommand \bibfield  [0]{\@secondoftwo}%
\providecommand \translation [1]{[#1]}%
\providecommand \BibitemOpen [0]{}%
\providecommand \bibitemStop [0]{}%
\providecommand \bibitemNoStop [0]{.\EOS\space}%
\providecommand \EOS [0]{\spacefactor3000\relax}%
\providecommand \BibitemShut  [1]{\csname bibitem#1\endcsname}%
\let\auto@bib@innerbib\@empty
\bibitem [{\citenamefont {Pascalutsa}\ \emph {et~al.}(2007)\citenamefont {Pascalutsa}, \citenamefont {Vanderhaeghen},\ and\ \citenamefont {Yang}}]{Pascalutsa:2006up}%
  \BibitemOpen
  \bibfield  {author} {\bibinfo {author} {\bibfnamefont {V.}~\bibnamefont {Pascalutsa}}, \bibinfo {author} {\bibfnamefont {M.}~\bibnamefont {Vanderhaeghen}}, \ and\ \bibinfo {author} {\bibfnamefont {S.~N.}\ \bibnamefont {Yang}},\ }\href {\doibase 10.1016/j.physrep.2006.09.006} {\bibfield  {journal} {\bibinfo  {journal} {Phys. Rept.}\ }\textbf {\bibinfo {volume} {437}},\ \bibinfo {pages} {125} (\bibinfo {year} {2007})},\ \Eprint {http://arxiv.org/abs/hep-ph/0609004} {arXiv:hep-ph/0609004 [hep-ph]} \BibitemShut {NoStop}%
\bibitem [{\citenamefont {Aznauryan}\ and\ \citenamefont {Burkert}(2012)}]{Aznauryan:2011qj}%
  \BibitemOpen
  \bibfield  {author} {\bibinfo {author} {\bibfnamefont {I.}~\bibnamefont {Aznauryan}}\ and\ \bibinfo {author} {\bibfnamefont {V.}~\bibnamefont {Burkert}},\ }\href {\doibase 10.1016/j.ppnp.2011.08.001} {\bibfield  {journal} {\bibinfo  {journal} {Prog. Part. Nucl. Phys.}\ }\textbf {\bibinfo {volume} {67}},\ \bibinfo {pages} {1} (\bibinfo {year} {2012})},\ \Eprint {http://arxiv.org/abs/1109.1720} {arXiv:1109.1720 [hep-ph]} \BibitemShut {NoStop}%
\bibitem [{\citenamefont {Punjabi}\ \emph {et~al.}(2015)\citenamefont {Punjabi}, \citenamefont {Perdrisat}, \citenamefont {Jones}, \citenamefont {Brash},\ and\ \citenamefont {Carlson}}]{Punjabi:2015bba}%
  \BibitemOpen
  \bibfield  {author} {\bibinfo {author} {\bibfnamefont {V.}~\bibnamefont {Punjabi}}, \bibinfo {author} {\bibfnamefont {C.~F.}\ \bibnamefont {Perdrisat}}, \bibinfo {author} {\bibfnamefont {M.~K.}\ \bibnamefont {Jones}}, \bibinfo {author} {\bibfnamefont {E.~J.}\ \bibnamefont {Brash}}, \ and\ \bibinfo {author} {\bibfnamefont {C.~E.}\ \bibnamefont {Carlson}},\ }\href {\doibase 10.1140/epja/i2015-15079-x} {\bibfield  {journal} {\bibinfo  {journal} {Eur. Phys. J.}\ }\textbf {\bibinfo {volume} {A51}},\ \bibinfo {pages} {79} (\bibinfo {year} {2015})},\ \Eprint {http://arxiv.org/abs/1503.01452} {arXiv:1503.01452 [nucl-ex]} \BibitemShut {NoStop}%
\bibitem [{\citenamefont {Eichmann}\ and\ \citenamefont {Ramalho}(2018)}]{Eichmann:2018ytt}%
  \BibitemOpen
  \bibfield  {author} {\bibinfo {author} {\bibfnamefont {G.}~\bibnamefont {Eichmann}}\ and\ \bibinfo {author} {\bibfnamefont {G.}~\bibnamefont {Ramalho}},\ }\href {\doibase 10.1103/PhysRevD.98.093007} {\bibfield  {journal} {\bibinfo  {journal} {Phys. Rev. D}\ }\textbf {\bibinfo {volume} {98}},\ \bibinfo {pages} {093007} (\bibinfo {year} {2018})},\ \Eprint {http://arxiv.org/abs/1806.04579} {arXiv:1806.04579 [hep-ph]} \BibitemShut {NoStop}%
\bibitem [{\citenamefont {Lin}\ \emph {et~al.}(2021)\citenamefont {Lin}, \citenamefont {Hammer},\ and\ \citenamefont {Mei\ss{}ner}}]{Lin:2021umz}%
  \BibitemOpen
  \bibfield  {author} {\bibinfo {author} {\bibfnamefont {Y.-H.}\ \bibnamefont {Lin}}, \bibinfo {author} {\bibfnamefont {H.-W.}\ \bibnamefont {Hammer}}, \ and\ \bibinfo {author} {\bibfnamefont {U.-G.}\ \bibnamefont {Mei\ss{}ner}},\ }\href {\doibase 10.1140/epja/s10050-021-00562-0} {\bibfield  {journal} {\bibinfo  {journal} {Eur. Phys. J. A}\ }\textbf {\bibinfo {volume} {57}},\ \bibinfo {pages} {255} (\bibinfo {year} {2021})},\ \Eprint {http://arxiv.org/abs/2106.06357} {arXiv:2106.06357 [hep-ph]} \BibitemShut {NoStop}%
\bibitem [{\citenamefont {Ramalho}\ and\ \citenamefont {Pe\~na}(2024)}]{Ramalho:2023hqd}%
  \BibitemOpen
  \bibfield  {author} {\bibinfo {author} {\bibfnamefont {G.}~\bibnamefont {Ramalho}}\ and\ \bibinfo {author} {\bibfnamefont {M.~T.}\ \bibnamefont {Pe\~na}},\ }\href {\doibase 10.1016/j.ppnp.2024.104097} {\bibfield  {journal} {\bibinfo  {journal} {Prog. Part. Nucl. Phys.}\ }\textbf {\bibinfo {volume} {136}},\ \bibinfo {pages} {104097} (\bibinfo {year} {2024})},\ \Eprint {http://arxiv.org/abs/2306.13900} {arXiv:2306.13900 [hep-ph]} \BibitemShut {NoStop}%
\bibitem [{\citenamefont {Peters}\ \emph {et~al.}(1998)\citenamefont {Peters}, \citenamefont {Post}, \citenamefont {Lenske}, \citenamefont {Leupold},\ and\ \citenamefont {Mosel}}]{Peters:1997va}%
  \BibitemOpen
  \bibfield  {author} {\bibinfo {author} {\bibfnamefont {W.}~\bibnamefont {Peters}}, \bibinfo {author} {\bibfnamefont {M.}~\bibnamefont {Post}}, \bibinfo {author} {\bibfnamefont {H.}~\bibnamefont {Lenske}}, \bibinfo {author} {\bibfnamefont {S.}~\bibnamefont {Leupold}}, \ and\ \bibinfo {author} {\bibfnamefont {U.}~\bibnamefont {Mosel}},\ }\href {\doibase 10.1016/S0375-9474(98)00803-3} {\bibfield  {journal} {\bibinfo  {journal} {Nucl. Phys. A}\ }\textbf {\bibinfo {volume} {632}},\ \bibinfo {pages} {109} (\bibinfo {year} {1998})},\ \Eprint {http://arxiv.org/abs/nucl-th/9708004} {arXiv:nucl-th/9708004} \BibitemShut {NoStop}%
\bibitem [{\citenamefont {Rapp}\ and\ \citenamefont {Wambach}(2000)}]{Rapp:1999ej}%
  \BibitemOpen
  \bibfield  {author} {\bibinfo {author} {\bibfnamefont {R.}~\bibnamefont {Rapp}}\ and\ \bibinfo {author} {\bibfnamefont {J.}~\bibnamefont {Wambach}},\ }\href {\doibase 10.1007/0-306-47101-9\_1} {\bibfield  {journal} {\bibinfo  {journal} {Adv. Nucl. Phys.}\ }\textbf {\bibinfo {volume} {25}},\ \bibinfo {pages} {1} (\bibinfo {year} {2000})},\ \Eprint {http://arxiv.org/abs/hep-ph/9909229} {arXiv:hep-ph/9909229} \BibitemShut {NoStop}%
\bibitem [{\citenamefont {Krivoruchenko}\ \emph {et~al.}(2002)\citenamefont {Krivoruchenko}, \citenamefont {Martemyanov}, \citenamefont {Faessler},\ and\ \citenamefont {Fuchs}}]{Krivoruchenko:2001jk}%
  \BibitemOpen
  \bibfield  {author} {\bibinfo {author} {\bibfnamefont {M.~I.}\ \bibnamefont {Krivoruchenko}}, \bibinfo {author} {\bibfnamefont {B.~V.}\ \bibnamefont {Martemyanov}}, \bibinfo {author} {\bibfnamefont {A.}~\bibnamefont {Faessler}}, \ and\ \bibinfo {author} {\bibfnamefont {C.}~\bibnamefont {Fuchs}},\ }\href {\doibase 10.1006/aphy.2002.6223} {\bibfield  {journal} {\bibinfo  {journal} {Annals Phys.}\ }\textbf {\bibinfo {volume} {296}},\ \bibinfo {pages} {299} (\bibinfo {year} {2002})},\ \Eprint {http://arxiv.org/abs/nucl-th/0110066} {arXiv:nucl-th/0110066} \BibitemShut {NoStop}%
\bibitem [{\citenamefont {Zetenyi}\ and\ \citenamefont {Wolf}(2003)}]{Zetenyi:2001fu}%
  \BibitemOpen
  \bibfield  {author} {\bibinfo {author} {\bibfnamefont {M.}~\bibnamefont {Zetenyi}}\ and\ \bibinfo {author} {\bibfnamefont {G.}~\bibnamefont {Wolf}},\ }\href {\doibase 10.1103/PhysRevC.67.044002} {\bibfield  {journal} {\bibinfo  {journal} {Phys. Rev. C}\ }\textbf {\bibinfo {volume} {67}},\ \bibinfo {pages} {044002} (\bibinfo {year} {2003})},\ \Eprint {http://arxiv.org/abs/nucl-th/0103062} {arXiv:nucl-th/0103062} \BibitemShut {NoStop}%
\bibitem [{\citenamefont {Post}\ \emph {et~al.}(2004)\citenamefont {Post}, \citenamefont {Leupold},\ and\ \citenamefont {Mosel}}]{Post:2003hu}%
  \BibitemOpen
  \bibfield  {author} {\bibinfo {author} {\bibfnamefont {M.}~\bibnamefont {Post}}, \bibinfo {author} {\bibfnamefont {S.}~\bibnamefont {Leupold}}, \ and\ \bibinfo {author} {\bibfnamefont {U.}~\bibnamefont {Mosel}},\ }\href {\doibase 10.1016/j.nuclphysa.2004.05.016} {\bibfield  {journal} {\bibinfo  {journal} {Nucl. Phys. A}\ }\textbf {\bibinfo {volume} {741}},\ \bibinfo {pages} {81} (\bibinfo {year} {2004})},\ \Eprint {http://arxiv.org/abs/nucl-th/0309085} {arXiv:nucl-th/0309085} \BibitemShut {NoStop}%
\bibitem [{\citenamefont {Rapp}\ \emph {et~al.}(2010)\citenamefont {Rapp}, \citenamefont {Wambach},\ and\ \citenamefont {van Hees}}]{Rapp:2009yu}%
  \BibitemOpen
  \bibfield  {author} {\bibinfo {author} {\bibfnamefont {R.}~\bibnamefont {Rapp}}, \bibinfo {author} {\bibfnamefont {J.}~\bibnamefont {Wambach}}, \ and\ \bibinfo {author} {\bibfnamefont {H.}~\bibnamefont {van Hees}},\ }\href {\doibase 10.1007/978-3-642-01539-7_6} {\bibfield  {journal} {\bibinfo  {journal} {Landolt-Bornstein}\ }\textbf {\bibinfo {volume} {23}},\ \bibinfo {pages} {134} (\bibinfo {year} {2010})},\ \Eprint {http://arxiv.org/abs/0901.3289} {arXiv:0901.3289 [hep-ph]} \BibitemShut {NoStop}%
\bibitem [{\citenamefont {Salabura}\ and\ \citenamefont {Stroth}(2021)}]{Salabura:2020tou}%
  \BibitemOpen
  \bibfield  {author} {\bibinfo {author} {\bibfnamefont {P.}~\bibnamefont {Salabura}}\ and\ \bibinfo {author} {\bibfnamefont {J.}~\bibnamefont {Stroth}},\ }\href {\doibase 10.1016/j.ppnp.2021.103869} {\bibfield  {journal} {\bibinfo  {journal} {Prog. Part. Nucl. Phys.}\ }\textbf {\bibinfo {volume} {120}},\ \bibinfo {pages} {103869} (\bibinfo {year} {2021})},\ \Eprint {http://arxiv.org/abs/2005.14589} {arXiv:2005.14589 [nucl-ex]} \BibitemShut {NoStop}%
\bibitem [{\citenamefont {Lepage}\ and\ \citenamefont {Brodsky}(1980)}]{brodskey}%
  \BibitemOpen
  \bibfield  {author} {\bibinfo {author} {\bibfnamefont {G.~P.}\ \bibnamefont {Lepage}}\ and\ \bibinfo {author} {\bibfnamefont {S.~J.}\ \bibnamefont {Brodsky}},\ }\href {\doibase 10.1103/PhysRevD.22.2157} {\bibfield  {journal} {\bibinfo  {journal} {Phys. Rev. D}\ }\textbf {\bibinfo {volume} {22}},\ \bibinfo {pages} {2157} (\bibinfo {year} {1980})}\BibitemShut {NoStop}%
\bibitem [{\citenamefont {Weinberg}(1979)}]{Weinberg:1978kz}%
  \BibitemOpen
  \bibfield  {author} {\bibinfo {author} {\bibfnamefont {S.}~\bibnamefont {Weinberg}},\ }\href {\doibase 10.1016/0378-4371(79)90223-1} {\bibfield  {journal} {\bibinfo  {journal} {Physica A}\ }\textbf {\bibinfo {volume} {96}},\ \bibinfo {pages} {327} (\bibinfo {year} {1979})}\BibitemShut {NoStop}%
\bibitem [{\citenamefont {Gasser}\ and\ \citenamefont {Leutwyler}(1985)}]{Gasser:1984gg}%
  \BibitemOpen
  \bibfield  {author} {\bibinfo {author} {\bibfnamefont {J.}~\bibnamefont {Gasser}}\ and\ \bibinfo {author} {\bibfnamefont {H.}~\bibnamefont {Leutwyler}},\ }\href {\doibase 10.1016/0550-3213(85)90492-4} {\bibfield  {journal} {\bibinfo  {journal} {Nucl. Phys.}\ }\textbf {\bibinfo {volume} {B250}},\ \bibinfo {pages} {465} (\bibinfo {year} {1985})}\BibitemShut {NoStop}%
\bibitem [{\citenamefont {Gasser}\ and\ \citenamefont {Leutwyler}(1984)}]{Gasser:1983yg}%
  \BibitemOpen
  \bibfield  {author} {\bibinfo {author} {\bibfnamefont {J.}~\bibnamefont {Gasser}}\ and\ \bibinfo {author} {\bibfnamefont {H.}~\bibnamefont {Leutwyler}},\ }\href {\doibase 10.1016/0003-4916(84)90242-2} {\bibfield  {journal} {\bibinfo  {journal} {Annals Phys.}\ }\textbf {\bibinfo {volume} {158}},\ \bibinfo {pages} {142} (\bibinfo {year} {1984})}\BibitemShut {NoStop}%
\bibitem [{\citenamefont {Scherer}\ and\ \citenamefont {Schindler}(2012)}]{Scherer:2012xha}%
  \BibitemOpen
  \bibfield  {author} {\bibinfo {author} {\bibfnamefont {S.}~\bibnamefont {Scherer}}\ and\ \bibinfo {author} {\bibfnamefont {M.~R.}\ \bibnamefont {Schindler}},\ }\href {\doibase 10.1007/978-3-642-19254-8} {\bibfield  {journal} {\bibinfo  {journal} {Lect. Notes Phys.}\ }\textbf {\bibinfo {volume} {830}} (\bibinfo {year} {2012}),\ 10.1007/978-3-642-19254-8}\BibitemShut {NoStop}%
\bibitem [{\citenamefont {Coleman}\ \emph {et~al.}(1969)\citenamefont {Coleman}, \citenamefont {Wess},\ and\ \citenamefont {Zumino}}]{Coleman:1969sm}%
  \BibitemOpen
  \bibfield  {author} {\bibinfo {author} {\bibfnamefont {S.~R.}\ \bibnamefont {Coleman}}, \bibinfo {author} {\bibfnamefont {J.}~\bibnamefont {Wess}}, \ and\ \bibinfo {author} {\bibfnamefont {B.}~\bibnamefont {Zumino}},\ }\href {\doibase 10.1103/PhysRev.177.2239} {\bibfield  {journal} {\bibinfo  {journal} {Phys. Rev.}\ }\textbf {\bibinfo {volume} {177}},\ \bibinfo {pages} {2239} (\bibinfo {year} {1969})}\BibitemShut {NoStop}%
\bibitem [{\citenamefont {Callan}\ \emph {et~al.}(1969)\citenamefont {Callan}, \citenamefont {Coleman}, \citenamefont {Wess},\ and\ \citenamefont {Zumino}}]{Callan:1969sn}%
  \BibitemOpen
  \bibfield  {author} {\bibinfo {author} {\bibfnamefont {C.~G.}\ \bibnamefont {Callan}, \bibfnamefont {Jr.}}, \bibinfo {author} {\bibfnamefont {S.~R.}\ \bibnamefont {Coleman}}, \bibinfo {author} {\bibfnamefont {J.}~\bibnamefont {Wess}}, \ and\ \bibinfo {author} {\bibfnamefont {B.}~\bibnamefont {Zumino}},\ }\href {\doibase 10.1103/PhysRev.177.2247} {\bibfield  {journal} {\bibinfo  {journal} {Phys. Rev.}\ }\textbf {\bibinfo {volume} {177}},\ \bibinfo {pages} {2247} (\bibinfo {year} {1969})}\BibitemShut {NoStop}%
\bibitem [{\citenamefont {Scherer}(2003)}]{Scherer:2002tk}%
  \BibitemOpen
  \bibfield  {author} {\bibinfo {author} {\bibfnamefont {S.}~\bibnamefont {Scherer}},\ }\href@noop {} {\bibfield  {journal} {\bibinfo  {journal} {Adv. Nucl. Phys.}\ }\textbf {\bibinfo {volume} {27}},\ \bibinfo {pages} {277} (\bibinfo {year} {2003})},\ \Eprint {http://arxiv.org/abs/hep-ph/0210398} {arXiv:hep-ph/0210398 [hep-ph]} \BibitemShut {NoStop}%
\bibitem [{\citenamefont {Gasser}\ \emph {et~al.}(1988)\citenamefont {Gasser}, \citenamefont {Sainio},\ and\ \citenamefont {Svarc}}]{Gasser:1987rb}%
  \BibitemOpen
  \bibfield  {author} {\bibinfo {author} {\bibfnamefont {J.}~\bibnamefont {Gasser}}, \bibinfo {author} {\bibfnamefont {M.~E.}\ \bibnamefont {Sainio}}, \ and\ \bibinfo {author} {\bibfnamefont {A.}~\bibnamefont {Svarc}},\ }\href {\doibase 10.1016/0550-3213(88)90108-3} {\bibfield  {journal} {\bibinfo  {journal} {Nucl. Phys. B}\ }\textbf {\bibinfo {volume} {307}},\ \bibinfo {pages} {779} (\bibinfo {year} {1988})}\BibitemShut {NoStop}%
\bibitem [{\citenamefont {Kubis}\ and\ \citenamefont {Mei{\ss}ner}(2001)}]{Kubis:2000aa}%
  \BibitemOpen
  \bibfield  {author} {\bibinfo {author} {\bibfnamefont {B.}~\bibnamefont {Kubis}}\ and\ \bibinfo {author} {\bibfnamefont {U.-G.}\ \bibnamefont {Mei{\ss}ner}},\ }\href {\doibase 10.1007/s100520100570} {\bibfield  {journal} {\bibinfo  {journal} {Eur. Phys. J.}\ }\textbf {\bibinfo {volume} {C18}},\ \bibinfo {pages} {747} (\bibinfo {year} {2001})},\ \Eprint {http://arxiv.org/abs/hep-ph/0010283} {arXiv:hep-ph/0010283 [hep-ph]} \BibitemShut {NoStop}%
\bibitem [{\citenamefont {Hilt}\ \emph {et~al.}(2018)\citenamefont {Hilt}, \citenamefont {Bauer}, \citenamefont {Scherer},\ and\ \citenamefont {Tiator}}]{Hilt:2017iup}%
  \BibitemOpen
  \bibfield  {author} {\bibinfo {author} {\bibfnamefont {M.}~\bibnamefont {Hilt}}, \bibinfo {author} {\bibfnamefont {T.}~\bibnamefont {Bauer}}, \bibinfo {author} {\bibfnamefont {S.}~\bibnamefont {Scherer}}, \ and\ \bibinfo {author} {\bibfnamefont {L.}~\bibnamefont {Tiator}},\ }\href {\doibase 10.1103/PhysRevC.97.035205} {\bibfield  {journal} {\bibinfo  {journal} {Phys. Rev. C}\ }\textbf {\bibinfo {volume} {97}},\ \bibinfo {pages} {035205} (\bibinfo {year} {2018})},\ \Eprint {http://arxiv.org/abs/1712.08904} {arXiv:1712.08904 [nucl-th]} \BibitemShut {NoStop}%
\bibitem [{\citenamefont {\"Unal}\ \emph {et~al.}(2021)\citenamefont {\"Unal}, \citenamefont {K\"u\c{c}\"ukarslan},\ and\ \citenamefont {Scherer}}]{Unal:2021byi}%
  \BibitemOpen
  \bibfield  {author} {\bibinfo {author} {\bibfnamefont {Y.}~\bibnamefont {\"Unal}}, \bibinfo {author} {\bibfnamefont {A.}~\bibnamefont {K\"u\c{c}\"ukarslan}}, \ and\ \bibinfo {author} {\bibfnamefont {S.}~\bibnamefont {Scherer}},\ }\href {\doibase 10.1103/PhysRevD.104.094014} {\bibfield  {journal} {\bibinfo  {journal} {Phys. Rev. D}\ }\textbf {\bibinfo {volume} {104}},\ \bibinfo {pages} {094014} (\bibinfo {year} {2021})},\ \Eprint {http://arxiv.org/abs/2104.03125} {arXiv:2104.03125 [hep-ph]} \BibitemShut {NoStop}%
\bibitem [{\citenamefont {Ecker}\ \emph {et~al.}(1989)\citenamefont {Ecker}, \citenamefont {Gasser}, \citenamefont {Pich},\ and\ \citenamefont {de~Rafael}}]{Ecker:1988te}%
  \BibitemOpen
  \bibfield  {author} {\bibinfo {author} {\bibfnamefont {G.}~\bibnamefont {Ecker}}, \bibinfo {author} {\bibfnamefont {J.}~\bibnamefont {Gasser}}, \bibinfo {author} {\bibfnamefont {A.}~\bibnamefont {Pich}}, \ and\ \bibinfo {author} {\bibfnamefont {E.}~\bibnamefont {de~Rafael}},\ }\href {\doibase 10.1016/0550-3213(89)90346-5} {\bibfield  {journal} {\bibinfo  {journal} {Nucl. Phys.}\ }\textbf {\bibinfo {volume} {B321}},\ \bibinfo {pages} {311} (\bibinfo {year} {1989})}\BibitemShut {NoStop}%
\bibitem [{\citenamefont {Donoghue}\ \emph {et~al.}(1989)\citenamefont {Donoghue}, \citenamefont {Ramirez},\ and\ \citenamefont {Valencia}}]{Donoghue:1988ed}%
  \BibitemOpen
  \bibfield  {author} {\bibinfo {author} {\bibfnamefont {J.~F.}\ \bibnamefont {Donoghue}}, \bibinfo {author} {\bibfnamefont {C.}~\bibnamefont {Ramirez}}, \ and\ \bibinfo {author} {\bibfnamefont {G.}~\bibnamefont {Valencia}},\ }\href {\doibase 10.1103/PhysRevD.39.1947} {\bibfield  {journal} {\bibinfo  {journal} {Phys. Rev. D}\ }\textbf {\bibinfo {volume} {39}},\ \bibinfo {pages} {1947} (\bibinfo {year} {1989})}\BibitemShut {NoStop}%
\bibitem [{\citenamefont {Anisovich}\ \emph {et~al.}(2005)\citenamefont {Anisovich}, \citenamefont {Klempt}, \citenamefont {Sarantsev},\ and\ \citenamefont {Thoma}}]{Anisovich:2004zz}%
  \BibitemOpen
  \bibfield  {author} {\bibinfo {author} {\bibfnamefont {A.}~\bibnamefont {Anisovich}}, \bibinfo {author} {\bibfnamefont {E.}~\bibnamefont {Klempt}}, \bibinfo {author} {\bibfnamefont {A.}~\bibnamefont {Sarantsev}}, \ and\ \bibinfo {author} {\bibfnamefont {U.}~\bibnamefont {Thoma}},\ }\href {\doibase 10.1140/epja/i2004-10125-6} {\bibfield  {journal} {\bibinfo  {journal} {Eur. Phys. J. A}\ }\textbf {\bibinfo {volume} {24}},\ \bibinfo {pages} {111} (\bibinfo {year} {2005})},\ \Eprint {http://arxiv.org/abs/hep-ph/0407211} {arXiv:hep-ph/0407211} \BibitemShut {NoStop}%
\bibitem [{\citenamefont {Hanhart}(2012)}]{Hanhart:2012wi}%
  \BibitemOpen
  \bibfield  {author} {\bibinfo {author} {\bibfnamefont {C.}~\bibnamefont {Hanhart}},\ }\href {\doibase 10.1016/j.physletb.2012.07.038} {\bibfield  {journal} {\bibinfo  {journal} {Phys. Lett.}\ }\textbf {\bibinfo {volume} {B715}},\ \bibinfo {pages} {170} (\bibinfo {year} {2012})},\ \Eprint {http://arxiv.org/abs/1203.6839} {arXiv:1203.6839 [hep-ph]} \BibitemShut {NoStop}%
\bibitem [{\citenamefont {Feynman}(1972)}]{feynmanVMD}%
  \BibitemOpen
  \bibfield  {author} {\bibinfo {author} {\bibfnamefont {R.~P.}\ \bibnamefont {Feynman}},\ }\href@noop {} {\emph {\bibinfo {title} {Photon-hadron Interactions}}}\ (\bibinfo  {publisher} {CRC Press},\ \bibinfo {address} {Boca Raton, USA},\ \bibinfo {year} {1972})\BibitemShut {NoStop}%
\bibitem [{\citenamefont {Sakurai}(1969)}]{sakuraiVMD}%
  \BibitemOpen
  \bibfield  {author} {\bibinfo {author} {\bibfnamefont {J.~J.}\ \bibnamefont {Sakurai}},\ }\href@noop {} {\emph {\bibinfo {title} {Currents and Mesons}}}\ (\bibinfo  {publisher} {University of Chicago Press},\ \bibinfo {address} {Chicago},\ \bibinfo {year} {1969})\BibitemShut {NoStop}%
\bibitem [{\citenamefont {Friman}\ and\ \citenamefont {Pirner}(1997)}]{Friman:1997tc}%
  \BibitemOpen
  \bibfield  {author} {\bibinfo {author} {\bibfnamefont {B.}~\bibnamefont {Friman}}\ and\ \bibinfo {author} {\bibfnamefont {H.~J.}\ \bibnamefont {Pirner}},\ }\href {\doibase 10.1016/S0375-9474(97)00050-X} {\bibfield  {journal} {\bibinfo  {journal} {Nucl. Phys. A}\ }\textbf {\bibinfo {volume} {617}},\ \bibinfo {pages} {496} (\bibinfo {year} {1997})},\ \Eprint {http://arxiv.org/abs/nucl-th/9701016} {arXiv:nucl-th/9701016} \BibitemShut {NoStop}%
\bibitem [{\citenamefont {Zetenyi}\ and\ \citenamefont {Wolf}(2012)}]{Zetenyi:2012hg}%
  \BibitemOpen
  \bibfield  {author} {\bibinfo {author} {\bibfnamefont {M.}~\bibnamefont {Zetenyi}}\ and\ \bibinfo {author} {\bibfnamefont {G.}~\bibnamefont {Wolf}},\ }\href {\doibase 10.1103/PhysRevC.86.065209} {\bibfield  {journal} {\bibinfo  {journal} {Phys. Rev. C}\ }\textbf {\bibinfo {volume} {86}},\ \bibinfo {pages} {065209} (\bibinfo {year} {2012})},\ \Eprint {http://arxiv.org/abs/1208.5671} {arXiv:1208.5671 [nucl-th]} \BibitemShut {NoStop}%
\bibitem [{\citenamefont {Braun}\ \emph {et~al.}(2006)\citenamefont {Braun}, \citenamefont {Lenz}, \citenamefont {Peters},\ and\ \citenamefont {Radyushkin}}]{sumrule}%
  \BibitemOpen
  \bibfield  {author} {\bibinfo {author} {\bibfnamefont {V.~M.}\ \bibnamefont {Braun}}, \bibinfo {author} {\bibfnamefont {A.}~\bibnamefont {Lenz}}, \bibinfo {author} {\bibfnamefont {G.}~\bibnamefont {Peters}}, \ and\ \bibinfo {author} {\bibfnamefont {A.~V.}\ \bibnamefont {Radyushkin}},\ }\href {\doibase 10.1103/PhysRevD.73.034020} {\bibfield  {journal} {\bibinfo  {journal} {Phys. Rev. D}\ }\textbf {\bibinfo {volume} {73}},\ \bibinfo {pages} {034020} (\bibinfo {year} {2006})},\ \Eprint {http://arxiv.org/abs/hep-ph/0510237} {arXiv:hep-ph/0510237} \BibitemShut {NoStop}%
\bibitem [{\citenamefont {Aliev}\ and\ \citenamefont {Ozpineci}(2006)}]{sumrule2}%
  \BibitemOpen
  \bibfield  {author} {\bibinfo {author} {\bibfnamefont {T.~M.}\ \bibnamefont {Aliev}}\ and\ \bibinfo {author} {\bibfnamefont {A.}~\bibnamefont {Ozpineci}},\ }\href {\doibase 10.1016/j.nuclphysb.2005.07.038} {\bibfield  {journal} {\bibinfo  {journal} {Nucl. Phys. B}\ }\textbf {\bibinfo {volume} {732}},\ \bibinfo {pages} {291} (\bibinfo {year} {2006})},\ \Eprint {http://arxiv.org/abs/hep-ph/0406331} {arXiv:hep-ph/0406331} \BibitemShut {NoStop}%
\bibitem [{\citenamefont {Wang}\ and\ \citenamefont {Lee}(2009)}]{sumrule3}%
  \BibitemOpen
  \bibfield  {author} {\bibinfo {author} {\bibfnamefont {L.}~\bibnamefont {Wang}}\ and\ \bibinfo {author} {\bibfnamefont {F.~X.}\ \bibnamefont {Lee}},\ }\href {\doibase 10.1103/PhysRevD.80.034003} {\bibfield  {journal} {\bibinfo  {journal} {Phys. Rev. D}\ }\textbf {\bibinfo {volume} {80}},\ \bibinfo {pages} {034003} (\bibinfo {year} {2009})},\ \Eprint {http://arxiv.org/abs/0905.1944} {arXiv:0905.1944 [hep-ph]} \BibitemShut {NoStop}%
\bibitem [{\citenamefont {Workman}\ \emph {et~al.}(2022)\citenamefont {Workman} \emph {et~al.}}]{PDG}%
  \BibitemOpen
  \bibfield  {author} {\bibinfo {author} {\bibfnamefont {R.~L.}\ \bibnamefont {Workman}} \emph {et~al.} (\bibinfo {collaboration} {Particle Data Group}),\ }\href {\doibase 10.1093/ptep/ptac097} {\bibfield  {journal} {\bibinfo  {journal} {PTEP}\ }\textbf {\bibinfo {volume} {2022}},\ \bibinfo {pages} {083C01} (\bibinfo {year} {2022})}\BibitemShut {NoStop}%
\bibitem [{\citenamefont {Warns}\ \emph {et~al.}(1990)\citenamefont {Warns}, \citenamefont {Schroder}, \citenamefont {Pfeil},\ and\ \citenamefont {Rollnik}}]{quarkmodelold1}%
  \BibitemOpen
  \bibfield  {author} {\bibinfo {author} {\bibfnamefont {M.}~\bibnamefont {Warns}}, \bibinfo {author} {\bibfnamefont {H.}~\bibnamefont {Schroder}}, \bibinfo {author} {\bibfnamefont {W.}~\bibnamefont {Pfeil}}, \ and\ \bibinfo {author} {\bibfnamefont {H.}~\bibnamefont {Rollnik}},\ }\href {\doibase 10.1007/BF01556276} {\bibfield  {journal} {\bibinfo  {journal} {Z. Phys. C}\ }\textbf {\bibinfo {volume} {45}},\ \bibinfo {pages} {627} (\bibinfo {year} {1990})}\BibitemShut {NoStop}%
\bibitem [{\citenamefont {Ramalho}\ and\ \citenamefont {Pe{\~ n}a}(2014)}]{Ramalho:2013mxa}%
  \BibitemOpen
  \bibfield  {author} {\bibinfo {author} {\bibfnamefont {G.}~\bibnamefont {Ramalho}}\ and\ \bibinfo {author} {\bibfnamefont {M.}~\bibnamefont {Pe{\~ n}a}},\ }\href {\doibase 10.1103/PhysRevD.89.094016} {\bibfield  {journal} {\bibinfo  {journal} {Phys. Rev. D}\ }\textbf {\bibinfo {volume} {89}},\ \bibinfo {pages} {094016} (\bibinfo {year} {2014})},\ \Eprint {http://arxiv.org/abs/1309.0730} {arXiv:1309.0730 [hep-ph]} \BibitemShut {NoStop}%
\bibitem [{\citenamefont {Ramalho}\ and\ \citenamefont {Pe{\~ n}a}(2017)}]{Ramalho:2016zgc}%
  \BibitemOpen
  \bibfield  {author} {\bibinfo {author} {\bibfnamefont {G.}~\bibnamefont {Ramalho}}\ and\ \bibinfo {author} {\bibfnamefont {M.}~\bibnamefont {Pe{\~ n}a}},\ }\href {\doibase 10.1103/PhysRevD.95.014003} {\bibfield  {journal} {\bibinfo  {journal} {Phys. Rev. D}\ }\textbf {\bibinfo {volume} {95}},\ \bibinfo {pages} {014003} (\bibinfo {year} {2017})},\ \Eprint {http://arxiv.org/abs/1610.08788} {arXiv:1610.08788 [nucl-th]} \BibitemShut {NoStop}%
\bibitem [{\citenamefont {Eichmann}\ \emph {et~al.}(2016)\citenamefont {Eichmann}, \citenamefont {Sanchis-Alepuz}, \citenamefont {Williams}, \citenamefont {Alkofer},\ and\ \citenamefont {Fischer}}]{Eichmann:2016yit}%
  \BibitemOpen
  \bibfield  {author} {\bibinfo {author} {\bibfnamefont {G.}~\bibnamefont {Eichmann}}, \bibinfo {author} {\bibfnamefont {H.}~\bibnamefont {Sanchis-Alepuz}}, \bibinfo {author} {\bibfnamefont {R.}~\bibnamefont {Williams}}, \bibinfo {author} {\bibfnamefont {R.}~\bibnamefont {Alkofer}}, \ and\ \bibinfo {author} {\bibfnamefont {C.~S.}\ \bibnamefont {Fischer}},\ }\href {\doibase 10.1016/j.ppnp.2016.07.001} {\bibfield  {journal} {\bibinfo  {journal} {Prog. Part. Nucl. Phys.}\ }\textbf {\bibinfo {volume} {91}},\ \bibinfo {pages} {1} (\bibinfo {year} {2016})},\ \Eprint {http://arxiv.org/abs/1606.09602} {arXiv:1606.09602 [hep-ph]} \BibitemShut {NoStop}%
\bibitem [{\citenamefont {Donoghue}\ \emph {et~al.}(1990)\citenamefont {Donoghue}, \citenamefont {Gasser},\ and\ \citenamefont {Leutwyler}}]{Donoghue:1990xh}%
  \BibitemOpen
  \bibfield  {author} {\bibinfo {author} {\bibfnamefont {J.~F.}\ \bibnamefont {Donoghue}}, \bibinfo {author} {\bibfnamefont {J.}~\bibnamefont {Gasser}}, \ and\ \bibinfo {author} {\bibfnamefont {H.}~\bibnamefont {Leutwyler}},\ }\href {\doibase 10.1016/0550-3213(90)90474-R} {\bibfield  {journal} {\bibinfo  {journal} {Nucl. Phys.}\ }\textbf {\bibinfo {volume} {B343}},\ \bibinfo {pages} {341} (\bibinfo {year} {1990})}\BibitemShut {NoStop}%
\bibitem [{\citenamefont {Donoghue}(1996)}]{Donoghue:1996kw}%
  \BibitemOpen
  \bibfield  {author} {\bibinfo {author} {\bibfnamefont {J.~F.}\ \bibnamefont {Donoghue}},\ }in\ \href@noop {} {\emph {\bibinfo {booktitle} {{Advanced School on Effective Theories Almunecar, Spain, June 25-July 1, 1995}}}}\ (\bibinfo {year} {1996})\ \Eprint {http://arxiv.org/abs/hep-ph/9607351} {arXiv:hep-ph/9607351 [hep-ph]} \BibitemShut {NoStop}%
\bibitem [{\citenamefont {Niecknig}\ \emph {et~al.}(2012)\citenamefont {Niecknig}, \citenamefont {Kubis},\ and\ \citenamefont {Schneider}}]{Niecknig:2012sj}%
  \BibitemOpen
  \bibfield  {author} {\bibinfo {author} {\bibfnamefont {F.}~\bibnamefont {Niecknig}}, \bibinfo {author} {\bibfnamefont {B.}~\bibnamefont {Kubis}}, \ and\ \bibinfo {author} {\bibfnamefont {S.~P.}\ \bibnamefont {Schneider}},\ }\href {\doibase 10.1140/epjc/s10052-012-2014-1} {\bibfield  {journal} {\bibinfo  {journal} {Eur. Phys. J.}\ }\textbf {\bibinfo {volume} {C72}},\ \bibinfo {pages} {2014} (\bibinfo {year} {2012})},\ \Eprint {http://arxiv.org/abs/1203.2501} {arXiv:1203.2501 [hep-ph]} \BibitemShut {NoStop}%
\bibitem [{\citenamefont {Kang}\ \emph {et~al.}(2014)\citenamefont {Kang}, \citenamefont {Kubis}, \citenamefont {Hanhart},\ and\ \citenamefont {Mei{\ss}ner}}]{Kang:2013jaa}%
  \BibitemOpen
  \bibfield  {author} {\bibinfo {author} {\bibfnamefont {X.-W.}\ \bibnamefont {Kang}}, \bibinfo {author} {\bibfnamefont {B.}~\bibnamefont {Kubis}}, \bibinfo {author} {\bibfnamefont {C.}~\bibnamefont {Hanhart}}, \ and\ \bibinfo {author} {\bibfnamefont {U.-G.}\ \bibnamefont {Mei{\ss}ner}},\ }\href {\doibase 10.1103/PhysRevD.89.053015} {\bibfield  {journal} {\bibinfo  {journal} {Phys. Rev.}\ }\textbf {\bibinfo {volume} {D89}},\ \bibinfo {pages} {053015} (\bibinfo {year} {2014})},\ \Eprint {http://arxiv.org/abs/1312.1193} {arXiv:1312.1193 [hep-ph]} \BibitemShut {NoStop}%
\bibitem [{\citenamefont {Leupold}(2018)}]{Leupold:2017ngs}%
  \BibitemOpen
  \bibfield  {author} {\bibinfo {author} {\bibfnamefont {S.}~\bibnamefont {Leupold}},\ }\href {\doibase 10.1140/epja/i2018-12447-0} {\bibfield  {journal} {\bibinfo  {journal} {Eur. Phys. J.}\ }\textbf {\bibinfo {volume} {A54}},\ \bibinfo {pages} {1} (\bibinfo {year} {2018})},\ \Eprint {http://arxiv.org/abs/1707.09210} {arXiv:1707.09210 [hep-ph]} \BibitemShut {NoStop}%
\bibitem [{\citenamefont {Colangelo}\ \emph {et~al.}(2001)\citenamefont {Colangelo}, \citenamefont {Gasser},\ and\ \citenamefont {Leutwyler}}]{Colangelo:2001df}%
  \BibitemOpen
  \bibfield  {author} {\bibinfo {author} {\bibfnamefont {G.}~\bibnamefont {Colangelo}}, \bibinfo {author} {\bibfnamefont {J.}~\bibnamefont {Gasser}}, \ and\ \bibinfo {author} {\bibfnamefont {H.}~\bibnamefont {Leutwyler}},\ }\href {\doibase 10.1016/S0550-3213(01)00147-X} {\bibfield  {journal} {\bibinfo  {journal} {Nucl. Phys.}\ }\textbf {\bibinfo {volume} {B603}},\ \bibinfo {pages} {125} (\bibinfo {year} {2001})},\ \Eprint {http://arxiv.org/abs/hep-ph/0103088} {arXiv:hep-ph/0103088 [hep-ph]} \BibitemShut {NoStop}%
\bibitem [{\citenamefont {Garcia-Martin}\ \emph {et~al.}(2011)\citenamefont {Garcia-Martin}, \citenamefont {Kaminski}, \citenamefont {Pelaez}, \citenamefont {Ruiz~de Elvira},\ and\ \citenamefont {Yndurain}}]{GarciaMartin:2011cn}%
  \BibitemOpen
  \bibfield  {author} {\bibinfo {author} {\bibfnamefont {R.}~\bibnamefont {Garcia-Martin}}, \bibinfo {author} {\bibfnamefont {R.}~\bibnamefont {Kaminski}}, \bibinfo {author} {\bibfnamefont {J.~R.}\ \bibnamefont {Pelaez}}, \bibinfo {author} {\bibfnamefont {J.}~\bibnamefont {Ruiz~de Elvira}}, \ and\ \bibinfo {author} {\bibfnamefont {F.~J.}\ \bibnamefont {Yndurain}},\ }\href {\doibase 10.1103/PhysRevD.83.074004} {\bibfield  {journal} {\bibinfo  {journal} {Phys. Rev.}\ }\textbf {\bibinfo {volume} {D83}},\ \bibinfo {pages} {074004} (\bibinfo {year} {2011})},\ \Eprint {http://arxiv.org/abs/1102.2183} {arXiv:1102.2183 [hep-ph]} \BibitemShut {NoStop}%
\bibitem [{\citenamefont {Gattringer}\ and\ \citenamefont {Lang}(2010)}]{Gattringer:2010zz}%
  \BibitemOpen
  \bibfield  {author} {\bibinfo {author} {\bibfnamefont {C.}~\bibnamefont {Gattringer}}\ and\ \bibinfo {author} {\bibfnamefont {C.~B.}\ \bibnamefont {Lang}},\ }\href {\doibase 10.1007/978-3-642-01850-3} {\bibfield  {journal} {\bibinfo  {journal} {Lect. Notes Phys.}\ }\textbf {\bibinfo {volume} {788}},\ \bibinfo {pages} {1} (\bibinfo {year} {2010})}\BibitemShut {NoStop}%
\bibitem [{\citenamefont {Leinweber}\ \emph {et~al.}(1993)\citenamefont {Leinweber}, \citenamefont {Draper},\ and\ \citenamefont {Woloshyn}}]{Leinweber:1992pv}%
  \BibitemOpen
  \bibfield  {author} {\bibinfo {author} {\bibfnamefont {D.~B.}\ \bibnamefont {Leinweber}}, \bibinfo {author} {\bibfnamefont {T.}~\bibnamefont {Draper}}, \ and\ \bibinfo {author} {\bibfnamefont {R.~M.}\ \bibnamefont {Woloshyn}},\ }\href {\doibase 10.1103/PhysRevD.48.2230} {\bibfield  {journal} {\bibinfo  {journal} {Phys. Rev. D}\ }\textbf {\bibinfo {volume} {48}},\ \bibinfo {pages} {2230} (\bibinfo {year} {1993})},\ \Eprint {http://arxiv.org/abs/hep-lat/9212016} {arXiv:hep-lat/9212016} \BibitemShut {NoStop}%
\bibitem [{\citenamefont {Alexandrou}\ \emph {et~al.}(2006)\citenamefont {Alexandrou}, \citenamefont {Edwards}, \citenamefont {Neff}, \citenamefont {Negele}, \citenamefont {Koutsou}, \citenamefont {Leontiou}, \citenamefont {Schroers},\ and\ \citenamefont {Tsapalis}}]{Alexandrou:2005em}%
  \BibitemOpen
  \bibfield  {author} {\bibinfo {author} {\bibfnamefont {C.}~\bibnamefont {Alexandrou}}, \bibinfo {author} {\bibfnamefont {R.}~\bibnamefont {Edwards}}, \bibinfo {author} {\bibfnamefont {H.}~\bibnamefont {Neff}}, \bibinfo {author} {\bibfnamefont {J.~W.}\ \bibnamefont {Negele}}, \bibinfo {author} {\bibfnamefont {G.}~\bibnamefont {Koutsou}}, \bibinfo {author} {\bibfnamefont {T.}~\bibnamefont {Leontiou}}, \bibinfo {author} {\bibfnamefont {W.}~\bibnamefont {Schroers}}, \ and\ \bibinfo {author} {\bibfnamefont {A.}~\bibnamefont {Tsapalis}},\ }\href {\doibase 10.22323/1.020.0091} {\bibfield  {journal} {\bibinfo  {journal} {PoS}\ }\textbf {\bibinfo {volume} {LAT2005}},\ \bibinfo {pages} {091} (\bibinfo {year} {2006})},\ \Eprint {http://arxiv.org/abs/hep-lat/0509140} {arXiv:hep-lat/0509140} \BibitemShut {NoStop}%
\bibitem [{\citenamefont {L{\"u}scher}(1991)}]{Luscher:1990ux}%
  \BibitemOpen
  \bibfield  {author} {\bibinfo {author} {\bibfnamefont {M.}~\bibnamefont {L{\"u}scher}},\ }\href {\doibase 10.1016/0550-3213(91)90366-6} {\bibfield  {journal} {\bibinfo  {journal} {Nucl. Phys. B}\ }\textbf {\bibinfo {volume} {354}},\ \bibinfo {pages} {531} (\bibinfo {year} {1991})}\BibitemShut {NoStop}%
\bibitem [{\citenamefont {Dudek}\ \emph {et~al.}(2013)\citenamefont {Dudek}, \citenamefont {Edwards},\ and\ \citenamefont {Thomas}}]{Dudek:2012xn}%
  \BibitemOpen
  \bibfield  {author} {\bibinfo {author} {\bibfnamefont {J.~J.}\ \bibnamefont {Dudek}}, \bibinfo {author} {\bibfnamefont {R.~G.}\ \bibnamefont {Edwards}}, \ and\ \bibinfo {author} {\bibfnamefont {C.~E.}\ \bibnamefont {Thomas}} (\bibinfo {collaboration} {Hadron Spectrum}),\ }\href {\doibase 10.1103/PhysRevD.87.034505} {\bibfield  {journal} {\bibinfo  {journal} {Phys. Rev. D}\ }\textbf {\bibinfo {volume} {87}},\ \bibinfo {pages} {034505} (\bibinfo {year} {2013})},\ \bibinfo {note} {[Erratum: Phys.Rev.D 90, 099902 (2014)]},\ \Eprint {http://arxiv.org/abs/1212.0830} {arXiv:1212.0830 [hep-ph]} \BibitemShut {NoStop}%
\bibitem [{\citenamefont {Polejaeva}\ and\ \citenamefont {Rusetsky}(2012)}]{Polejaeva:2012ut}%
  \BibitemOpen
  \bibfield  {author} {\bibinfo {author} {\bibfnamefont {K.}~\bibnamefont {Polejaeva}}\ and\ \bibinfo {author} {\bibfnamefont {A.}~\bibnamefont {Rusetsky}},\ }\href {\doibase 10.1140/epja/i2012-12067-8} {\bibfield  {journal} {\bibinfo  {journal} {Eur. Phys. J. A}\ }\textbf {\bibinfo {volume} {48}},\ \bibinfo {pages} {67} (\bibinfo {year} {2012})},\ \Eprint {http://arxiv.org/abs/1203.1241} {arXiv:1203.1241 [hep-lat]} \BibitemShut {NoStop}%
\bibitem [{\citenamefont {Mai}\ \emph {et~al.}(2021)\citenamefont {Mai}, \citenamefont {D\"oring},\ and\ \citenamefont {Rusetsky}}]{Mai:2021lwb}%
  \BibitemOpen
  \bibfield  {author} {\bibinfo {author} {\bibfnamefont {M.}~\bibnamefont {Mai}}, \bibinfo {author} {\bibfnamefont {M.}~\bibnamefont {D\"oring}}, \ and\ \bibinfo {author} {\bibfnamefont {A.}~\bibnamefont {Rusetsky}},\ }\href {\doibase 10.1140/epjs/s11734-021-00146-5} {\bibfield  {journal} {\bibinfo  {journal} {Eur. Phys. J. ST}\ }\textbf {\bibinfo {volume} {230}},\ \bibinfo {pages} {1623} (\bibinfo {year} {2021})},\ \Eprint {http://arxiv.org/abs/2103.00577} {arXiv:2103.00577 [hep-lat]} \BibitemShut {NoStop}%
\bibitem [{\citenamefont {Granados}\ \emph {et~al.}(2017)\citenamefont {Granados}, \citenamefont {Leupold},\ and\ \citenamefont {Perotti}}]{Granados:2017cib}%
  \BibitemOpen
  \bibfield  {author} {\bibinfo {author} {\bibfnamefont {C.}~\bibnamefont {Granados}}, \bibinfo {author} {\bibfnamefont {S.}~\bibnamefont {Leupold}}, \ and\ \bibinfo {author} {\bibfnamefont {E.}~\bibnamefont {Perotti}},\ }\href {\doibase 10.1140/epja/i2017-12324-4} {\bibfield  {journal} {\bibinfo  {journal} {Eur. Phys. J.}\ }\textbf {\bibinfo {volume} {A53}},\ \bibinfo {pages} {117} (\bibinfo {year} {2017})},\ \Eprint {http://arxiv.org/abs/1701.09130} {arXiv:1701.09130 [hep-ph]} \BibitemShut {NoStop}%
\bibitem [{\citenamefont {Junker}\ \emph {et~al.}(2020)\citenamefont {Junker}, \citenamefont {Leupold}, \citenamefont {Perotti},\ and\ \citenamefont {Vitos}}]{Junker:2019vvy}%
  \BibitemOpen
  \bibfield  {author} {\bibinfo {author} {\bibfnamefont {O.}~\bibnamefont {Junker}}, \bibinfo {author} {\bibfnamefont {S.}~\bibnamefont {Leupold}}, \bibinfo {author} {\bibfnamefont {E.}~\bibnamefont {Perotti}}, \ and\ \bibinfo {author} {\bibfnamefont {T.}~\bibnamefont {Vitos}},\ }\href {\doibase 10.1103/PhysRevC.101.015206} {\bibfield  {journal} {\bibinfo  {journal} {Phys. Rev. C}\ }\textbf {\bibinfo {volume} {101}},\ \bibinfo {pages} {015206} (\bibinfo {year} {2020})},\ \Eprint {http://arxiv.org/abs/1910.07396} {arXiv:1910.07396 [hep-ph]} \BibitemShut {NoStop}%
\bibitem [{\citenamefont {Alvarado}\ \emph {et~al.}(2023)\citenamefont {Alvarado}, \citenamefont {An}, \citenamefont {Alvarez-Ruso},\ and\ \citenamefont {Leupold}}]{Alvarado:2023loi}%
  \BibitemOpen
  \bibfield  {author} {\bibinfo {author} {\bibfnamefont {F.}~\bibnamefont {Alvarado}}, \bibinfo {author} {\bibfnamefont {D.}~\bibnamefont {An}}, \bibinfo {author} {\bibfnamefont {L.}~\bibnamefont {Alvarez-Ruso}}, \ and\ \bibinfo {author} {\bibfnamefont {S.}~\bibnamefont {Leupold}},\ }\href {\doibase 10.1103/PhysRevD.108.114021} {\bibfield  {journal} {\bibinfo  {journal} {Phys. Rev. D}\ }\textbf {\bibinfo {volume} {108}},\ \bibinfo {pages} {114021} (\bibinfo {year} {2023})},\ \Eprint {http://arxiv.org/abs/2310.07796} {arXiv:2310.07796 [hep-ph]} \BibitemShut {NoStop}%
\bibitem [{\citenamefont {Aung}\ \emph {et~al.}(2024)\citenamefont {Aung}, \citenamefont {Leupold}, \citenamefont {Perotti},\ and\ \citenamefont {Yan}}]{Aung:2024qmf}%
  \BibitemOpen
  \bibfield  {author} {\bibinfo {author} {\bibfnamefont {M.~M.}\ \bibnamefont {Aung}}, \bibinfo {author} {\bibfnamefont {S.}~\bibnamefont {Leupold}}, \bibinfo {author} {\bibfnamefont {E.}~\bibnamefont {Perotti}}, \ and\ \bibinfo {author} {\bibfnamefont {Y.}~\bibnamefont {Yan}},\ }\href@noop {} {\  (\bibinfo {year} {2024})},\ \Eprint {http://arxiv.org/abs/2401.17756} {arXiv:2401.17756 [hep-ph]} \BibitemShut {NoStop}%
\bibitem [{\citenamefont {Kaewsnod}\ \emph {et~al.}(2022)\citenamefont {Kaewsnod}, \citenamefont {Xu}, \citenamefont {Zhao}, \citenamefont {Liu}, \citenamefont {Srisuphaphon}, \citenamefont {Limphirat},\ and\ \citenamefont {Yan}}]{tailandgroup}%
  \BibitemOpen
  \bibfield  {author} {\bibinfo {author} {\bibfnamefont {A.}~\bibnamefont {Kaewsnod}}, \bibinfo {author} {\bibfnamefont {K.}~\bibnamefont {Xu}}, \bibinfo {author} {\bibfnamefont {Z.}~\bibnamefont {Zhao}}, \bibinfo {author} {\bibfnamefont {X.}~\bibnamefont {Liu}}, \bibinfo {author} {\bibfnamefont {S.}~\bibnamefont {Srisuphaphon}}, \bibinfo {author} {\bibfnamefont {A.}~\bibnamefont {Limphirat}}, \ and\ \bibinfo {author} {\bibfnamefont {Y.}~\bibnamefont {Yan}},\ }\href {\doibase 10.1140/epja/s10050-022-00837-0} {\bibfield  {journal} {\bibinfo  {journal} {Eur. Phys. J. A}\ }\textbf {\bibinfo {volume} {58}},\ \bibinfo {pages} {185} (\bibinfo {year} {2022})},\ \Eprint {http://arxiv.org/abs/2204.09528} {arXiv:2204.09528 [hep-ph]} \BibitemShut {NoStop}%
\bibitem [{\citenamefont {Santopinto}\ and\ \citenamefont {Giannini}(2012)}]{Santopinto:2012nq}%
  \BibitemOpen
  \bibfield  {author} {\bibinfo {author} {\bibfnamefont {E.}~\bibnamefont {Santopinto}}\ and\ \bibinfo {author} {\bibfnamefont {M.~M.}\ \bibnamefont {Giannini}},\ }\href {\doibase 10.1103/PhysRevC.86.065202} {\bibfield  {journal} {\bibinfo  {journal} {Phys. Rev. C}\ }\textbf {\bibinfo {volume} {86}},\ \bibinfo {pages} {065202} (\bibinfo {year} {2012})},\ \Eprint {http://arxiv.org/abs/1506.01207} {arXiv:1506.01207 [nucl-th]} \BibitemShut {NoStop}%
\bibitem [{\citenamefont {Aiello}\ \emph {et~al.}(1998)\citenamefont {Aiello}, \citenamefont {Giannini},\ and\ \citenamefont {Santopinto}}]{Aiello:1998xq}%
  \BibitemOpen
  \bibfield  {author} {\bibinfo {author} {\bibfnamefont {M.}~\bibnamefont {Aiello}}, \bibinfo {author} {\bibfnamefont {M.~M.}\ \bibnamefont {Giannini}}, \ and\ \bibinfo {author} {\bibfnamefont {E.}~\bibnamefont {Santopinto}},\ }\href {\doibase 10.1088/0954-3899/24/4/007} {\bibfield  {journal} {\bibinfo  {journal} {J. Phys. G}\ }\textbf {\bibinfo {volume} {24}},\ \bibinfo {pages} {753} (\bibinfo {year} {1998})},\ \Eprint {http://arxiv.org/abs/nucl-th/9801013} {arXiv:nucl-th/9801013} \BibitemShut {NoStop}%
\bibitem [{\citenamefont {Aznauryan}\ and\ \citenamefont {Burkert}(2017)}]{Aznauryan:2017nkz}%
  \BibitemOpen
  \bibfield  {author} {\bibinfo {author} {\bibfnamefont {I.~G.}\ \bibnamefont {Aznauryan}}\ and\ \bibinfo {author} {\bibfnamefont {V.}~\bibnamefont {Burkert}},\ }\href {\doibase 10.1103/PhysRevC.95.065207} {\bibfield  {journal} {\bibinfo  {journal} {Phys. Rev. C}\ }\textbf {\bibinfo {volume} {95}},\ \bibinfo {pages} {065207} (\bibinfo {year} {2017})},\ \Eprint {http://arxiv.org/abs/1703.01751} {arXiv:1703.01751 [nucl-th]} \BibitemShut {NoStop}%
\bibitem [{\citenamefont {Aznauryan}\ \emph {et~al.}(2013)\citenamefont {Aznauryan} \emph {et~al.}}]{Aznauryan:2012ba}%
  \BibitemOpen
  \bibfield  {author} {\bibinfo {author} {\bibfnamefont {I.~G.}\ \bibnamefont {Aznauryan}} \emph {et~al.},\ }\href {\doibase 10.1142/S0218301313300154} {\bibfield  {journal} {\bibinfo  {journal} {Int. J. Mod. Phys. E}\ }\textbf {\bibinfo {volume} {22}},\ \bibinfo {pages} {1330015} (\bibinfo {year} {2013})},\ \Eprint {http://arxiv.org/abs/1212.4891} {arXiv:1212.4891 [nucl-th]} \BibitemShut {NoStop}%
\bibitem [{\citenamefont {Aznauryan}\ \emph {et~al.}(2009)\citenamefont {Aznauryan} \emph {et~al.}}]{CLAS:2009ces}%
  \BibitemOpen
  \bibfield  {author} {\bibinfo {author} {\bibfnamefont {I.~G.}\ \bibnamefont {Aznauryan}} \emph {et~al.} (\bibinfo {collaboration} {CLAS}),\ }\href {\doibase 10.1103/PhysRevC.80.055203} {\bibfield  {journal} {\bibinfo  {journal} {Phys. Rev. C}\ }\textbf {\bibinfo {volume} {80}},\ \bibinfo {pages} {055203} (\bibinfo {year} {2009})},\ \Eprint {http://arxiv.org/abs/0909.2349} {arXiv:0909.2349 [nucl-ex]} \BibitemShut {NoStop}%
\bibitem [{\citenamefont {Mokeev}\ \emph {et~al.}(2016)\citenamefont {Mokeev} \emph {et~al.}}]{Mokeev:2015lda}%
  \BibitemOpen
  \bibfield  {author} {\bibinfo {author} {\bibfnamefont {V.~I.}\ \bibnamefont {Mokeev}} \emph {et~al.},\ }\href {\doibase 10.1103/PhysRevC.93.025206} {\bibfield  {journal} {\bibinfo  {journal} {Phys. Rev. C}\ }\textbf {\bibinfo {volume} {93}},\ \bibinfo {pages} {025206} (\bibinfo {year} {2016})},\ \Eprint {http://arxiv.org/abs/1509.05460} {arXiv:1509.05460 [nucl-ex]} \BibitemShut {NoStop}%
\bibitem [{\citenamefont {Mokeev}\ \emph {et~al.}(2012)\citenamefont {Mokeev} \emph {et~al.}}]{jlabdata3}%
  \BibitemOpen
  \bibfield  {author} {\bibinfo {author} {\bibfnamefont {V.~I.}\ \bibnamefont {Mokeev}} \emph {et~al.} (\bibinfo {collaboration} {CLAS}),\ }\href {\doibase 10.1103/PhysRevC.86.035203} {\bibfield  {journal} {\bibinfo  {journal} {Phys. Rev. C}\ }\textbf {\bibinfo {volume} {86}},\ \bibinfo {pages} {035203} (\bibinfo {year} {2012})},\ \Eprint {http://arxiv.org/abs/1205.3948} {arXiv:1205.3948 [nucl-ex]} \BibitemShut {NoStop}%
\bibitem [{\citenamefont {Bartholomy}\ \emph {et~al.}(2005)\citenamefont {Bartholomy} \emph {et~al.}}]{CB-ELSA:2004sqg}%
  \BibitemOpen
  \bibfield  {author} {\bibinfo {author} {\bibfnamefont {O.}~\bibnamefont {Bartholomy}} \emph {et~al.} (\bibinfo {collaboration} {CB-ELSA}),\ }\href {\doibase 10.1103/PhysRevLett.94.012003} {\bibfield  {journal} {\bibinfo  {journal} {Phys. Rev. Lett.}\ }\textbf {\bibinfo {volume} {94}},\ \bibinfo {pages} {012003} (\bibinfo {year} {2005})},\ \Eprint {http://arxiv.org/abs/hep-ex/0407022} {arXiv:hep-ex/0407022} \BibitemShut {NoStop}%
\bibitem [{\citenamefont {Ahrens}\ \emph {et~al.}(2002)\citenamefont {Ahrens} \emph {et~al.}}]{GDH:2002pkk}%
  \BibitemOpen
  \bibfield  {author} {\bibinfo {author} {\bibfnamefont {J.}~\bibnamefont {Ahrens}} \emph {et~al.} (\bibinfo {collaboration} {GDH, A2}),\ }\href {\doibase 10.1103/PhysRevLett.88.232002} {\bibfield  {journal} {\bibinfo  {journal} {Phys. Rev. Lett.}\ }\textbf {\bibinfo {volume} {88}},\ \bibinfo {pages} {232002} (\bibinfo {year} {2002})},\ \Eprint {http://arxiv.org/abs/hep-ex/0203006} {arXiv:hep-ex/0203006} \BibitemShut {NoStop}%
\bibitem [{\citenamefont {Sokhoyan}\ \emph {et~al.}(2015)\citenamefont {Sokhoyan} \emph {et~al.}}]{CBELSATAPS:2015kka}%
  \BibitemOpen
  \bibfield  {author} {\bibinfo {author} {\bibfnamefont {V.}~\bibnamefont {Sokhoyan}} \emph {et~al.} (\bibinfo {collaboration} {CBELSA/TAPS}),\ }\href {\doibase 10.1140/epja/i2015-15187-7} {\bibfield  {journal} {\bibinfo  {journal} {Eur. Phys. J. A}\ }\textbf {\bibinfo {volume} {51}},\ \bibinfo {pages} {95} (\bibinfo {year} {2015})},\ \bibinfo {note} {[Erratum: Eur.Phys.J.A 51, 187 (2015)]},\ \Eprint {http://arxiv.org/abs/1507.02488} {arXiv:1507.02488 [nucl-ex]} \BibitemShut {NoStop}%
\bibitem [{\citenamefont {Dugger}\ \emph {et~al.}(2007)\citenamefont {Dugger} \emph {et~al.}}]{Dugger:2007bt}%
  \BibitemOpen
  \bibfield  {author} {\bibinfo {author} {\bibfnamefont {M.}~\bibnamefont {Dugger}} \emph {et~al.},\ }\href {\doibase 10.1103/PhysRevC.76.025211} {\bibfield  {journal} {\bibinfo  {journal} {Phys. Rev. C}\ }\textbf {\bibinfo {volume} {76}},\ \bibinfo {pages} {025211} (\bibinfo {year} {2007})},\ \Eprint {http://arxiv.org/abs/0705.0816} {arXiv:0705.0816 [hep-ex]} \BibitemShut {NoStop}%
\bibitem [{\citenamefont {Ciepa\l{}}(2024)}]{Ciepal:2024yub}%
  \BibitemOpen
  \bibfield  {author} {\bibinfo {author} {\bibfnamefont {I.}~\bibnamefont {Ciepa\l{}}} (\bibinfo {collaboration} {HADES}),\ }\href {\doibase 10.1051/epjconf/202429101011} {\bibfield  {journal} {\bibinfo  {journal} {EPJ Web Conf.}\ }\textbf {\bibinfo {volume} {291}},\ \bibinfo {pages} {01011} (\bibinfo {year} {2024})}\BibitemShut {NoStop}%
\bibitem [{\citenamefont {Ablikim}\ \emph {et~al.}(2020)\citenamefont {Ablikim} \emph {et~al.}}]{BESIII:2020nme}%
  \BibitemOpen
  \bibfield  {author} {\bibinfo {author} {\bibfnamefont {M.}~\bibnamefont {Ablikim}} \emph {et~al.} (\bibinfo {collaboration} {BESIII}),\ }\href {\doibase 10.1088/1674-1137/44/4/040001} {\bibfield  {journal} {\bibinfo  {journal} {Chin. Phys. C}\ }\textbf {\bibinfo {volume} {44}},\ \bibinfo {pages} {040001} (\bibinfo {year} {2020})},\ \Eprint {http://arxiv.org/abs/1912.05983} {arXiv:1912.05983 [hep-ex]} \BibitemShut {NoStop}%
\bibitem [{\citenamefont {Altmannshofer}\ \emph {et~al.}(2019)\citenamefont {Altmannshofer} \emph {et~al.}}]{Belle-II:2018jsg}%
  \BibitemOpen
  \bibfield  {author} {\bibinfo {author} {\bibfnamefont {W.}~\bibnamefont {Altmannshofer}} \emph {et~al.} (\bibinfo {collaboration} {Belle-II}),\ }\href {\doibase 10.1093/ptep/ptz106} {\bibfield  {journal} {\bibinfo  {journal} {PTEP}\ }\textbf {\bibinfo {volume} {2019}},\ \bibinfo {pages} {123C01} (\bibinfo {year} {2019})},\ \bibinfo {note} {[Erratum: PTEP 2020, 029201 (2020)]},\ \Eprint {http://arxiv.org/abs/1808.10567} {arXiv:1808.10567 [hep-ex]} \BibitemShut {NoStop}%
\bibitem [{\citenamefont {{K\"orner}}\ and\ \citenamefont {Kuroda}(1977)}]{Korner:1976hv}%
  \BibitemOpen
  \bibfield  {author} {\bibinfo {author} {\bibfnamefont {J.~G.}\ \bibnamefont {{K\"orner}}}\ and\ \bibinfo {author} {\bibfnamefont {M.}~\bibnamefont {Kuroda}},\ }\href {\doibase 10.1103/PhysRevD.16.2165} {\bibfield  {journal} {\bibinfo  {journal} {Phys. Rev.}\ }\textbf {\bibinfo {volume} {D16}},\ \bibinfo {pages} {2165} (\bibinfo {year} {1977})}\BibitemShut {NoStop}%
\bibitem [{\citenamefont {Holmberg}\ and\ \citenamefont {Leupold}(2019)}]{Holmberg:2019ltw}%
  \BibitemOpen
  \bibfield  {author} {\bibinfo {author} {\bibfnamefont {M.}~\bibnamefont {Holmberg}}\ and\ \bibinfo {author} {\bibfnamefont {S.}~\bibnamefont {Leupold}},\ }\href {\doibase 10.1103/PhysRevD.100.114001} {\bibfield  {journal} {\bibinfo  {journal} {Phys. Rev. D}\ }\textbf {\bibinfo {volume} {100}},\ \bibinfo {pages} {114001} (\bibinfo {year} {2019})},\ \Eprint {http://arxiv.org/abs/1909.13562} {arXiv:1909.13562 [hep-ph]} \BibitemShut {NoStop}%
\bibitem [{\citenamefont {Salone}\ and\ \citenamefont {Leupold}(2021)}]{Salone:2021bvx}%
  \BibitemOpen
  \bibfield  {author} {\bibinfo {author} {\bibfnamefont {N.}~\bibnamefont {Salone}}\ and\ \bibinfo {author} {\bibfnamefont {S.}~\bibnamefont {Leupold}},\ }\href {\doibase 10.1140/epja/s10050-021-00493-w} {\bibfield  {journal} {\bibinfo  {journal} {Eur. Phys. J. A}\ }\textbf {\bibinfo {volume} {57}},\ \bibinfo {pages} {183} (\bibinfo {year} {2021})},\ \Eprint {http://arxiv.org/abs/2104.05675} {arXiv:2104.05675 [hep-ph]} \BibitemShut {NoStop}%
\bibitem [{\citenamefont {Hoferichter}\ \emph {et~al.}(2014)\citenamefont {Hoferichter}, \citenamefont {Kubis}, \citenamefont {Leupold}, \citenamefont {Niecknig},\ and\ \citenamefont {Schneider}}]{Hoferichter:2014vra}%
  \BibitemOpen
  \bibfield  {author} {\bibinfo {author} {\bibfnamefont {M.}~\bibnamefont {Hoferichter}}, \bibinfo {author} {\bibfnamefont {B.}~\bibnamefont {Kubis}}, \bibinfo {author} {\bibfnamefont {S.}~\bibnamefont {Leupold}}, \bibinfo {author} {\bibfnamefont {F.}~\bibnamefont {Niecknig}}, \ and\ \bibinfo {author} {\bibfnamefont {S.~P.}\ \bibnamefont {Schneider}},\ }\href {\doibase 10.1140/epjc/s10052-014-3180-0} {\bibfield  {journal} {\bibinfo  {journal} {Eur. Phys. J.}\ }\textbf {\bibinfo {volume} {C74}},\ \bibinfo {pages} {3180} (\bibinfo {year} {2014})},\ \Eprint {http://arxiv.org/abs/1410.4691} {arXiv:1410.4691 [hep-ph]} \BibitemShut {NoStop}%
\bibitem [{\citenamefont {Bardeen}\ and\ \citenamefont {Tung}(1968)}]{Bardeen:1969aw}%
  \BibitemOpen
  \bibfield  {author} {\bibinfo {author} {\bibfnamefont {W.~A.}\ \bibnamefont {Bardeen}}\ and\ \bibinfo {author} {\bibfnamefont {W.~K.}\ \bibnamefont {Tung}},\ }\href {\doibase 10.1103/physrevd.4.3229.2, 10.1103/PhysRev.173.1423} {\bibfield  {journal} {\bibinfo  {journal} {Phys. Rev.}\ }\textbf {\bibinfo {volume} {173}},\ \bibinfo {pages} {1423} (\bibinfo {year} {1968})},\ \bibinfo {note} {[Erratum: Phys. Rev. D4, 3229 (1971)]}\BibitemShut {NoStop}%
\bibitem [{\citenamefont {Tarrach}(1975)}]{Tarrach:1975tu}%
  \BibitemOpen
  \bibfield  {author} {\bibinfo {author} {\bibfnamefont {R.}~\bibnamefont {Tarrach}},\ }\href {\doibase 10.1007/BF02894857} {\bibfield  {journal} {\bibinfo  {journal} {Nuovo Cim.}\ }\textbf {\bibinfo {volume} {A28}},\ \bibinfo {pages} {409} (\bibinfo {year} {1975})}\BibitemShut {NoStop}%
\bibitem [{\citenamefont {Jacob}\ and\ \citenamefont {Wick}(1959)}]{Jacob:1959at}%
  \BibitemOpen
  \bibfield  {author} {\bibinfo {author} {\bibfnamefont {M.}~\bibnamefont {Jacob}}\ and\ \bibinfo {author} {\bibfnamefont {G.}~\bibnamefont {Wick}},\ }\href {\doibase 10.1016/0003-4916(59)90051-X} {\bibfield  {journal} {\bibinfo  {journal} {Annals Phys.}\ }\textbf {\bibinfo {volume} {7}},\ \bibinfo {pages} {404} (\bibinfo {year} {1959})}\BibitemShut {NoStop}%
\bibitem [{\citenamefont {Carlson}(1986)}]{Carlson:1985mm}%
  \BibitemOpen
  \bibfield  {author} {\bibinfo {author} {\bibfnamefont {C.~E.}\ \bibnamefont {Carlson}},\ }\href {\doibase 10.1103/PhysRevD.34.2704} {\bibfield  {journal} {\bibinfo  {journal} {Phys. Rev.}\ }\textbf {\bibinfo {volume} {D34}},\ \bibinfo {pages} {2704} (\bibinfo {year} {1986})}\BibitemShut {NoStop}%
\bibitem [{\citenamefont {Ramalho}(2016)}]{Ramalho:2016zzo}%
  \BibitemOpen
  \bibfield  {author} {\bibinfo {author} {\bibfnamefont {G.}~\bibnamefont {Ramalho}},\ }\href {\doibase 10.1103/PhysRevD.93.113012} {\bibfield  {journal} {\bibinfo  {journal} {Phys. Rev. D}\ }\textbf {\bibinfo {volume} {93}},\ \bibinfo {pages} {113012} (\bibinfo {year} {2016})},\ \Eprint {http://arxiv.org/abs/1602.03832} {arXiv:1602.03832 [hep-ph]} \BibitemShut {NoStop}%
\bibitem [{\citenamefont {Schneider}\ \emph {et~al.}(2012)\citenamefont {Schneider}, \citenamefont {Kubis},\ and\ \citenamefont {Niecknig}}]{Schneider:2012ez}%
  \BibitemOpen
  \bibfield  {author} {\bibinfo {author} {\bibfnamefont {S.~P.}\ \bibnamefont {Schneider}}, \bibinfo {author} {\bibfnamefont {B.}~\bibnamefont {Kubis}}, \ and\ \bibinfo {author} {\bibfnamefont {F.}~\bibnamefont {Niecknig}},\ }\href {\doibase 10.1103/PhysRevD.86.054013} {\bibfield  {journal} {\bibinfo  {journal} {Phys. Rev.}\ }\textbf {\bibinfo {volume} {D86}},\ \bibinfo {pages} {054013} (\bibinfo {year} {2012})},\ \Eprint {http://arxiv.org/abs/1206.3098} {arXiv:1206.3098 [hep-ph]} \BibitemShut {NoStop}%
\bibitem [{\citenamefont {Tiator}\ \emph {et~al.}(2009)\citenamefont {Tiator}, \citenamefont {Drechsel}, \citenamefont {Kamalov},\ and\ \citenamefont {Vanderhaeghen}}]{Tiator:2009mt}%
  \BibitemOpen
  \bibfield  {author} {\bibinfo {author} {\bibfnamefont {L.}~\bibnamefont {Tiator}}, \bibinfo {author} {\bibfnamefont {D.}~\bibnamefont {Drechsel}}, \bibinfo {author} {\bibfnamefont {S.}~\bibnamefont {Kamalov}}, \ and\ \bibinfo {author} {\bibfnamefont {M.}~\bibnamefont {Vanderhaeghen}},\ }\href {\doibase 10.1088/1674-1137/33/12/005} {\bibfield  {journal} {\bibinfo  {journal} {Chin. Phys. C}\ }\textbf {\bibinfo {volume} {33}},\ \bibinfo {pages} {1069} (\bibinfo {year} {2009})},\ \Eprint {http://arxiv.org/abs/0909.2335} {arXiv:0909.2335 [nucl-th]} \BibitemShut {NoStop}%
\bibitem [{\citenamefont {Tiator}\ \emph {et~al.}(2011)\citenamefont {Tiator}, \citenamefont {Drechsel}, \citenamefont {Kamalov},\ and\ \citenamefont {Vanderhaeghen}}]{Tiator:2011pw}%
  \BibitemOpen
  \bibfield  {author} {\bibinfo {author} {\bibfnamefont {L.}~\bibnamefont {Tiator}}, \bibinfo {author} {\bibfnamefont {D.}~\bibnamefont {Drechsel}}, \bibinfo {author} {\bibfnamefont {S.}~\bibnamefont {Kamalov}}, \ and\ \bibinfo {author} {\bibfnamefont {M.}~\bibnamefont {Vanderhaeghen}},\ }\href {\doibase 10.1140/epjst/e2011-01488-9} {\bibfield  {journal} {\bibinfo  {journal} {Eur. Phys. J. ST}\ }\textbf {\bibinfo {volume} {198}},\ \bibinfo {pages} {141} (\bibinfo {year} {2011})},\ \Eprint {http://arxiv.org/abs/1109.6745} {arXiv:1109.6745 [nucl-th]} \BibitemShut {NoStop}%
\bibitem [{\citenamefont {Aznauryan}\ \emph {et~al.}(2008)\citenamefont {Aznauryan}, \citenamefont {Burkert},\ and\ \citenamefont {Lee}}]{Aznauryan2008u0defhelicity}%
  \BibitemOpen
  \bibfield  {author} {\bibinfo {author} {\bibfnamefont {I.~G.}\ \bibnamefont {Aznauryan}}, \bibinfo {author} {\bibfnamefont {V.~D.}\ \bibnamefont {Burkert}}, \ and\ \bibinfo {author} {\bibfnamefont {T.~S.~H.}\ \bibnamefont {Lee}},\ }\href@noop {} {\  (\bibinfo {year} {2008})},\ \Eprint {http://arxiv.org/abs/0810.0997} {arXiv:0810.0997 [nucl-th]} \BibitemShut {NoStop}%
\bibitem [{\citenamefont {Eden}\ \emph {et~al.}(1966)\citenamefont {Eden}, \citenamefont {Landshoff}, \citenamefont {Olive},\ and\ \citenamefont {Polkinghorne}}]{Eden:1966dnq}%
  \BibitemOpen
  \bibfield  {author} {\bibinfo {author} {\bibfnamefont {R.~J.}\ \bibnamefont {Eden}}, \bibinfo {author} {\bibfnamefont {P.~V.}\ \bibnamefont {Landshoff}}, \bibinfo {author} {\bibfnamefont {D.~I.}\ \bibnamefont {Olive}}, \ and\ \bibinfo {author} {\bibfnamefont {J.~C.}\ \bibnamefont {Polkinghorne}},\ }\href@noop {} {\emph {\bibinfo {title} {{The analytic S-matrix}}}}\ (\bibinfo  {publisher} {Cambridge Univ. Press},\ \bibinfo {address} {Cambridge},\ \bibinfo {year} {1966})\BibitemShut {NoStop}%
\bibitem [{\citenamefont {Watson}(1954)}]{Watson:1954uc}%
  \BibitemOpen
  \bibfield  {author} {\bibinfo {author} {\bibfnamefont {K.~M.}\ \bibnamefont {Watson}},\ }\href {\doibase 10.1103/PhysRev.95.228} {\bibfield  {journal} {\bibinfo  {journal} {Phys. Rev.}\ }\textbf {\bibinfo {volume} {95}},\ \bibinfo {pages} {228} (\bibinfo {year} {1954})}\BibitemShut {NoStop}%
\bibitem [{\citenamefont {Hoferichter}\ \emph {et~al.}(2016{\natexlab{a}})\citenamefont {Hoferichter}, \citenamefont {Kubis}, \citenamefont {Ruiz~de Elvira}, \citenamefont {Hammer},\ and\ \citenamefont {Mei{\ss}ner}}]{Hoferichter:2016duk}%
  \BibitemOpen
  \bibfield  {author} {\bibinfo {author} {\bibfnamefont {M.}~\bibnamefont {Hoferichter}}, \bibinfo {author} {\bibfnamefont {B.}~\bibnamefont {Kubis}}, \bibinfo {author} {\bibfnamefont {J.}~\bibnamefont {Ruiz~de Elvira}}, \bibinfo {author} {\bibfnamefont {H.~W.}\ \bibnamefont {Hammer}}, \ and\ \bibinfo {author} {\bibfnamefont {U.-G.}\ \bibnamefont {Mei{\ss}ner}},\ }\href {\doibase 10.1140/epja/i2016-16331-7} {\bibfield  {journal} {\bibinfo  {journal} {Eur. Phys. J.}\ }\textbf {\bibinfo {volume} {A52}},\ \bibinfo {pages} {331} (\bibinfo {year} {2016}{\natexlab{a}})},\ \Eprint {http://arxiv.org/abs/1609.06722} {arXiv:1609.06722 [hep-ph]} \BibitemShut {NoStop}%
\bibitem [{\citenamefont {Fujikawa}\ \emph {et~al.}(2008)\citenamefont {Fujikawa} \emph {et~al.}}]{Fujikawa:2008ma}%
  \BibitemOpen
  \bibfield  {author} {\bibinfo {author} {\bibfnamefont {M.}~\bibnamefont {Fujikawa}} \emph {et~al.} (\bibinfo {collaboration} {Belle}),\ }\href {\doibase 10.1103/PhysRevD.78.072006} {\bibfield  {journal} {\bibinfo  {journal} {Phys. Rev.}\ }\textbf {\bibinfo {volume} {D78}},\ \bibinfo {pages} {072006} (\bibinfo {year} {2008})},\ \Eprint {http://arxiv.org/abs/0805.3773} {arXiv:0805.3773 [hep-ex]} \BibitemShut {NoStop}%
\bibitem [{\citenamefont {Muskhelishvili}(1958)}]{zbMATH03081975}%
  \BibitemOpen
  \bibfield  {author} {\bibinfo {author} {\bibfnamefont {N.}~\bibnamefont {Muskhelishvili}},\ }\href@noop {} {\emph {\bibinfo {title} {Singular integral equations: Boundary problems of function theory and their application to mathematical physics}}}\ (\bibinfo  {publisher} {Springer},\ \bibinfo {year} {1958})\BibitemShut {NoStop}%
\bibitem [{\citenamefont {Omnes}(1958)}]{Omnes:1958hv}%
  \BibitemOpen
  \bibfield  {author} {\bibinfo {author} {\bibfnamefont {R.}~\bibnamefont {Omnes}},\ }\href {\doibase 10.1007/BF02747746} {\bibfield  {journal} {\bibinfo  {journal} {Nuovo Cim.}\ }\textbf {\bibinfo {volume} {8}},\ \bibinfo {pages} {316} (\bibinfo {year} {1958})}\BibitemShut {NoStop}%
\bibitem [{\citenamefont {Donoghue}\ and\ \citenamefont {Na}(1997)}]{Donoghue:1996bt}%
  \BibitemOpen
  \bibfield  {author} {\bibinfo {author} {\bibfnamefont {J.~F.}\ \bibnamefont {Donoghue}}\ and\ \bibinfo {author} {\bibfnamefont {E.~S.}\ \bibnamefont {Na}},\ }\href {\doibase 10.1103/PhysRevD.56.7073} {\bibfield  {journal} {\bibinfo  {journal} {Phys. Rev. D}\ }\textbf {\bibinfo {volume} {56}},\ \bibinfo {pages} {7073} (\bibinfo {year} {1997})},\ \Eprint {http://arxiv.org/abs/hep-ph/9611418} {arXiv:hep-ph/9611418} \BibitemShut {NoStop}%
\bibitem [{\citenamefont {Garcia-Martin}\ and\ \citenamefont {Moussallam}(2010)}]{GarciaMartin:2010cw}%
  \BibitemOpen
  \bibfield  {author} {\bibinfo {author} {\bibfnamefont {R.}~\bibnamefont {Garcia-Martin}}\ and\ \bibinfo {author} {\bibfnamefont {B.}~\bibnamefont {Moussallam}},\ }\href {\doibase 10.1140/epjc/s10052-010-1471-7} {\bibfield  {journal} {\bibinfo  {journal} {Eur. Phys. J.}\ }\textbf {\bibinfo {volume} {C70}},\ \bibinfo {pages} {155} (\bibinfo {year} {2010})},\ \Eprint {http://arxiv.org/abs/1006.5373} {arXiv:1006.5373 [hep-ph]} \BibitemShut {NoStop}%
\bibitem [{\citenamefont {Granados}\ and\ \citenamefont {Weiss}(2014)}]{Granados:2013moa}%
  \BibitemOpen
  \bibfield  {author} {\bibinfo {author} {\bibfnamefont {C.}~\bibnamefont {Granados}}\ and\ \bibinfo {author} {\bibfnamefont {C.}~\bibnamefont {Weiss}},\ }\href {\doibase 10.1007/JHEP01(2014)092} {\bibfield  {journal} {\bibinfo  {journal} {JHEP}\ }\textbf {\bibinfo {volume} {1401}},\ \bibinfo {pages} {092} (\bibinfo {year} {2014})},\ \Eprint {http://arxiv.org/abs/1308.1634} {arXiv:1308.1634 [hep-ph]} \BibitemShut {NoStop}%
\bibitem [{\citenamefont {Hoferichter}\ \emph {et~al.}(2016{\natexlab{b}})\citenamefont {Hoferichter}, \citenamefont {Ruiz~de Elvira}, \citenamefont {Kubis},\ and\ \citenamefont {Mei{\ss}ner}}]{Hoferichter:2015hva}%
  \BibitemOpen
  \bibfield  {author} {\bibinfo {author} {\bibfnamefont {M.}~\bibnamefont {Hoferichter}}, \bibinfo {author} {\bibfnamefont {J.}~\bibnamefont {Ruiz~de Elvira}}, \bibinfo {author} {\bibfnamefont {B.}~\bibnamefont {Kubis}}, \ and\ \bibinfo {author} {\bibfnamefont {U.-G.}\ \bibnamefont {Mei{\ss}ner}},\ }\href {\doibase 10.1016/j.physrep.2016.02.002} {\bibfield  {journal} {\bibinfo  {journal} {Phys. Rept.}\ }\textbf {\bibinfo {volume} {625}},\ \bibinfo {pages} {1} (\bibinfo {year} {2016}{\natexlab{b}})},\ \Eprint {http://arxiv.org/abs/1510.06039} {arXiv:1510.06039 [hep-ph]} \BibitemShut {NoStop}%
\bibitem [{\citenamefont {Holmberg}\ and\ \citenamefont {Leupold}(2018)}]{Holmberg:2018dtv}%
  \BibitemOpen
  \bibfield  {author} {\bibinfo {author} {\bibfnamefont {M.}~\bibnamefont {Holmberg}}\ and\ \bibinfo {author} {\bibfnamefont {S.}~\bibnamefont {Leupold}},\ }\href {\doibase 10.1140/epja/i2018-12533-3} {\bibfield  {journal} {\bibinfo  {journal} {Eur. Phys. J.}\ }\textbf {\bibinfo {volume} {A54}},\ \bibinfo {pages} {103} (\bibinfo {year} {2018})},\ \Eprint {http://arxiv.org/abs/1802.05168} {arXiv:1802.05168 [hep-ph]} \BibitemShut {NoStop}%
\bibitem [{\citenamefont {Jenkins}\ and\ \citenamefont {Manohar}(1991)}]{Jenkins:1991es}%
  \BibitemOpen
  \bibfield  {author} {\bibinfo {author} {\bibfnamefont {E.~E.}\ \bibnamefont {Jenkins}}\ and\ \bibinfo {author} {\bibfnamefont {A.~V.}\ \bibnamefont {Manohar}},\ }\href {\doibase 10.1016/0370-2693(91)90840-M} {\bibfield  {journal} {\bibinfo  {journal} {Phys. Lett.}\ }\textbf {\bibinfo {volume} {B259}},\ \bibinfo {pages} {353} (\bibinfo {year} {1991})}\BibitemShut {NoStop}%
\bibitem [{\citenamefont {Ledwig}\ \emph {et~al.}(2014)\citenamefont {Ledwig}, \citenamefont {Martin~Camalich}, \citenamefont {Geng},\ and\ \citenamefont {Vicente~Vacas}}]{Ledwig:2014rfa}%
  \BibitemOpen
  \bibfield  {author} {\bibinfo {author} {\bibfnamefont {T.}~\bibnamefont {Ledwig}}, \bibinfo {author} {\bibfnamefont {J.}~\bibnamefont {Martin~Camalich}}, \bibinfo {author} {\bibfnamefont {L.~S.}\ \bibnamefont {Geng}}, \ and\ \bibinfo {author} {\bibfnamefont {M.~J.}\ \bibnamefont {Vicente~Vacas}},\ }\href {\doibase 10.1103/PhysRevD.90.054502} {\bibfield  {journal} {\bibinfo  {journal} {Phys. Rev.}\ }\textbf {\bibinfo {volume} {D90}},\ \bibinfo {pages} {054502} (\bibinfo {year} {2014})},\ \Eprint {http://arxiv.org/abs/1405.5456} {arXiv:1405.5456 [hep-ph]} \BibitemShut {NoStop}%
\bibitem [{\citenamefont {Mommers}\ and\ \citenamefont {Leupold}(2022)}]{Mommers:2022dgw}%
  \BibitemOpen
  \bibfield  {author} {\bibinfo {author} {\bibfnamefont {C.~J.~G.}\ \bibnamefont {Mommers}}\ and\ \bibinfo {author} {\bibfnamefont {S.}~\bibnamefont {Leupold}},\ }\href {\doibase 10.1103/PhysRevD.106.093001} {\bibfield  {journal} {\bibinfo  {journal} {Phys. Rev. D}\ }\textbf {\bibinfo {volume} {106}},\ \bibinfo {pages} {093001} (\bibinfo {year} {2022})},\ \Eprint {http://arxiv.org/abs/2208.11078} {arXiv:2208.11078 [hep-ph]} \BibitemShut {NoStop}%
\bibitem [{\citenamefont {Pascalutsa}\ and\ \citenamefont {Timmermans}(1999)}]{Pascalutsa:1999zz}%
  \BibitemOpen
  \bibfield  {author} {\bibinfo {author} {\bibfnamefont {V.}~\bibnamefont {Pascalutsa}}\ and\ \bibinfo {author} {\bibfnamefont {R.}~\bibnamefont {Timmermans}},\ }\href {\doibase 10.1103/PhysRevC.60.042201} {\bibfield  {journal} {\bibinfo  {journal} {Phys. Rev.}\ }\textbf {\bibinfo {volume} {C60}},\ \bibinfo {pages} {042201} (\bibinfo {year} {1999})},\ \Eprint {http://arxiv.org/abs/nucl-th/9905065} {arXiv:nucl-th/9905065 [nucl-th]} \BibitemShut {NoStop}%
\bibitem [{\citenamefont {Pascalutsa}\ and\ \citenamefont {Vanderhaeghen}(2006)}]{Pascalutsa:2005nd}%
  \BibitemOpen
  \bibfield  {author} {\bibinfo {author} {\bibfnamefont {V.}~\bibnamefont {Pascalutsa}}\ and\ \bibinfo {author} {\bibfnamefont {M.}~\bibnamefont {Vanderhaeghen}},\ }\href {\doibase 10.1016/j.physletb.2006.03.023} {\bibfield  {journal} {\bibinfo  {journal} {Phys. Lett.}\ }\textbf {\bibinfo {volume} {B636}},\ \bibinfo {pages} {31} (\bibinfo {year} {2006})},\ \Eprint {http://arxiv.org/abs/hep-ph/0511261} {arXiv:hep-ph/0511261 [hep-ph]} \BibitemShut {NoStop}%
\bibitem [{\citenamefont {Ledwig}\ \emph {et~al.}(2012)\citenamefont {Ledwig}, \citenamefont {Martin-Camalich}, \citenamefont {Pascalutsa},\ and\ \citenamefont {Vanderhaeghen}}]{Ledwig:2011cx}%
  \BibitemOpen
  \bibfield  {author} {\bibinfo {author} {\bibfnamefont {T.}~\bibnamefont {Ledwig}}, \bibinfo {author} {\bibfnamefont {J.}~\bibnamefont {Martin-Camalich}}, \bibinfo {author} {\bibfnamefont {V.}~\bibnamefont {Pascalutsa}}, \ and\ \bibinfo {author} {\bibfnamefont {M.}~\bibnamefont {Vanderhaeghen}},\ }\href {\doibase 10.1103/PhysRevD.85.034013} {\bibfield  {journal} {\bibinfo  {journal} {Phys. Rev.}\ }\textbf {\bibinfo {volume} {D85}},\ \bibinfo {pages} {034013} (\bibinfo {year} {2012})},\ \Eprint {http://arxiv.org/abs/1108.2523} {arXiv:1108.2523 [hep-ph]} \BibitemShut {NoStop}%
\bibitem [{\citenamefont {Peskin}\ and\ \citenamefont {Schroeder}(1995)}]{pesschr}%
  \BibitemOpen
  \bibfield  {author} {\bibinfo {author} {\bibfnamefont {M.~E.}\ \bibnamefont {Peskin}}\ and\ \bibinfo {author} {\bibfnamefont {D.~V.}\ \bibnamefont {Schroeder}},\ }\href@noop {} {\emph {\bibinfo {title} {An Introduction to Quantum Field Theory}}}\ (\bibinfo  {publisher} {Westview Press},\ \bibinfo {year} {1995})\BibitemShut {NoStop}%
\bibitem [{\citenamefont {Bjorken}\ and\ \citenamefont {Drell}(1965)}]{bjorken-drell}%
  \BibitemOpen
  \bibfield  {author} {\bibinfo {author} {\bibfnamefont {J.~D.}\ \bibnamefont {Bjorken}}\ and\ \bibinfo {author} {\bibfnamefont {S.~D.}\ \bibnamefont {Drell}},\ }\href@noop {} {\emph {\bibinfo {title} {Relativistic Quantum Fields}}}\ (\bibinfo  {publisher} {Mc Graw-Hill},\ \bibinfo {year} {1965})\BibitemShut {NoStop}%
\bibitem [{\citenamefont {Shrestha}\ and\ \citenamefont {Manley}(2012)}]{Shrestha:2012ep}%
  \BibitemOpen
  \bibfield  {author} {\bibinfo {author} {\bibfnamefont {M.}~\bibnamefont {Shrestha}}\ and\ \bibinfo {author} {\bibfnamefont {D.~M.}\ \bibnamefont {Manley}},\ }\href {\doibase 10.1103/PhysRevC.86.055203} {\bibfield  {journal} {\bibinfo  {journal} {Phys. Rev. C}\ }\textbf {\bibinfo {volume} {86}},\ \bibinfo {pages} {055203} (\bibinfo {year} {2012})},\ \Eprint {http://arxiv.org/abs/1208.2710} {arXiv:1208.2710 [hep-ph]} \BibitemShut {NoStop}%
\bibitem [{\citenamefont {Koniuk}\ and\ \citenamefont {Isgur}(1980)}]{Koniuk:1979vy}%
  \BibitemOpen
  \bibfield  {author} {\bibinfo {author} {\bibfnamefont {R.}~\bibnamefont {Koniuk}}\ and\ \bibinfo {author} {\bibfnamefont {N.}~\bibnamefont {Isgur}},\ }\href {\doibase 10.1103/PhysRevD.21.1868} {\bibfield  {journal} {\bibinfo  {journal} {Phys. Rev. D}\ }\textbf {\bibinfo {volume} {21}},\ \bibinfo {pages} {1868} (\bibinfo {year} {1980})},\ \bibinfo {note} {[Erratum: Phys.Rev.D 23, 818 (1981)]}\BibitemShut {NoStop}%
\bibitem [{\citenamefont {Bijnens}\ \emph {et~al.}(1999)\citenamefont {Bijnens}, \citenamefont {Colangelo},\ and\ \citenamefont {Ecker}}]{Bijnens:1999sh}%
  \BibitemOpen
  \bibfield  {author} {\bibinfo {author} {\bibfnamefont {J.}~\bibnamefont {Bijnens}}, \bibinfo {author} {\bibfnamefont {G.}~\bibnamefont {Colangelo}}, \ and\ \bibinfo {author} {\bibfnamefont {G.}~\bibnamefont {Ecker}},\ }\href {\doibase 10.1088/1126-6708/1999/02/020} {\bibfield  {journal} {\bibinfo  {journal} {JHEP}\ }\textbf {\bibinfo {volume} {02}},\ \bibinfo {pages} {020} (\bibinfo {year} {1999})},\ \Eprint {http://arxiv.org/abs/hep-ph/9902437} {arXiv:hep-ph/9902437 [hep-ph]} \BibitemShut {NoStop}%
\bibitem [{\citenamefont {Stoica}\ \emph {et~al.}(2011)\citenamefont {Stoica}, \citenamefont {Lutz},\ and\ \citenamefont {Scholten}}]{Stoica:2011cy}%
  \BibitemOpen
  \bibfield  {author} {\bibinfo {author} {\bibfnamefont {S.}~\bibnamefont {Stoica}}, \bibinfo {author} {\bibfnamefont {M.~F.~M.}\ \bibnamefont {Lutz}}, \ and\ \bibinfo {author} {\bibfnamefont {O.}~\bibnamefont {Scholten}},\ }\href {\doibase 10.1103/PhysRevD.84.125001} {\bibfield  {journal} {\bibinfo  {journal} {Phys. Rev.}\ }\textbf {\bibinfo {volume} {D84}},\ \bibinfo {pages} {125001} (\bibinfo {year} {2011})},\ \Eprint {http://arxiv.org/abs/1106.5619} {arXiv:1106.5619 [hep-ph]} \BibitemShut {NoStop}%
\bibitem [{\citenamefont {Karplus}\ \emph {et~al.}(1958)\citenamefont {Karplus}, \citenamefont {Sommerfield},\ and\ \citenamefont {Wichmann}}]{Karplus:1958zz}%
  \BibitemOpen
  \bibfield  {author} {\bibinfo {author} {\bibfnamefont {R.}~\bibnamefont {Karplus}}, \bibinfo {author} {\bibfnamefont {C.~M.}\ \bibnamefont {Sommerfield}}, \ and\ \bibinfo {author} {\bibfnamefont {E.~H.}\ \bibnamefont {Wichmann}},\ }\href {\doibase 10.1103/PhysRev.111.1187} {\bibfield  {journal} {\bibinfo  {journal} {Phys. Rev.}\ }\textbf {\bibinfo {volume} {111}},\ \bibinfo {pages} {1187} (\bibinfo {year} {1958})}\BibitemShut {NoStop}%
\bibitem [{\citenamefont {'t~Hooft}\ and\ \citenamefont {Veltman}(1979)}]{tHooft:1978jhc}%
  \BibitemOpen
  \bibfield  {author} {\bibinfo {author} {\bibfnamefont {G.}~\bibnamefont {'t~Hooft}}\ and\ \bibinfo {author} {\bibfnamefont {M.~J.~G.}\ \bibnamefont {Veltman}},\ }\href {\doibase 10.1016/0550-3213(79)90605-9} {\bibfield  {journal} {\bibinfo  {journal} {Nucl. Phys.}\ }\textbf {\bibinfo {volume} {B153}},\ \bibinfo {pages} {365} (\bibinfo {year} {1979})}\BibitemShut {NoStop}%
\bibitem [{\citenamefont {Bronzan}\ and\ \citenamefont {Kacser}(1963)}]{Bronzan:1963mby}%
  \BibitemOpen
  \bibfield  {author} {\bibinfo {author} {\bibfnamefont {J.~B.}\ \bibnamefont {Bronzan}}\ and\ \bibinfo {author} {\bibfnamefont {C.}~\bibnamefont {Kacser}},\ }\href {\doibase 10.1103/PhysRev.132.2703} {\bibfield  {journal} {\bibinfo  {journal} {Phys. Rev.}\ }\textbf {\bibinfo {volume} {132}},\ \bibinfo {pages} {2703} (\bibinfo {year} {1963})}\BibitemShut {NoStop}%
\bibitem [{cou()}]{couplings}%
  \BibitemOpen
  \href@noop {} {}\bibinfo {howpublished} {\url{https://userweb.jlab.org/~isupov/couplings/}}\BibitemShut {NoStop}%
\bibitem [{\citenamefont {Ramalho}(2019)}]{Ramalho:2019ocp}%
  \BibitemOpen
  \bibfield  {author} {\bibinfo {author} {\bibfnamefont {G.}~\bibnamefont {Ramalho}},\ }\href {\doibase 10.1103/PhysRevD.100.114014} {\bibfield  {journal} {\bibinfo  {journal} {Phys. Rev. D}\ }\textbf {\bibinfo {volume} {100}},\ \bibinfo {pages} {114014} (\bibinfo {year} {2019})},\ \Eprint {http://arxiv.org/abs/1909.00013} {arXiv:1909.00013 [hep-ph]} \BibitemShut {NoStop}%
\bibitem [{\citenamefont {Adamczewski-Musch}\ \emph {et~al.}(2020)\citenamefont {Adamczewski-Musch} \emph {et~al.}}]{HADES:2020kce}%
  \BibitemOpen
  \bibfield  {author} {\bibinfo {author} {\bibfnamefont {J.}~\bibnamefont {Adamczewski-Musch}} \emph {et~al.} (\bibinfo {collaboration} {HADES}),\ }\href {\doibase 10.1103/PhysRevC.102.024001} {\bibfield  {journal} {\bibinfo  {journal} {Phys. Rev. C}\ }\textbf {\bibinfo {volume} {102}},\ \bibinfo {pages} {024001} (\bibinfo {year} {2020})},\ \Eprint {http://arxiv.org/abs/2004.08265} {arXiv:2004.08265 [nucl-ex]} \BibitemShut {NoStop}%
\bibitem [{\citenamefont {Burkert}\ \emph {et~al.}(2023)\citenamefont {Burkert} \emph {et~al.}}]{Burkert:2022bqo}%
  \BibitemOpen
  \bibfield  {author} {\bibinfo {author} {\bibfnamefont {V.}~\bibnamefont {Burkert}} \emph {et~al.},\ }\href {\doibase 10.1016/j.physletb.2023.138070} {\bibfield  {journal} {\bibinfo  {journal} {Phys. Lett. B}\ }\textbf {\bibinfo {volume} {844}},\ \bibinfo {pages} {138070} (\bibinfo {year} {2023})},\ \Eprint {http://arxiv.org/abs/2207.08472} {arXiv:2207.08472 [hep-ph]} \BibitemShut {NoStop}%
\bibitem [{\citenamefont {Gomez~Tejedor}\ \emph {et~al.}(1996)\citenamefont {Gomez~Tejedor}, \citenamefont {Cano},\ and\ \citenamefont {Oset}}]{GomezTejedor:1995kj}%
  \BibitemOpen
  \bibfield  {author} {\bibinfo {author} {\bibfnamefont {J.~A.}\ \bibnamefont {Gomez~Tejedor}}, \bibinfo {author} {\bibfnamefont {F.}~\bibnamefont {Cano}}, \ and\ \bibinfo {author} {\bibfnamefont {E.}~\bibnamefont {Oset}},\ }\href {\doibase 10.1016/0370-2693(96)00404-2} {\bibfield  {journal} {\bibinfo  {journal} {Phys. Lett. B}\ }\textbf {\bibinfo {volume} {379}},\ \bibinfo {pages} {39} (\bibinfo {year} {1996})},\ \Eprint {http://arxiv.org/abs/nucl-th/9510007} {arXiv:nucl-th/9510007} \BibitemShut {NoStop}%
\bibitem [{\citenamefont {Behnke}\ \emph {et~al.}(2013)\citenamefont {Behnke}, \citenamefont {Kr\"oninger}, \citenamefont {Sch\"orner-Sadenius},\ and\ \citenamefont {Schott}}]{Behnke:2013pga}%
  \BibitemOpen
  \bibinfo {editor} {\bibfnamefont {O.}~\bibnamefont {Behnke}}, \bibinfo {editor} {\bibfnamefont {K.}~\bibnamefont {Kr\"oninger}}, \bibinfo {editor} {\bibfnamefont {T.}~\bibnamefont {Sch\"orner-Sadenius}}, \ and\ \bibinfo {editor} {\bibfnamefont {G.}~\bibnamefont {Schott}},\ eds.,\ \href@noop {} {\emph {\bibinfo {title} {{Data analysis in high energy physics}: {A practical guide to statistical methods}}}}\ (\bibinfo  {publisher} {Wiley-VCH},\ \bibinfo {address} {Weinheim, Germany},\ \bibinfo {year} {2013})\BibitemShut {NoStop}%
\bibitem [{\citenamefont {Kubis}(2003)}]{KubisPhD}%
  \BibitemOpen
  \bibfield  {author} {\bibinfo {author} {\bibfnamefont {B.}~\bibnamefont {Kubis}},\ }\emph {\bibinfo {title} {Strong interactions and electromagnetism in low-energy hadron physics}},\ \href@noop {} {Ph.D. thesis},\ \bibinfo  {school} {Forschungszentrum J\"ulich} (\bibinfo {year} {2003})\BibitemShut {NoStop}%
\bibitem [{fai()}]{fair-homepage}%
  \BibitemOpen
  \href@noop {} {}\bibinfo {howpublished} {\url{http://www.fair-center.eu/}}\BibitemShut {NoStop}%
\bibitem [{\citenamefont {An}(2023)}]{An:2023dqx}%
  \BibitemOpen
  \bibfield  {author} {\bibinfo {author} {\bibfnamefont {D.}~\bibnamefont {An}},\ }in\ \href@noop {} {\emph {\bibinfo {booktitle} {{17th International Workshop on Meson Physics}}}}\ (\bibinfo {year} {2023})\ \Eprint {http://arxiv.org/abs/2309.15992} {arXiv:2309.15992 [hep-ph]} \BibitemShut {NoStop}%
\bibitem [{\citenamefont {Rarita}\ and\ \citenamefont {Schwinger}(1941)}]{Rarita:1941mf}%
  \BibitemOpen
  \bibfield  {author} {\bibinfo {author} {\bibfnamefont {W.}~\bibnamefont {Rarita}}\ and\ \bibinfo {author} {\bibfnamefont {J.}~\bibnamefont {Schwinger}},\ }\href {\doibase 10.1103/PhysRev.60.61} {\bibfield  {journal} {\bibinfo  {journal} {Phys. Rev.}\ }\textbf {\bibinfo {volume} {60}},\ \bibinfo {pages} {61} (\bibinfo {year} {1941})}\BibitemShut {NoStop}%
\bibitem [{\citenamefont {de~Jong}\ and\ \citenamefont {Malfliet}(1992)}]{deJong:1992wm}%
  \BibitemOpen
  \bibfield  {author} {\bibinfo {author} {\bibfnamefont {F.}~\bibnamefont {de~Jong}}\ and\ \bibinfo {author} {\bibfnamefont {R.}~\bibnamefont {Malfliet}},\ }\href {\doibase 10.1103/PhysRevC.46.2567} {\bibfield  {journal} {\bibinfo  {journal} {Phys. Rev.}\ }\textbf {\bibinfo {volume} {C46}},\ \bibinfo {pages} {2567} (\bibinfo {year} {1992})}\BibitemShut {NoStop}%
\bibitem [{\citenamefont {Penner}(1999)}]{Penner:1999jia}%
  \BibitemOpen
  \bibfield  {author} {\bibinfo {author} {\bibfnamefont {G.}~\bibnamefont {Penner}},\ }\emph {\bibinfo {title} {{Vector Meson Production and Nucleon Resonance Analysis in a Coupled Channel Approach}}},\ \href@noop {} {Ph.D. thesis},\ \bibinfo  {school} {Giessen U.} (\bibinfo {year} {1999})\BibitemShut {NoStop}%
\bibitem [{\citenamefont {Penner}\ and\ \citenamefont {Mosel}(2002)}]{Penner:2002md}%
  \BibitemOpen
  \bibfield  {author} {\bibinfo {author} {\bibfnamefont {G.}~\bibnamefont {Penner}}\ and\ \bibinfo {author} {\bibfnamefont {U.}~\bibnamefont {Mosel}},\ }\href {\doibase 10.1103/PhysRevC.66.055212} {\bibfield  {journal} {\bibinfo  {journal} {Phys. Rev. C}\ }\textbf {\bibinfo {volume} {66}},\ \bibinfo {pages} {055212} (\bibinfo {year} {2002})},\ \Eprint {http://arxiv.org/abs/nucl-th/0207069} {arXiv:nucl-th/0207069} \BibitemShut {NoStop}%
\bibitem [{\citenamefont {Dobado}\ and\ \citenamefont {Pelaez}(1997)}]{Dobado:1996ps}%
  \BibitemOpen
  \bibfield  {author} {\bibinfo {author} {\bibfnamefont {A.}~\bibnamefont {Dobado}}\ and\ \bibinfo {author} {\bibfnamefont {J.~R.}\ \bibnamefont {Pelaez}},\ }\href {\doibase 10.1103/PhysRevD.56.3057} {\bibfield  {journal} {\bibinfo  {journal} {Phys. Rev. D}\ }\textbf {\bibinfo {volume} {56}},\ \bibinfo {pages} {3057} (\bibinfo {year} {1997})},\ \Eprint {http://arxiv.org/abs/hep-ph/9604416} {arXiv:hep-ph/9604416} \BibitemShut {NoStop}%
\bibitem [{\citenamefont {Dax}\ \emph {et~al.}(2018)\citenamefont {Dax}, \citenamefont {Isken},\ and\ \citenamefont {Kubis}}]{Dax:2018rvs}%
  \BibitemOpen
  \bibfield  {author} {\bibinfo {author} {\bibfnamefont {M.}~\bibnamefont {Dax}}, \bibinfo {author} {\bibfnamefont {T.}~\bibnamefont {Isken}}, \ and\ \bibinfo {author} {\bibfnamefont {B.}~\bibnamefont {Kubis}},\ }\href {\doibase 10.1140/epjc/s10052-018-6346-3} {\bibfield  {journal} {\bibinfo  {journal} {Eur. Phys. J.}\ }\textbf {\bibinfo {volume} {C78}},\ \bibinfo {pages} {859} (\bibinfo {year} {2018})},\ \Eprint {http://arxiv.org/abs/1808.08957} {arXiv:1808.08957 [hep-ph]} \BibitemShut {NoStop}%
\bibitem [{\citenamefont {Chung}(1971)}]{Chung:1971ri}%
  \BibitemOpen
  \bibfield  {author} {\bibinfo {author} {\bibfnamefont {S.~U.}\ \bibnamefont {Chung}},\ }\href {\doibase 10.5170/CERN-1971-008} {\  (\bibinfo {year} {1971}),\ 10.5170/CERN-1971-008}\BibitemShut {NoStop}%
\bibitem [{\citenamefont {Chung}(2008)}]{Chung:1102240}%
  \BibitemOpen
  \bibfield  {author} {\bibinfo {author} {\bibfnamefont {S.~U.}\ \bibnamefont {Chung}},\ }\href {https://cds.cern.ch/record/1102240} {\emph {\bibinfo {title} {{Spin formalisms; 1st updated version}}}}\ (\bibinfo  {publisher} {Brookhaven Nat. Lab.},\ \bibinfo {address} {Upton, NY},\ \bibinfo {year} {2008})\ \bibinfo {note} {this report is an updated version of CERN-1971-008}\BibitemShut {NoStop}%
\end{thebibliography}%
\bibliographystyle{apsrev4-1}

\end{document}